\documentclass[journal,onecolumn,12pt, draftcls]{IEEEtran}
\usepackage{enumitem}   
\usepackage{algorithm}
\usepackage{algorithmic}

\usepackage[normalem]{ulem}
\useunder{\uline}{\ul}{}

\usepackage{cite}

 \usepackage[table,xcdraw]{xcolor}	

\ifCLASSINFOpdf
   \usepackage[pdftex]{graphicx}
   \graphicspath{{../pdf/}{../jpeg/}}
   \DeclareGraphicsExtensions{.pdf,.jpeg,.png}
\else
   \usepackage[dvips]{graphicx}
   \graphicspath{{../eps/}}
   \DeclareGraphicsExtensions{.eps}
\fi

\usepackage{amsmath}
\newtheorem{theorem}{Theorem}
\newtheorem{lemma}{Lemma}
\newtheorem{proposition}{Proposition}

\newtheorem{definition}{Definition}
\newtheorem{example}{Example}

\usepackage{algorithm}
\usepackage{algorithmic}

\ifCLASSOPTIONcompsoc
  \usepackage[caption=false,font=normalsize,labelfont=sf,textfont=sf]{subfig}
\else
  \usepackage[caption=false,font=footnotesize]{subfig}
\fi

\begin{document}

\title{New Two-Stage Automorphism Group Decoders for Cyclic Codes in the Erasure Channel }

\author{Chanki~Kim,~\IEEEmembership{Student Member,~IEEE} and
        Jong-Seon~No,~\IEEEmembership{Fellow,~IEEE}}

\maketitle

\begin{abstract}
Recently, error correcting codes in the erasure channel have drawn great attention for various applications such as distributed storage systems and wireless sensor networks, but many of their decoding algorithms are not practical because they have higher decoding complexity and longer delay. Thus, the automorphism group decoder (AGD) for cyclic codes in the erasure channel was introduced, which has good erasure decoding performance with low decoding complexity. In this paper, we propose new two-stage AGDs (TS-AGDs) for cyclic codes in the erasure channel by modifying the parity check matrix and introducing the preprocessing stage to the AGD scheme. The proposed TS-AGD has been analyzed for the perfect codes, BCH codes, and maximum distance separable (MDS) codes. Through numerical analysis, it is shown that the proposed decoding algorithm has good erasure decoding performance with lower decoding complexity and delay than the conventional AGD. For some cyclic codes, it is shown that the proposed TS-AGD achieves the perfect decoding in the erasure channel, that is, the same decoding performance as the maximum likelihood (ML) decoder. For MDS codes, TS-AGDs with the expanded parity check matrix and the submatrix inversion are also proposed and analyzed.

\end{abstract}


\IEEEpeerreviewmaketitle

\section{Introduction}
\IEEEPARstart{R}{esearch} on error correcting codes in the erasure channel is one of the major subjects in information theory. Erasure channel is a typical channel model for wireless sensor networks and distributed storage systems, where the locations of symbol errors are known.
  
Algebraic codes have a long history from Hamming codes to algebraic geometry codes. The decoders of algebraic codes are designed using the mathematical properties of the codes, which have difficulty in implementing practical decoders. However, lots of research works for their decoding algorithms have been done to reduce the decoding complexity and delay. In cyclic codes, one-step majority decoding \cite{OSMD} and permutation decoding\cite{PD} schemes are exemplary methods which can be practically implemented using their cyclic property in the error channel. However, these decoding schemes are applicable only to the limited parameters of the error correcting codes.
  
An iterative decoder can be one of the solution as an implementable decoder and thus, the iterative decoding algorithms and error correcting codes with iterative decoder such as turbo and low-density parity-check (LDPC) codes have been widely studied. Iterative decoders have various implementation methods according to the error correcting codes, their decoding performance, and complexity. One of these is the belief propagation decoder for LDPC codes, which is based on log-likelihood ratio (LLR)-based computation and which shows Shannon capacity-approaching decoding performance. In addition, there have been lots of researches to apply iterative decoding to algebraic codes in error channels. In \cite{RS1} and \cite{RS2}, the iterative decoding of Reed-Solomon (RS) codes with sparse parity check matrix and belief-propagation decoding algorithm is proposed. Iterative erasure decoder (IED) of algebraic codes \cite{Hollmann} has also been studied. However, IED has inherently inferior decoding performance compared to the maximum likelihood (ML) decoder and the gap between the decoding performances becomes larger in the algebraic codes, because the sparseness of their parity check matrices is not guaranteed contrary to the LDPC codes. Thus a possible solution for algebraic codes is to modify the structure of the decoder in the erasure channel.

Recently, one approach to overcome the inferior decoding performance of IED for the algebraic codes in the erasure channel was proposed, called the automorphism group decoder (AGD) for cyclic codes \cite{Hehn}. AGD uses the permutation of the automorphism group in the middle of the IED procedure. For cyclic codes, the permutation operation can be substituted by the cyclic shift operation for codewords, which are also codewords. In fact, many similar concepts have been proposed for cyclic LDPC codes in the error channel such as multiple-bases belief-propagation (MBBP) \cite{Hehn2} and revolving iterative decoding (RID) \cite{Chen}, \cite{Liu}. It was shown that for some cyclic codes, AGD improves the decoding performance but it requires higher decoding complexity and delay, because the average number of iterations for this decoding scheme is increased by the cyclic shift operation.

In this paper, new AGD algorithms for the cyclic codes in erasure channels are proposed to improve the decoding performance and reduce the decoding complexity and delay. First, the parity check matrix of the $(n,k)$ cyclic code is modified such that some of the $(n-k)$-tuple column vectors in the parity check matrix are standard basis vectors in the appropriate column indices and Hamming weight of the row vectors in the parity check matrix becomes as low as possible. Then, the proposed decoding process is done in two decoding stages, referred to as a two-stage AGD (TS-AGD), that is, the first decoding stage finds the cyclic shift values of the received codeword for the successful erasure decoding while in the second decoding stage, the erasure decoding process is done for the received codewords cyclically shifted by the cyclic shift values found in the first decoding stage. The numerical analysis shows that the proposed TS-AGD algorithm outperfoms the conventional AGD algorithm and further it reduces decoding complexity and delay. For some cyclic codes, it is shown that the proposed TS-AGD achieves the perfect decoding, showing the decoding performance identical to that of the ML decoding.

Generally, each check equation in the IED has its own erasure decoding capability. For some algebraic codes, it is known that $n-k$ check equations are not sufficient to achieve good decoding performance. Thus, stopping redundancy was proposed \cite{SR_first}, which increases the number of check equations and it guarantees the successful decoding for all the erasure symbols up to $d_{min}-1$. However, stopping redundancy implausibly increases the decoding complexity. Stopping redundancy has been studied for several algebraic codes such as maximum distance separable (MDS) codes \cite{SR1}, Reed-Muller codes \cite{SR2}, and algebraic geometry codes\cite{SR3}.

MDS codes are the algebraic codes which satisfy the Singleton bound, that is, for $(n,k,d)$ MDS codes, we have
\begin{displaymath}
d \le n-k+1.
\end{displaymath}
It is known that MDS codes have the optimal decoding performance in the erasure channel. The RS code is a well-known cyclic MDS code that has been widely applied to compact disc, satellite communication, and distributed storage systems. The decoding complexity of RS codes is high in general because the decoding process requires lots of computations in the finite field.

We also propose another two-stage decoding scheme to decode cyclic MDS codes, by modifying the TS-AGD by stopping redundancy. The modified TS-AGD with stopping redundancy for the MDS codes has the same decoding performance as that of the ML decoder but the decoding complexity of the proposed decoder is dramatically reduced compared to that of the ML decoder. Further, several lower bounds on the stopping redundancy for the perfect decoding of cyclic MDS codes have been derived. In order to further improve the performance of the proposed decoder, the proposed TS-AGD with submatrix inversion of the parity check matrix is also considered.

This paper is organized as follows. In Section \ref{Pre}, AGD is reviewed and compared to IED. In Section \ref{Prop}, the proposed decoding algorithm for the binary cyclic codes in the erasure channel is introduced by modifying the parity check matrix and the AGD algorithm, called TS-AGD. For some cyclic codes, the proposed TS-AGD achieves the perfect decoding in the erasure channel. The numerical analysis of the performance of the proposed decoding algorithm is also given. In Section \ref{Prop_MDS}, the proposed TS-AGD algorithms are modified for the cyclic MDS codes by using stopping redundancy and submatrix inversion. Then, several lower bounds on the stopping redundancy and submatrix inversion for the perfect decoding are derived. Finally, the conclusion is given in Section \ref{Conc}.

\section{Preliminary}
\label{Pre}
In this section, the decoding procedures of IED and AGD are explained and compared and several definitions are presented.

\subsection{IED and AGD}
An $(n,k)$ error correcting code has an $(n-k) \times n$ parity check matrix $H$, which can be represented by a bipartite graph $\cal{G}$ with $n$ variable nodes and $n-k$ check nodes. Let $V$ and $U$ be sets of variable nodes and check nodes and let $d_v$ and $d_c$ be the degrees of variable nodes and check nodes, respectively. The bipartite graph is then denoted by $\cal{G}$$=$$(V,U,H)$. In the erasure channel, the variable nodes have two different states, i.e., erasure and non-erasure states, while the check nodes have three states, decodable, non-decodable, and non-erasure states. The decoding procedure of IED consists of several iterations, where each iteration performs check node update (CNU) and variable node update (VNU) operations sequentially. It is assumed that the check nodes deriving the decoding procedure are operated in a parallel way, known as flooding decoding.

\begin{figure*}[!t]
\centering
\subfloat[CNU operation]{\includegraphics[width=4.5in]{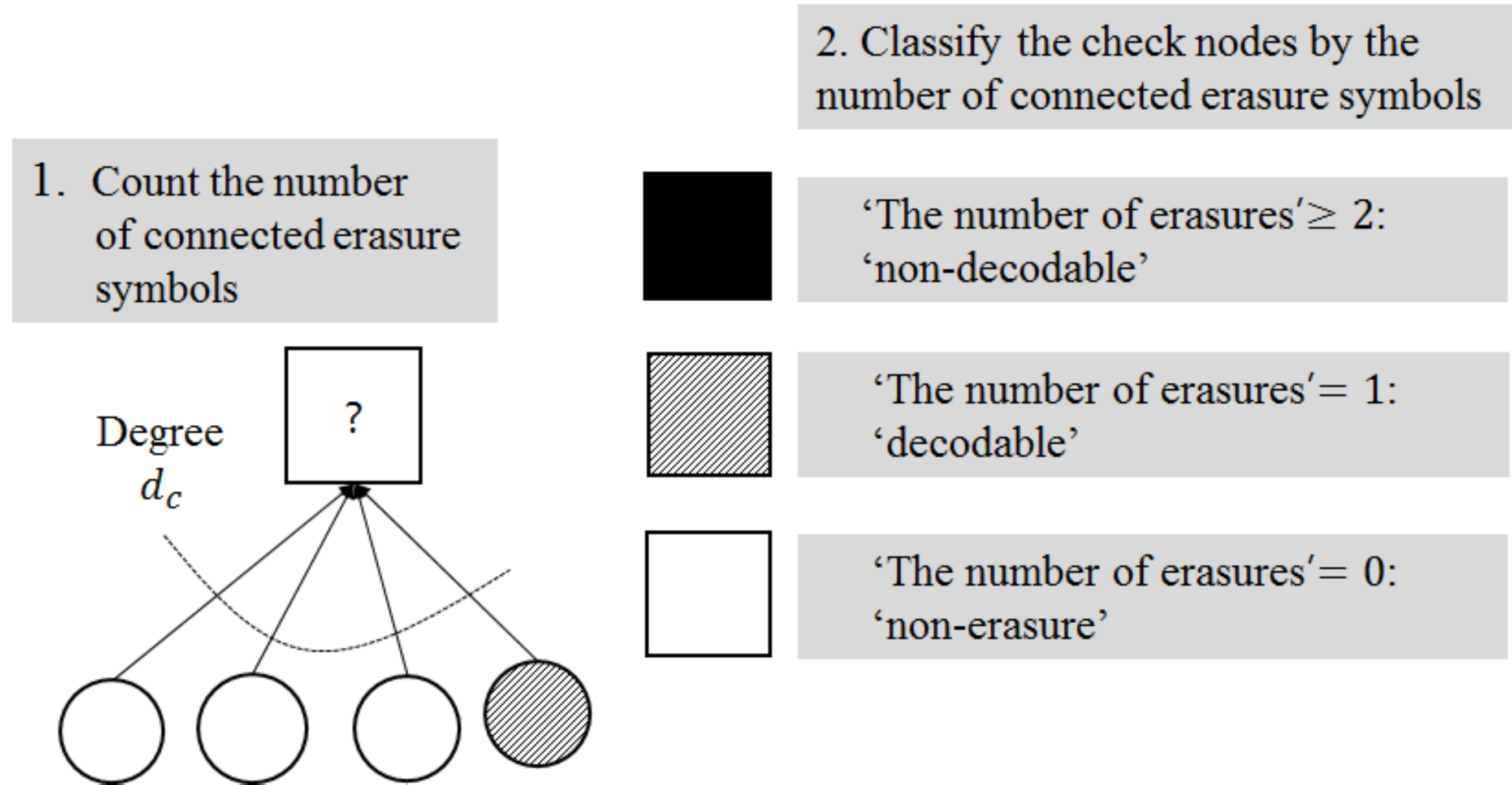}%
\label{CUO}}
\hfil
\subfloat[VNU operation]{\includegraphics[width=5in]{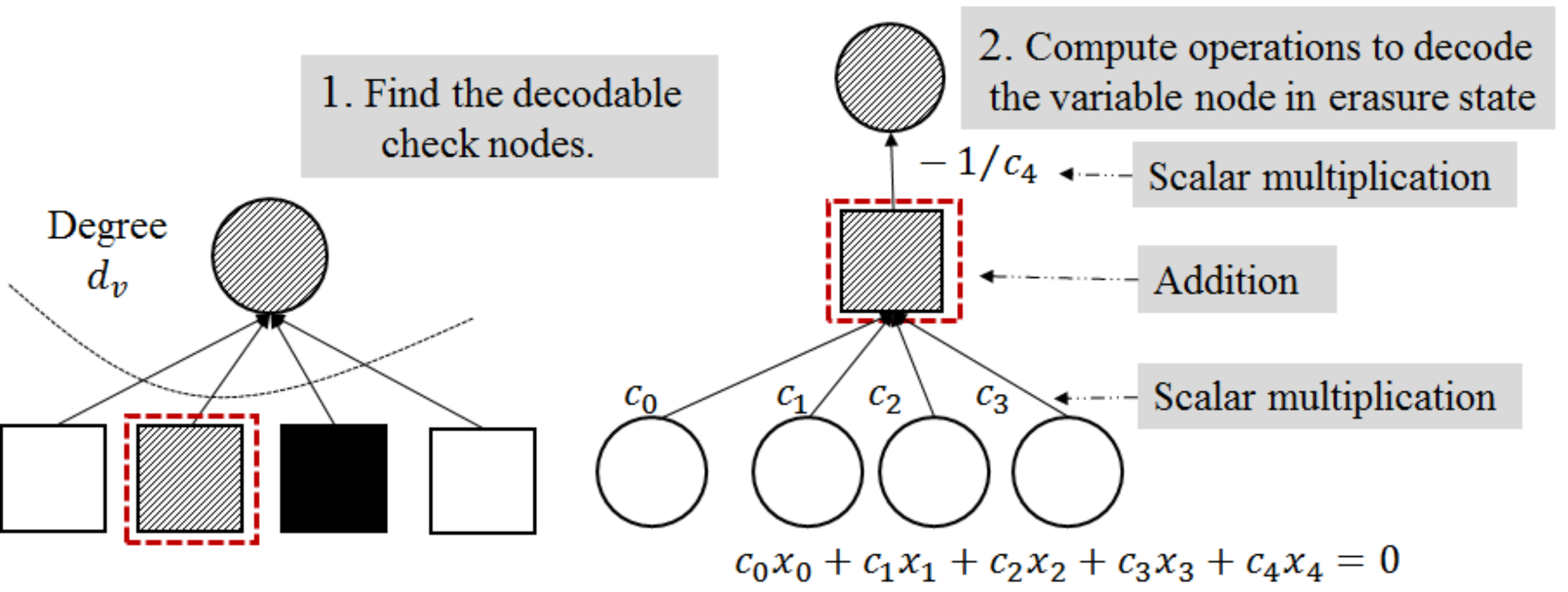}%
\label{VUO}}
\caption{Node operations of IED.}
\label{IED_Op}
\end{figure*}

The CNU operation is the procedure that each check node finds its state by counting the number of the erasure states of the variable nodes connected to itself. A decodable state of a check node is declared when the number of the connected variable nodes in the erasure state is 1, because check equation of IED can decode the variable node in the erasure state. If the check node has the variable nodes in the erasure state larger than or equal to 2, a non-decodable state is declared for the check node. The check nodes with no connected variable nodes in the erasure state are called non-erasure states. In this way, all of the check nodes simultaneously update their states and IED proceeds to the VNU operation. The VNU operation is a procedure by which variable nodes in the erasure state are decoded using connected decodable check nodes. Fig. \ref{IED_Op} shows the CNU and VNU operations. Note that after the decodable check nodes decode the corresponding variable nodes in the erasure state, then state of the check node is changed to the non-erasure state, which will be removed.

AGD can be applied to cyclic codes, where AGD consists of the repeated IED and cyclic shift operations of the received codewords. That is, if there is no decodable check node, then the received codeword is cyclically shifted until decodable check nodes are found. If it is found, the IED algorithm is  repeatedly applied to the cyclically shifted received codewords.

It is known that the cyclic shift operation is easy to implement with negligible complexity and delay. In the AGD, IED should be performed for each cyclically shifted received codeword until the decoding is successful or the number of cyclic shifts is equal to the length of codeword. Although the decoding complexity and delay of the AGD are much higher than those of the IED, the decoding performance of the AGD is much better than that of the IED.

\subsection{Some Definitions}
In this subsection, several mathematical notations and abbreviations are defined. First, $\text{wt}(v)$ denotes Hamming weight of vector $v$ and $\text{supp}(v)$ denotes the set of indices of the nonzero elements in $v$. The $i$-th standard basis vector $u_i$ is the basis vector, where the $i$-th element of $u_i$ is equal to 1 and the other elements are equal to 0. 

There are several definitions of a binary sequence as follows. Let $s_{D}(t)$ denote a characteristic sequence of index set $D$ such that $s_D(t)=1$ if $t\in D$ and otherwise, $s_D(t)=0$. Two binary sequences are frequently used in this paper, that is, the erasure sequence and the parity check sequence defined as follows. 

\begin{definition}[Erasure sequence]
Erasure sequence $s_{e}(t)$ is defined as a characteristic sequence of the erasure set $S_e$, which is the set of indices of erasure symbols in the received codeword transmitted through the erasure channel.
\end{definition}

\begin{definition}[Parity check sequence]
\label{parity_check_sequence}
Parity check sequence $s_{p}(t)$ of the $ (n-k) \times n$ parity check matrix $H$ of the $(n,k)$ cyclic code is a binary sequence of length $n$ defined as
\begin{displaymath}
s_{p}(t)= \left\{ \begin{array}{ll}
0, & \textrm{if the \textit{t}-th column of \textit{H} is a standard basis vector}\\
1, & \textrm{otherwise}
\end{array} \right.
\end{displaymath}
where the standard basis vector means that only one element of the vector is 1 and the others are zero. Furthermore, let $S_p$ denote the support set of $1-s_{p}(t)$, i.e., the set of column indices of the standard basis vectors.
\end{definition}

For column indices of the parity check matrix $H$, the elements of $S_p$ are called standard basis indices and otherwise, non-standard basis indices. Thus, the number of 1's in a period of $s_p(t)$ is larger than or equal to $k$. The Hamming cross-correlation of two binary $\{0,1\}$ sequences, $s_e(t)$ and $s_p(t)$, is defined as
\begin{displaymath}
R_H(\tau)=\sum_{t=0}^{n-1}{s_e(t) s_p (t+\tau)}
\end{displaymath}
where $R_H(\tau)$ takes values in $\{0,1,2,...,n-k\}$. The stopping redundancy for the parity check matrix is defined as follows.

\begin{definition}[Stopping redundancy $\rho$ \cite{SR1}]
Stopping redundancy $\rho$ of the code $C$ is the minimum number of check equations that the decoder can correct all of the erasure patterns with erasure symbols less than or equal to $d-1$, where $d$ is the minimum distance of the code $C$.
\end{definition}

A mask is a useful notation to represent the parity check matrix of MDS codes because only the location of nonzero values in the parity check matrix is of our interests, which is defined as follows.

\begin{definition}[Mask]
Mask $A$ is an $(n-k)\times n $ binary matrix whose element $a_{i,j}$ is 1 if the $(i,j)$ element of matrix $H$ is nonzero and otherwise, zero.
\end{definition}

It is known that the ML decoding performance in the erasure channel is best, that is, a practical decoder in the erasure channel can have the same or inferior erasure decoding performance to that of the ML decoder. At this point, the perfect decoding is defined as follows.
\begin{definition}[Perfect decoding]
It is called perfect decoding in the erasure channel if its erasure decoding performance is the same as that of ML decoder.
\end{definition}

In general, perfect decoding is not common because it is rarely possible to show it. In this paper, the perfect decoding is shown by checking all cases of erasure patterns for some of the cyclic codes with small values of $n$ and $k$.

There is an example of $(8,4)$ RS code, where an ML decoder can correct any four erasure symbols regardless of their locations. 

\begin{figure}[!t]
\centering
\includegraphics[width=3.3in]{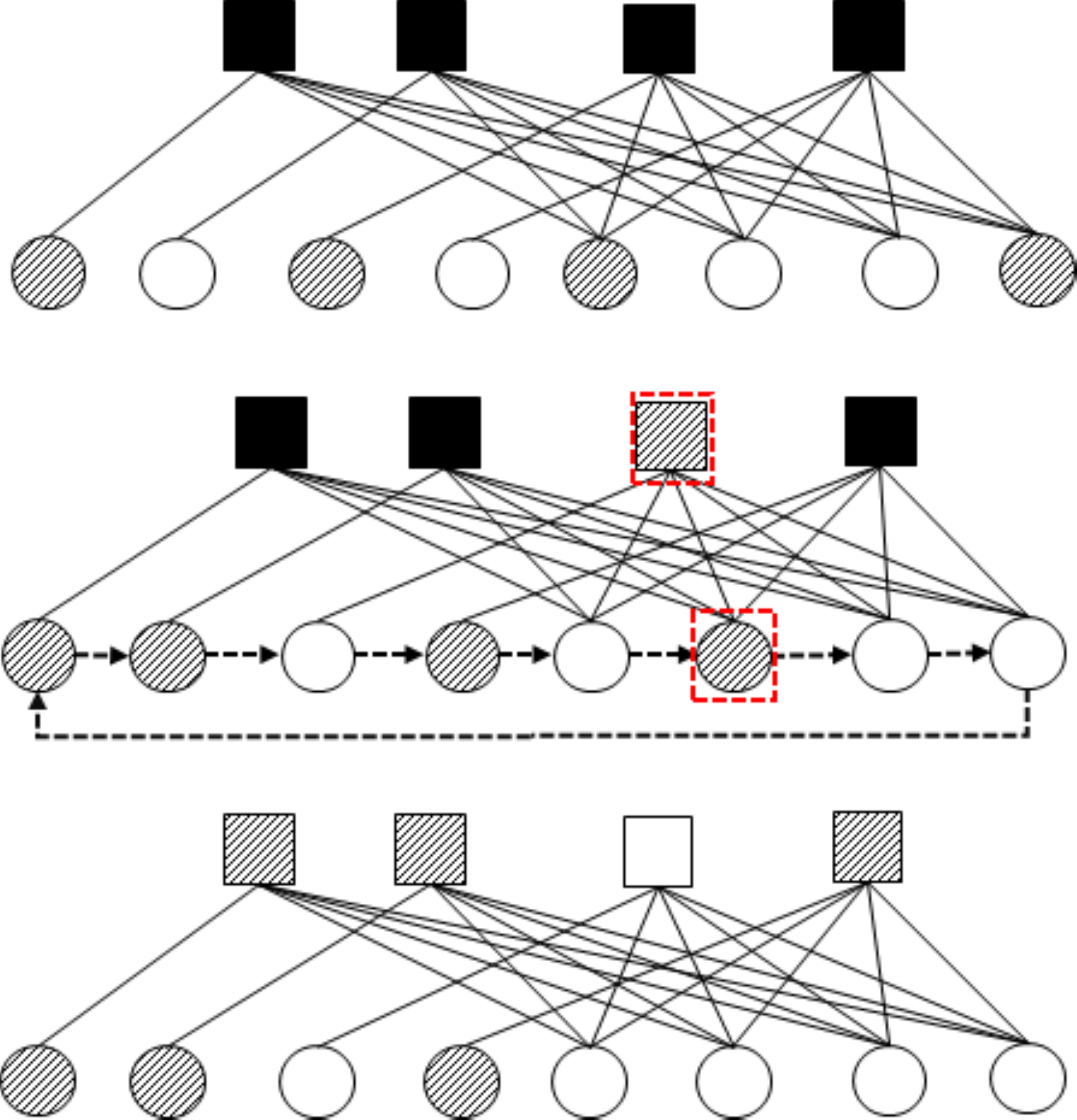}
\caption{AGD for $(8,4)$ RS code in Example 1.}
\label{Fig_Exa_1}
\end{figure}

\begin{example}[AGD]
Suppose that an (8,4) RS code is defined in $F_{17}$ and its generator polynomial is given as
\begin{displaymath}
g(x)=\prod_{i=0}^{k-1}(x-2^i)=x^4+2x^3+2x^2+16x+13.
\end{displaymath}
Then, the corresponding parity check polynomial is obtained as\\
\begin{displaymath}
h(x)=\frac{(x^8-1)}{g(x)} =x^4+15x^3+2x^2+x+13
\end{displaymath}
and the systematic parity check matrix can be constructed from $h(x)$ as

\begin{displaymath}
\mathbf{H} =
\left( \begin{array}{cccccccc}
1&0&0&0&4&9&8&4\\0&1&0&0&1&2&11&9\\0&0&1&0&15&5&15&9\\0&0&0&1&15&2&1&13
\end{array} \right).
\end{displaymath}

The parity check matrix $H$ can be described by a bipartite graph as shown in Fig. \ref{Fig_Exa_1}. Circles represent variable nodes and squares represent check nodes and the edge between the $i$-th circle and the $j$-th square indicates that the $(i,j)$ element of $H$ is nonzero. In Fig. \ref{Fig_Exa_1}, the circles with dashed line are variable nodes in erasure states, where there are four erasure symbols. Then, the erasure sequence $s_{e}(t)$ is $(1,0,1,0,1,0,0,1)$ and the parity check sequence $s_{p}(t)$ is $(0,0,0,0,1,1,1,1)$.

The AGD procedure is described in Fig. \ref{Fig_Exa_1}. In the first bipartite graph Fig. \ref{Fig_Exa_1}, the decoder performs CNU operations and confirms that there is no check node in a decodable state. The IED declares a decoding failure, whereas the AGD proceeds to the next decoding procedure by cyclic shifting the received codeword. In the second bipartite graph of Fig. \ref{Fig_Exa_1}, one right cyclic shift operation for the received codeword and CNU operation are done. After four CNU operations, it is found that one check node is in a decodable state, which can proceed to a VNU operation to correct the fourth erasure symbol. In the third bipartite graph of Fig. \ref{Fig_Exa_1}, after three CNU operations are performed, the decoder finds that the other check nodes are all in decodable states, which can proceed to three VNU operations to correct the three remaining erasure symbols and then the decoding procedure is completed. In the above decoding procedure for AGD, three IED operations and one cyclic shift are performed.

\end{example}

Example 1 shows that the AGD has superior performance to the IED. However, successive IEDs are needed for each cyclic shift operation, which is the main issue of the decoding complexity and delay in the AGD. This example shows that it is needed to select cyclic shift values and construct parity check matrix carefully to reduce the number of iterations. In the next section, we propose a new decoding scheme for the binary cyclic codes, which reduces decoding complexity and delay without sacrificing the decoding performance.

\section{A New Two-Stage AGD With Reduced Complexity and Delay}
\label{Prop}
Decoding complexity and delay can be reduced in the AGD if the decoder knows the cyclic shift values of the received codeword for successful decoding. In this section, we propose a modification method for the parity check matrix and a new two-stage decoding algorithm, and the result of a numerical analysis for the proposed decoding algorithm is discussed.

\subsection{Modification of the Parity Check Matrix}

First, we briefly review the parity check matrix of cyclic codes proposed by Hehn \cite{Hehn}, where a cyclic orbit generator (cog) and the cog family are used. A cog is cyclically distinguishable codeword of a dual code with minimum Hamming weight, which can be used as a row of the parity check matrix and the cog family is a set of cogs which share the same Hamming autocorrelation property. Hehn \cite{Hehn} proposed a construction method of the parity check matrix, in which the rows are composed of cogs in the cog family as many as possible. 
 
First, we propose a method to modify the parity check matrix for the proposed two-stage decoding algorithm because the decoding performance of the proposed two-stage decoding algorithm depends on the structure of the parity check matrix. The following criteria are used for the modification of the parity check matrix using Definition \ref{parity_check_sequence}.

\begin{enumerate}
\item Modify the parity check matrix such that as many of its column vectors as possible are the standard basis vectors.
\item The parity check sequence of the parity check matrix has out-of-phase Hamming autocorrelation values as low as possible.
\item Each row of the parity check matrix has as low Hamming weight as possible.
\end{enumerate}
In fact, the best criteria for the parity check matrix of $(n,k)$ cyclic codes can be described as:
\begin{enumerate}
\item $n-k$ columns of the parity check matrix are standard basis vectors.
\item All out-of-phase Hamming autocorrelation values of the parity check sequence of the parity check matrix are equal.
\item The Hamming weights of all rows of the parity check matrix is equal to the minimum Hamming weight of its dual code.
\end{enumerate}
It is easy to check that in order for the parity check sequences to satisfy the second criterion, they should be the characteristic sequences of cyclic difference sets with parameters $(n,k,\lambda)$ for $(n,k)$ cyclic codes, if their parameters are allowed for the cyclic difference sets. It is known that some cyclic codes satisfy the above best criteria. The other criteria can be compromised if one criterion cannot be achieved due to the other criteria. The reason why we propose the above criteria for the modification of the parity check matrix will be explained in the next subsection.

\subsection{A New Two-Stage AGD}

Using AGD and IED algorithms, we propose a new two-stage AGD of $(n,k)$ cyclic codes in the erasure channel as follows.
\subsubsection{Preprocessing Stage (First Decoding Stage)}
Find a $\{0,1\}$ parity check sequence $s_p(t)$ of length $n$ from the parity check matrix $H$ of an $(n,k)$ cyclic code. Find a $\{0,1\}$ erasure sequence $s_e(t)$ of length $n$ from the received codeword $\mathbf{r}=(r_0,r_1,...,r_{n-1})$ transmitted through the erasure channel. Then, calculate the Hamming cross-correlation as
\begin{displaymath}
R_H(\tau)=\sum_{t=0}^{n-1}{s_p(t)s_e(t+\tau)},\text{ } 0\le\tau\le n-1.
\end{displaymath}
Clearly, $R_H(\tau)$ takes values of the nonnegative integers less than or equal to $\min\{|S_e|,n-|S_p|\}$ because $|S_e|$ is the number of erasure symbols and $n-|S_p|$ is the number of non-standard basis vectors of the parity check matrix. It can be assumed that the decoding complexity of the preprocessing stage for each $\tau$ is analogous to the CNU of one check node. If there exists $\tau$ such that $R_H(\tau)=0$, then proceed to the second decoding stage. If not found, cyclically shift the received codeword and proceed to the second decoding stage for $\mathbf{r}^{(\tau)}$ in the order of $\tau$'s such that values of $R_H(\tau)$ are increasing.
\subsubsection{IED Decoding Stage (Second Decoding Stage)}
In the second decoding stage, the IED algorithm is used for the decoding of the cyclically shifted received codeword according to the values of $R_H(\tau)$. Recall that $S_p$ is the support set of $1-s_p(t)$. Let $\mathbf{r}^{(\tau)}=(r_{n-\tau},r_{n-\tau+1},...,r_{n-\tau-1})$ be a received codeword cyclically shifted by $\tau$, where erasure symbols are located in the indices in $S_e^{(\tau)}=\{t | s_e(t-\tau)=1, 0\le t \le n-1\}$.

\begin{figure}[!t]
\centering
\includegraphics[width=6in]{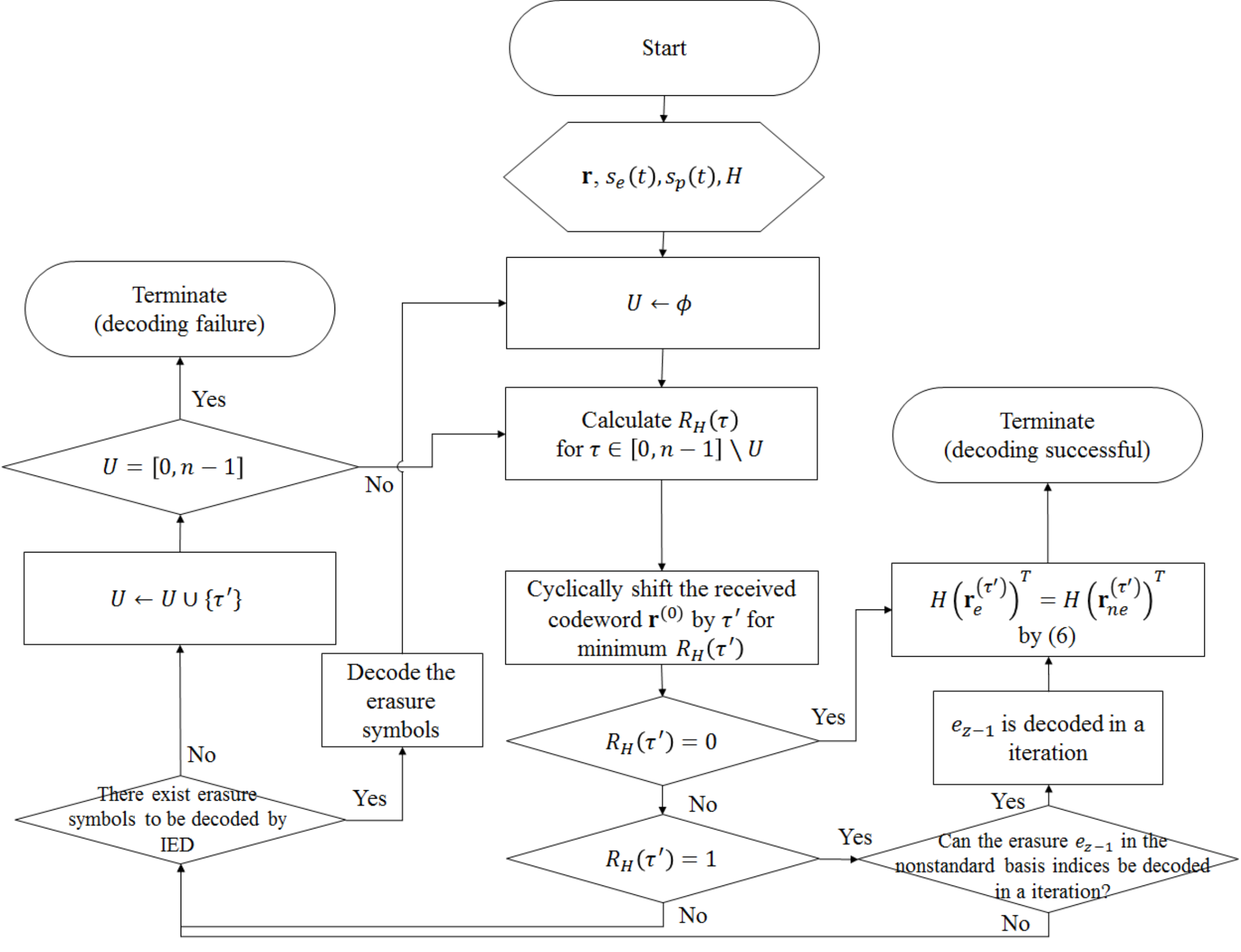}
\caption{Flowchart of the TS-AGD algorithm.}
\label{TS_AGD_FlowChart_lb}
\end{figure}

\begin{enumerate}[label=(\roman*)]
\item 
\begin{figure}[!t]
\centering
\includegraphics[width=5in]{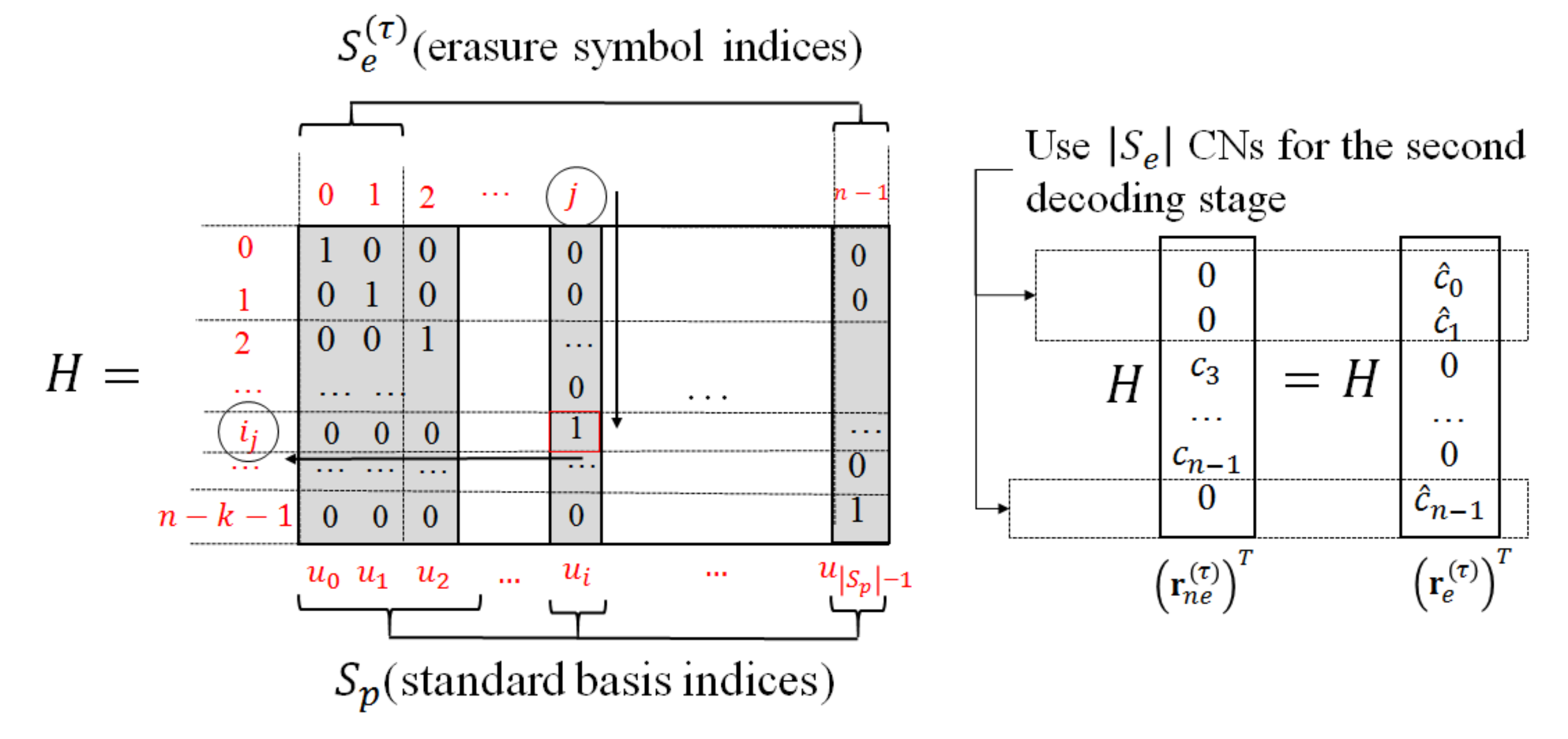}
\caption{The second stage decoding procedure of the TS-AGD of $\tau$ such that $R_H{(\tau)}=0.$}
\label{TS_AGD_Tau_0_case}
\end{figure}
For $\tau$ such that $R_H(\tau)=0$: It is clear that $S_{e}^{(\tau)}\subseteq S_{p}$, that is, all of the erasure symbols in $\mathbf{r}^{(\tau)}$ are located in the indices of standard basis vectors. Note that the $i$-th element of the received vector $\mathbf{r}$ is expressed as the transmitted symbol $c_i$ for a non-erasure symbol and $\hat{c}_i$ for an erasure symbol. Suppose that $\mathbf{r}^{(\tau)}$ can be split into two $n$-tuple vectors as
\begin{displaymath}
\mathbf{r}^{(\tau)}=\mathbf{r}_{e}^{(\tau)}+\mathbf{r}_{ne}^{(\tau)}
\end{displaymath}
where the $j$-th component of $\mathbf{r}_e^{(\tau)}$ is denoted as $\hat{c_j}$ for $j\in S_e^{(\tau)}$ and otherwise, 0 and the $j$-th component of $\mathbf{r}_{ne}^{(\tau)}$ is equal to the $j$-th component of $\mathbf{r}^{(\tau)}$ for $j\notin S_{e}^{(\tau)}$ and otherwise, 0. In general, the syndrome vector should be zero as
\begin{displaymath}
S=H {(\mathbf{r}^{(\tau)})}^T=H {(\mathbf{r}_{e}^{(\tau)})}^T+ H {(\mathbf{r}_{ne}^{(\tau)})}^T=0
\end{displaymath}
and thus
\begin{displaymath}
\label{Corr_Eqn}
H {(\mathbf{r}_{e}^{(\tau)})}^T = H {(\mathbf{r}_{ne}^{(\tau)})}^T.
\end{displaymath}
If the $j$-th column vector of $H$ is the $i$-th standard basis vector $u_i$, $\hat{c_j}$ is equal to the $i$-th component of $H {(\mathbf{r}_{ne}^{(\tau)})}^T$ because $R_H(\tau)=0$. Therefore, each $j$-th column for $j\in S_e \subset S_p$ has a different standard basis vector $u_i$. In this case, we can recover all of the erasure symbols by $H{(\mathbf{r}_{ne}^{(\tau)})}^T$ in one iteration, which is described in Fig. \ref{TS_AGD_Tau_0_case}. The decoding complexity in the second decoding stage with $\tau$ for $R_H(\tau)=0$ is identical to that in $|S_e|$ CNU operations because the decoder knows the $|S_e|$ check nodes for the CNU operation from the preprocessing stage.
\item For $\tau$ such that $R_H(\tau)=1$: In this case, we have one erasure symbol in the non-standard basis vector of $H$ and the other erasure symbols are located in the column indices in $S_p$. Here, the decoding process is done in two steps, that is, one for one erasure symbol in the non-standard basis vector of $H$ and the other for the other erasure symbols with indices in $S_p$. Suppose that the set of erasure symbol indices is given as $\{ e_0, e_1, e_2,...,e_{z-1}\}$, where $z$ is the number of erasure symbols. Suppose that the $e_j$-th column is the $i_j$-th standard basis vector $u_{i_j}$, $0\le j \le z-2$, and the $e_{z-1}$-th column of $H$ is a non-standard basis vector. We also have $|S_p|-z+1$ standard basis vectors in $H$, where non-erasure symbols are located. In the first decoding step, assume that for $z\le i \le |S_p|$, some $i_j$-th component of the $e_{z-1}$-th column of $H$ is equal to 1. Then, using the $i_j$-th row of $H$, the erasure symbol $\hat{c}_{e_{z-1}}$ can be recovered because there is no erasure symbol except for $\hat{c}_{e_{z-1}}$ at the positions of element 1 in the $i_j$-th row of $H$. Then, we go to the second decoding stage, which is the same as that of $R(\tau)=0$. If the $i_j$-th component of the $e_{z-1}$-th column of $H$ is 0, decoding of the first step cannot be successful because $e_{z-1}$ disappears in the IED procedure. If the first decoding step is not successful, then we try to decode it for other $\tau$ values such that $R_H(\tau)=1$. The second decoding procedure is described in Fig. \ref{TS_AGD_Tau_1_case}.

\begin{figure}[!t]
\centering
\includegraphics[width=5in]{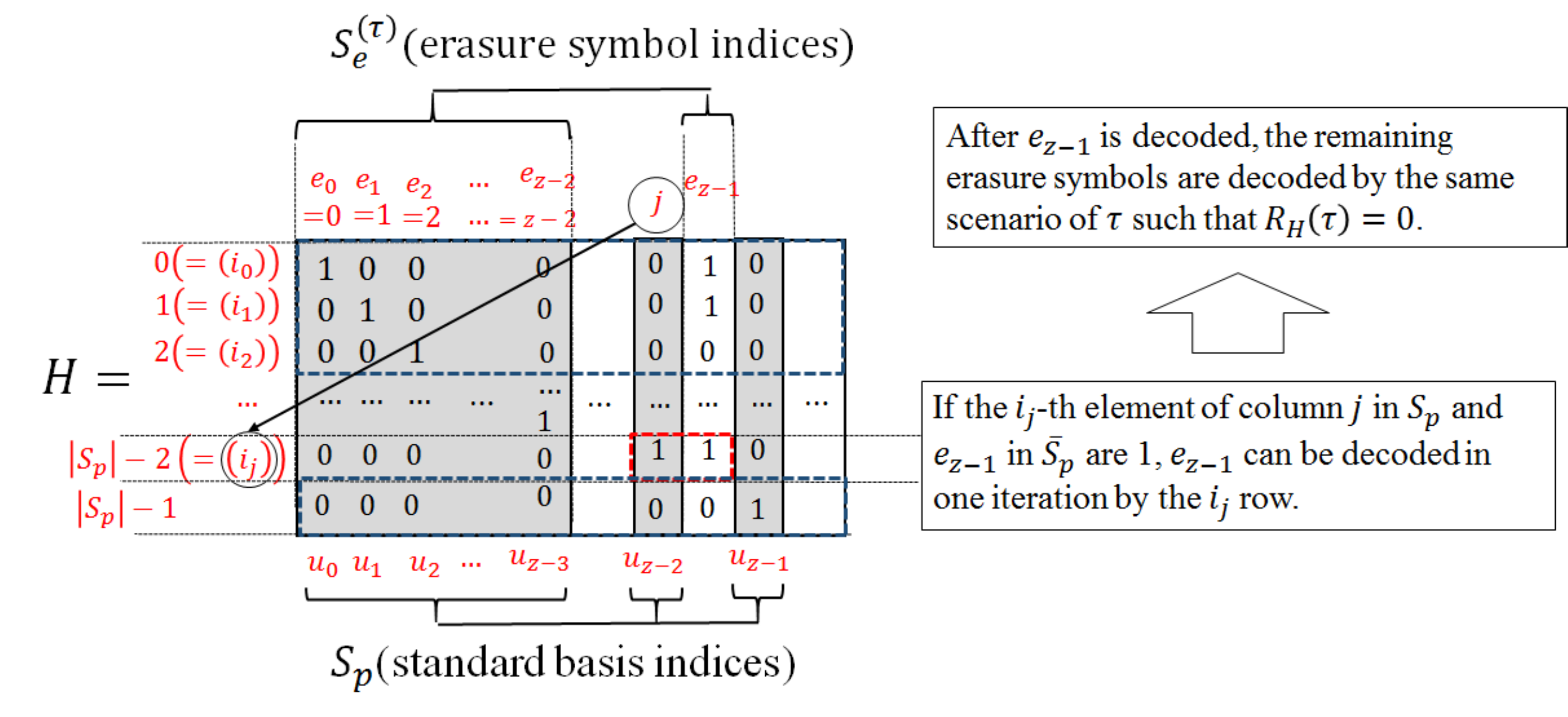}

\caption{The second stage decoding procedure of the TS-AGD of $\tau$ such that $R_H(\tau)=1$.}
\label{TS_AGD_Tau_1_case}
\end{figure}
\item For $\tau$ such that $R_H(\tau)\ge 2$: Let $\bar{S_p}=\{t|s_p(t)=1\}$, i.e., the complement of $S_p$. Let $S_p=S_{p_e}\cup S_{p_{ne}}$, where $S_{p_e}$ is a subset of indices such that the erasure symbols exist and $S_{p_{ne}}=S_p \setminus S_{p_e}$. Similarly, let $\bar{S_p}= \bar{S}_{p_e} \cup \bar{S}_{p_{ne}}$ and then clearly, $|\bar{S}_{p_e}|=R_H(\tau)$. For a $j\in S_{p_{ne}}$, suppose that the $j$-th element of the $e_{i}$-th column of $H$ with $e_{i}\in\bar{S}_{p_e}$ is 1 and that the $j$-th elements of the other columns with indices in $\bar{S}_{p_e}\setminus\{e_{j}\}$ of $H$ are all zero and further, there exists $u_{j}$ in the columns with indices in $S_{p_{ne}}$. Then, we can recover the erasure symbol with index $e_{i}$. That is, all erasure symbols except for $\hat{c}_{e_{i}}$ are disappeared in the inner product of the $j$-th row of $H$ and the received codeword cyclically shifted by $\tau$ and thus $\hat{c}_{e_{i}}$ can be recovered. To decode the remaining erasure symbols, it is needed to return to the preprocessing stage to find the values of $\tau$'s with lower values of $R_H(\tau)$. The second decoding stage of the proposed two-stage decoding algorithm is described in Fig. \ref{TS_AGD_Tau_2_case}.
\begin{figure}[!t]
\centering
\includegraphics[width=5in]{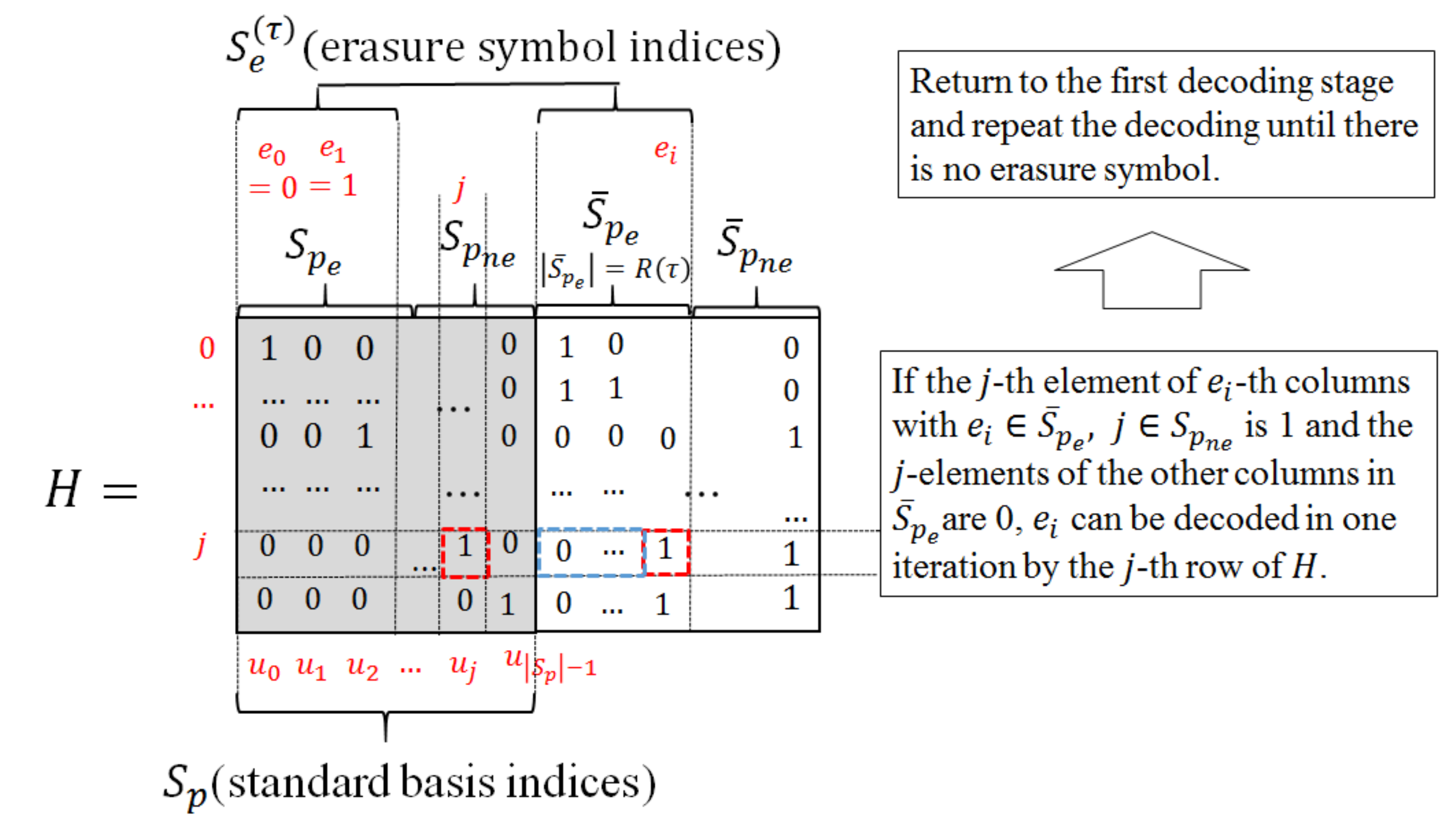}

\caption{The second stage decoding procedure of the TS-AGD of $\tau$ such that $R_H(\tau)\ge2$.}
\label{TS_AGD_Tau_2_case}
\end{figure}

The overall decoding procedure is described in Algorithm \ref{TwostAGD} and Fig. \ref{TS_AGD_FlowChart_lb} shows a flowchart to summarize the proposed decoding algorithm.

\end{enumerate}

\begin{algorithm}
\caption{Two-stage AGD}
\label{TwostAGD}
\begin{algorithmic} 
\REQUIRE received codeword $\mathbf{r}$, $s_p(t)$, modified $H$, and IED
\STATE $U = \phi, V\gets \mathbf{0}$; \COMMENT{$V$:$n$-tuple vectors}
\STATE $\mathbf{r}^{(0)} \gets \mathbf{r}$
\STATE a;
	\FOR{$ \tau = 0 $ \TO $ n-1 $}
	\STATE $s_e(t) \gets \mathbf{r}^{(0)}$; \COMMENT{obtain $s_e(t)$ from $\mathbf{r}^{(0)}$}
	\STATE Calculate $R_H(\tau)=\sum_{t=0}^{n-1}{s_e(t)s_p(t+\tau)}$
		\IF{$R_H(\tau) = 0$}
		\STATE Obtain $\mathbf{r}^{(\tau)}$ by cyclic shifting $\mathbf{r}^{(0)}$ by $\tau$
			\STATE Do IED for $\mathbf{r}^{(\tau)}$
			\STATE \textbf{STOP}
		\ENDIF
		\STATE $V_{\tau}\gets R_H(\tau)$; \COMMENT{$V_{\tau}$: $\tau$-th component of $V$}
	\ENDFOR
\FOR{$ i=1$ \TO $n$}

\STATE $\tau' \gets \text{argmin}_{\tau\in [0,n-1]\setminus U}V_{\tau}, U\gets U \cup \{\tau'\}$
\STATE Obtain $\mathbf{r}^{(\tau')}$ by cyclic shifting $\mathbf{r}^{(0)}$ by $\tau'$
\STATE Do IED for $\mathbf{r}^{(\tau')}$ 
\IF{$V_{\tau'}=1$ and the erasures in the non-standard basis indices are decoded by IED}
\STATE {Do IED to decode the remaining erasure symbols}
\STATE {\textbf{STOP}}
\ELSIF{There exist the decoded erasure symbols by IED}
\STATE $U\gets \phi$
\STATE Obtain $\mathbf{r}^{(0)}$ by cyclic shifting $\mathbf{r}^{(\tau')}$ by $n-\tau'$
\STATE \textbf{Goto} a;
\ENDIF
\STATE Obtain $\mathbf{r}^{(0)}$ by cyclic shifting $\mathbf{r}^{(\tau')}$ by $n-\tau'$
\ENDFOR
\end{algorithmic}
\end{algorithm}

In the proposed TS-AGD algorithm, the decoding complexity of the first decoding stage is negligible whereas the decoding complexity and delay are remarkably reduced in the second decoding stage, because the first decoding stage provides the cyclic shift values of the received codeword for successful decoding. 

\subsection{Analysis of Modification Criteria for the Parity Check Matrix}
This subsection analyzes the modification criteria of $H$ for $(n,k)$ cyclic codes. The first criterion is related to the number of standard basis vectors, that is, the number of $t$'s such that $s_p(t)=0$, which is less than or equal to $n-k$. As described in the previous subsection, the proposed TS-AGD procedure can be done for the cyclically shifted received codeword $\mathbf{r}^{(\tau)}$ such that $R_H(\tau)$ has low values. As the number of zeros in $s_p(t)$ increases, it is more probable for $R_H(\tau)$ to have low values.

The second criterion is how to locate the standard basis vectors in the parity check matrix. It is not easy to prove the second criterion and thus the following theorem replaces the proof of the second criterion. First, we need the lemma for proof of the following theorem.

\begin{lemma}[Bonferroni inequality \cite{Bonferroni}]
\label{Bonferroni_lb}
Let $E_i$, $i\in A$, be sets of elements. Then we have the following inequality as
\begin{multline}
\label{Bonferroni_ineq}
{\sum_{I\subset A, |I|=1}|E_i|}-{\sum_{I\subset A, |I|=2}\left|\bigcap_{i \in I}{E_i}\right|} \le \\ \left|\bigcup_{i\in A} E_i \right|\le{\sum_{I\subset A, |I|=1}|E_i|}-\frac{2}{|A|}{\sum_{I\subset A, |I|=2}\left|\bigcap_{i \in I}{E_i}\right|}.
\end{multline}
\end{lemma}
\begin{theorem}
\label{selectionpcps}
The upper bound on the number of occurrences of $R_H(\tau)\le 1$ for $0 \le \tau \le n-1$ is maximized if the parity check sequence of the modified parity check matrix has a constant out-of-phase autocorrelation value.
\end{theorem}

\begin{IEEEproof}
First, it is desirable for the proposed decoding algorithm to successfully decode more erasure patterns, which is possible if $R_H(\tau)\le 1$. Thus, we have to modify the parity check matrix, for which $R_H(\tau) \le 1$ is most common for as many shift values $\tau$ as possible. The following two cases are considered.

\begin{enumerate}[label=(\roman*)]
\item $R_H(\tau) = 0$:

\begin{figure}[!t]
\centering
\includegraphics[width=3.5in]{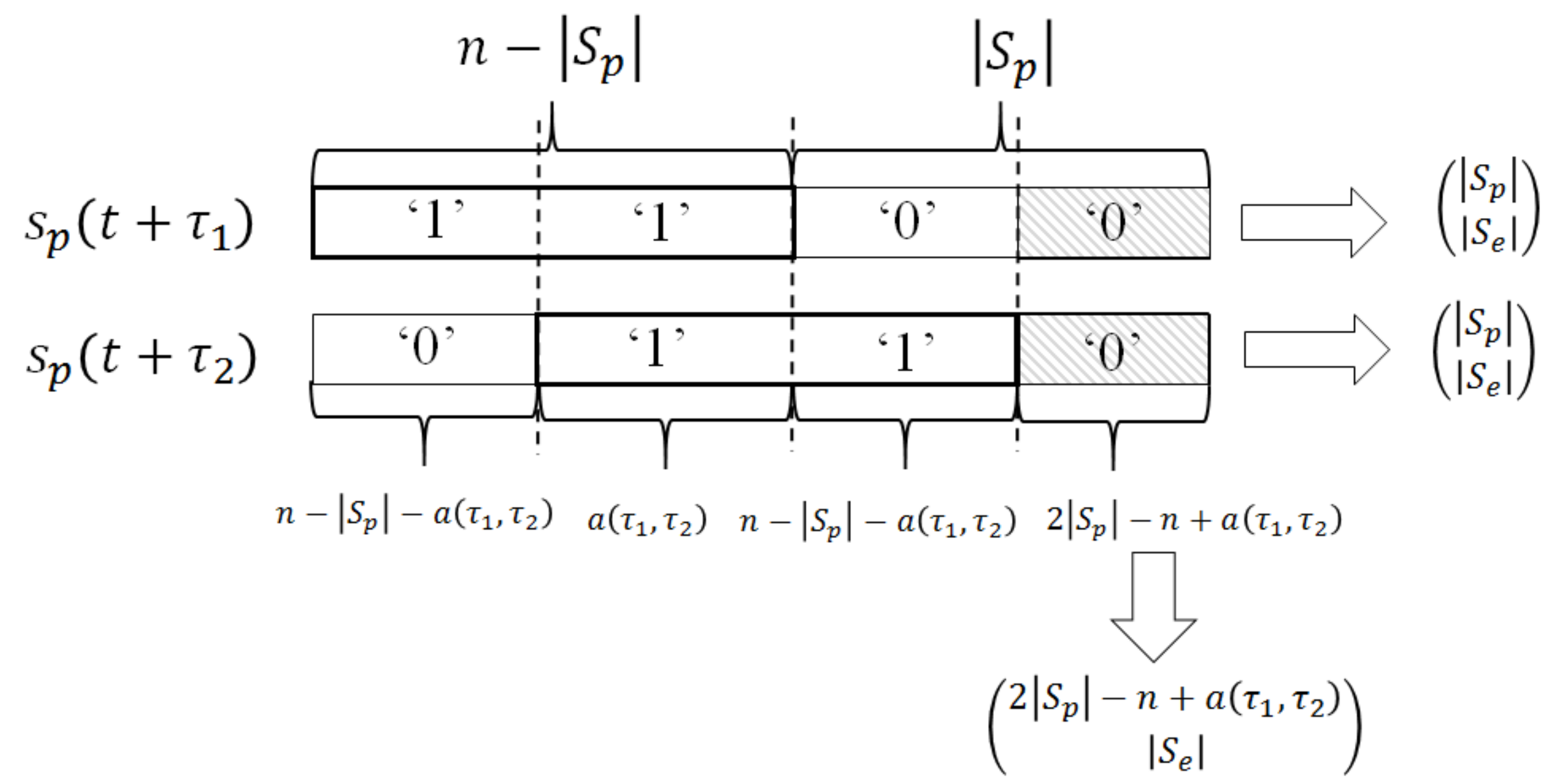}
\caption{The number of doubly counted erasure patterns for $\tau$ such that $R_H(\tau)=0$.}
\label{TS_AGD_Proof_Tau_0_Lb}
\end{figure}
This means that $S_e^{(\tau)}\subseteq S_p$. It is easy to check that in $R_H(\tau)$, it is equivalent to cyclically shift $s_p(t)$ instead of $s_e(t)$. Let $S_p^{(\tau)}$ be the support set of $1-s_p(t+\tau)$. Let $E_{\tau}$ be the set of erasure patterns which can be successfully recovered by $s_p(t+\tau)$. Then, we have $|E_{\tau}|= {|S_p| \choose |S_e|}$, which leads to
\begin{equation}
\sum_{\tau=0}^{n-1}|E_\tau|\le n { | S_p | \choose |S_e|}.
\label{First_term_tau_0}
\end{equation}
It is easy to check that doubly counted erasure patterns are included in (\ref{First_term_tau_0}), which should be excluded. If the shaded parts in Fig. \ref{TS_AGD_Proof_Tau_0_Lb} include all the erasure symbols, those erasure patterns are doubly counted, where $a(\tau_1, \tau_2)$ denotes the number of pairs $\left(s_p(t+\tau_1),s_p(t+\tau_2)\right)=(1,1)$. Thus we have $ 2|S_p|+a(\tau_1,\tau_2)-n \choose |S_e|$ doubly counted erasure patterns. Using Lemma 1, the number of erasure patterns which are successfully decoded by $s_p(t)$ is bounded as
\begin{multline}
\label{Tau_0_pf_eqn}
\left|\bigcup_{\tau=0}^{n-1} E_{\tau} \right| \le {\sum_{\tau=0}^{n-1}|E_{\tau}|}-\frac{2}{n}{\sum_{\tau_1,\tau_2}\left|E_{\tau_1} \cap E_{\tau_2}\right|} \le \\ n {|S_p| \choose |S_e|}-\frac{2}{n}{\sum_{\tau_1,\tau_2}{2|S_p|+a(\tau_1,\tau_2)-n \choose |S_e|}}.
\end{multline}

\item $R_H(\tau) = 1$: 
\begin{figure}[!t]
\centering
\includegraphics[width=3.5in]{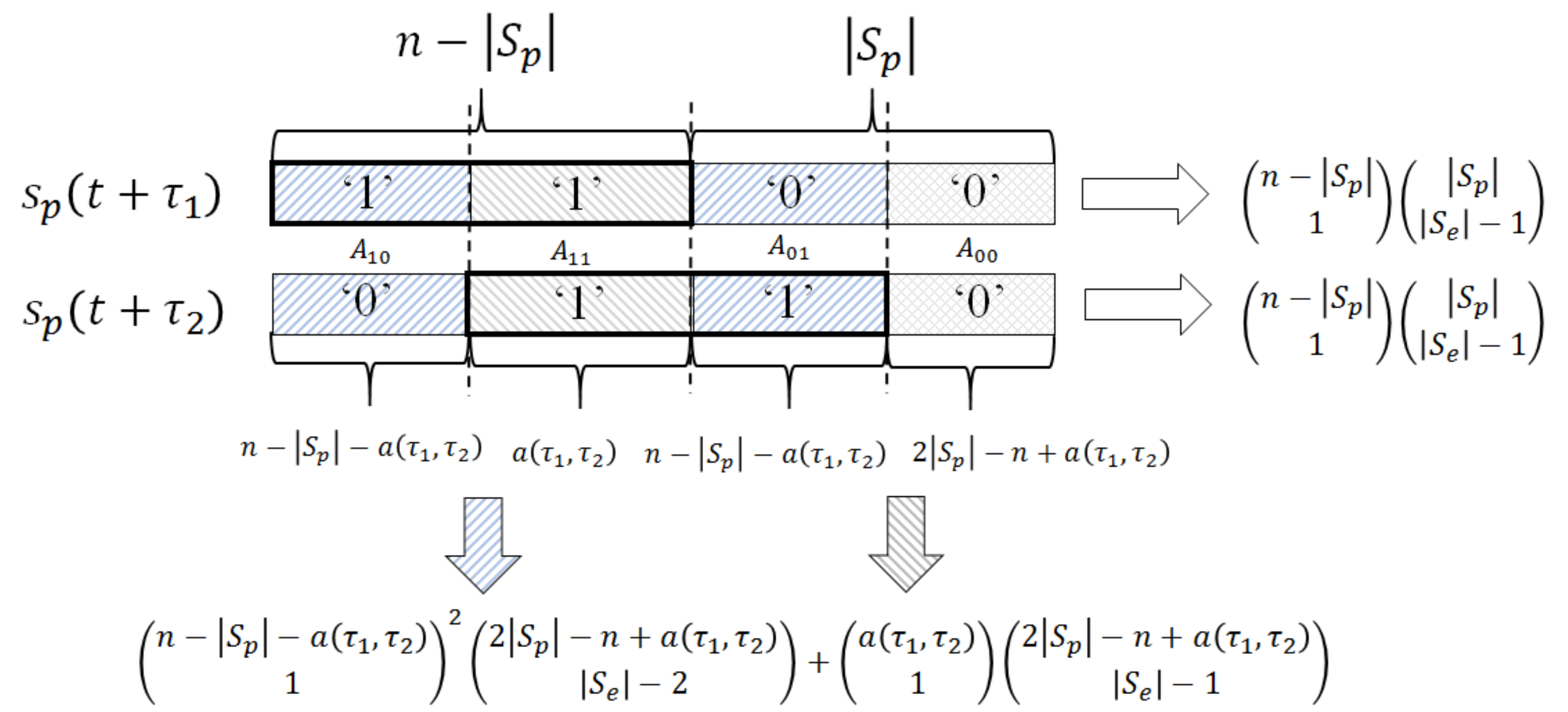}
\caption{The number of doubly counted erasure patterns for $\tau$ such that $R_H(\tau)=1$.}
\label{TS_AGD_Proof_Tau_1_Lb}
\end{figure}

In this case, the index of one erasure symbol is in $\bar{S}_p$ and the indices of the other erasure symbols are in $S_p$. Thus, the total number of such erasure patterns is ${n-|S_p| \choose 1}{|S_p| \choose |S_e|-1}$, where doubly counted erasure patterns are included. There are two cases of doubly counted erasure patterns as shown in Fig. \ref{TS_AGD_Proof_Tau_1_Lb}.
\begin{enumerate}
\item Each of two erasure symbols is located in $A_{10}$ and $A_{01}$, respectively and the other erasure symbols are located in $A_{00}$, which are counted as ${n-|S_p|-a(\tau_1,\tau_2) \choose 1}^2{2|S_p|-n+a(\tau_1,\tau_2) \choose |S_e|-2}$.

\item One erasure symbol is located in $A_{11}$ and the other erasure symbols are located in $A_{00}$, which are counted as ${a(\tau_1,\tau_2) \choose 1} {2|S_p|+a(\tau_1,\tau_2)-n \choose |S_e|-1}$. Similarly, from Lemma 1, the number of erasure patterns which are successfully decoded by $s_p(t)$ is given as
\begin{multline}
\label{Tau_1_pf_eqn}
\left|\bigcup_{\tau=0}^{n-1} E_{\tau} \right| \le {\sum_{\tau=0}^{n-1}|E_{\tau}|}-\frac{2}{n}{\sum_{\tau_1,\tau_2}\left|E_{\tau_1} \cap E_{\tau_2}\right|} \le 
n {n-|S_p| \choose 1}{|S_p| \choose |S_e|-1}-\frac{2}{n}\sum_{\tau_1,\tau_2 \in[0,n-1]}{}\Bigg(	
\\{n-|S_p|-a(\tau_1,\tau_2) \choose 1}^2{2|S_p|+a(\tau_1,\tau_2)-n \choose |S_e|-2}+				
{a(\tau_1,\tau_2) \choose 1} {2|S_p|+a(\tau_1,\tau_2)-n \choose |S_e|-1}\Bigg).					
\end{multline}
\end{enumerate}
\end{enumerate}
In order to maximize the upper bounds in (\ref{Tau_0_pf_eqn}) and (\ref{Tau_1_pf_eqn}), the second terms of the right hand sides should be minimized, which can be solved by the convex optimization as described in Appendix A. That is, it is derived in Appendix A that maximizing the upper bound on the number of occurrences of $R_H(\tau) \le 1$ for $0\le \tau \le n-1$ by convex optimization occurs when the out-of-phase autocorrelation values of $s_p(t)$ are constant. Thus, we prove the theorem.
\end{IEEEproof}

The third criterion is related to the performance of the decoder, that is, $H$ with the minimum Hamming weight of rows can have better decoding performance in IED as mentioned in \cite{Hehn} as cog, because more erasure symbols are removed in the inner product of the received codeword and the rows with the minimum Hamming weight of $H$.

\subsection{Proposed TS-AGD for Some Cyclic Codes}

In this subsection, the proposed TS-AGD is applied to several cyclic codes in the erasure channel, that is, the perfect codes such as Hamming codes, Golay codes, and extended Golay codes, and BCH codes. Surprisingly, some cyclic codes such as perfect codes can achieve the decoding performance identical to that of the ML decoder, known as \textit{perfect decoding}. In general, it is desirable for algebraic decoders to be designed to decode all erasure symbols up to the minimum distance $d$. Some decoding algorithm such as the AGD \cite{Hehn} has been proposed to decode some of received codewords with erasure symbols up to $n-k$. 

To analyze the decoding complexity and delay, the number of iterations and decoding complexity are described as graphs. The number of iterations counts parallelized CNU and VNU operations as $\frac{1}{2}$, respectively as in \cite{Hehn}. Note that AGD and TS-AGD use CNU operations more than VNU operations due to cyclic shifts. In the proposed TS-AGD, the preprocessing can slightly increase the decoding complexity but it reduces the number of iterations for IED. The number of VNU operations for the TS-AGD is identical to that for AGD because there is no difference in terms of the erasure decoding performance for the fixed parity check matrix $H$. Therefore, with regard to decoding complexity, the Hamming correlation calculation for each $\tau$ in the preprocessing stage and the CNU operation of one check node in the second decoding stage are counted as 1 in both cases, respectively but the decoding complexity of the VNU operations is ignored. 

\subsubsection{Proposed Decoding Algorithms for Perfect Codes}

\begin{enumerate}[label=(\roman*)]
\item $(2^m-1,2^m-1-m,3)$ Hamming codes:
Clearly, Hamming codes have only one cog and thus one row in $H$ is needed to achieve the ML decoding performance as in the following proposition.
\begin{proposition}
\label{prop1_lb}
For an $(n,k,d)$ linear code $\cal{C}$, the IED of $2^{n-k}\times n$ expanded parity check matrix whose rows consist of all of the codewords of its dual code $\cal{C}^{\perp}$ can achieve ML decoding performance.
\end{proposition}
\begin{IEEEproof}
Note that the ML decoder can decode only if $S_e$ of the erasure pattern does not include the support of any codeword. Let $H$ be an $(n-k)\times n$ submatrix with full rank by selecting rows from the expanded parity check matrix. Let $H_{S_e}$ be an $(n-k) \times |S_e|$ submatrix generated by selecting the columns with indices in $S_e$ from $H$. Let $\epsilon$ be an $|S_e|$-tuple erasure vector, that is, $\epsilon$ consists of the elements with indices in $S_e$ of the transmitted codeword. Then, we have the syndrome of
\begin{displaymath}
S=H {(\mathbf{r}^{(\tau)})}^T=H {(\mathbf{r}_{e}^{(\tau)})}^T+ H {(\mathbf{r}_{ne}^{(\tau)})}^T=0,
\end{displaymath}
which can be modified as
\begin{equation}
H_{S_e}\epsilon^T=H {(\mathbf{r}_{e}^{(\tau)})}^T = H {(\mathbf{r}_{ne}^{(\tau)})}^T.
\label{prop1_eqn2}
\end{equation}
If the rank of $H_{S_e}$ is lower than $|S_e|$, (\ref{prop1_eqn2}) has multiple solutions, implying that the decoder cannot decode the codeword. Thus, $H_{S_e}$ should have full rank and then there exist $|S_e|$ linearly independent rows in $H_{S_e}$. Let $H'$ and $H'_{S_e}$ be $|S_e|\times n$ and $|S_e|\times|S_e|$ matrices constructed from $H$ and $H_{S_e}$ by selecting $|S_e|$ linearly independent rows, respectively. By selecting the elements with the same row indices as those of $H'_{S_e}$ from $H{(\mathbf{r}_{ne}^{(\tau)})}^T$ in (\ref{prop1_eqn2}), we can find $\epsilon$ by inverting $H'_{S_e}$. From the properties of linear codes, each row of ${H'}_{S_e}^{-1} H'$ is actually the codeword of the dual code $\cal{C}^{\perp}$. Then, the IED of $H$ whose rows consist of the codewords of $\cal{C}^{\perp}$ can correct the erasure patterns, which do not include the codeword. Clearly, this corresponds to the ML decoder.
\end{IEEEproof}

From the above proposition, the Hamming codes which have only one cog can achieve the ML decoding performance, because they have the same performance as a $2^{n-k}\times n$ expanded parity check matrix.

Below, the decoding performance of $(23,12)$ Golay, $(24,12)$ extended Golay, and $(31,21,5)$ BCH codes is presented in the table, which compares the exact ML erasure decoding performance with the AGD and the TS-AGD decoding performances. Proposition \ref{prop1_lb} is useful in that it can be used to calculate the erasure decoding performance of the ML decoder based on IED.

\item For the $(23,12,7)$ binary Golay code:
Using the proposed modification criteria, the parity check matrix of the $(23,12,7)$ binary Golay code can be modified as
\begin{equation}
\label{H_m_lb}
H_{m} =
\left( \begin{array}{c}
1	0	0	0	1	0	0	0	0	0	0	0	0	1	1	0	0	0	1	1	1	0	1	\\
0	1	0	0	1	0	1	0	0	1	0	0	0	0	1	0	1	0	1	0	0	0	1	\\
0	0	1	0	0	0	0	0	0	0	1	0	0	1	1	0	1	0	1	0	0	1	1	\\
0	0	0	1	0	0	1	0	0	1	1	0	0	1	1	0	0	0	0	1	0	0	1	\\
0	0	0	0	1	1	1	0	0	0	1	0	0	0	1	0	0	0	0	0	1	1	1	\\
0	0	0	0	1	0	1	1	0	0	0	0	0	1	0	0	1	0	0	1	0	1	1	\\
0	0	0	0	1	0	0	0	1	1	1	0	0	0	0	0	0	0	1	1	0	1	1	\\
0	0	0	0	0	0	1	0	0	1	0	1	0	1	0	0	0	0	1	0	1	1	1	\\
0	0	0	0	0	0	1	0	0	0	1	0	1	0	0	0	1	0	1	1	1	0	1	\\
0	0	0	0	1	0	0	0	0	1	1	0	0	1	0	1	1	0	0	0	1	0	1	\\
0	0	0	0	0	0	0	0	0	1	0	0	0	0	1	0	1	1	0	1	1	1	1

\end{array} \right)
\end{equation}

and its parity check sequence is given as

\begin{displaymath}
s_{p}(t)=\left(\begin{array}{c}0	0	0	0	1	0	1	0	0	1	1	0	0	1	1	0	1	0	1	1	1	1	1\end{array} \right),
\end{displaymath}

which corresponds to the characteristic sequence of cyclic difference set with parameters $(23,12,6)$. Here, the modified parity check matrix has 12 standard basis vectors and its rows have the minimum Hamming weights. Thus, (\ref{H_m_lb}) satisfies the three modification criteria for the parity check matrix. The numerical analysis shows in Table \ref{Golay_EP} that the proposed TS-AGD with $H_m$ can achieve the same performance as the ML decoder and outperform the AGD with $H_{sys}$, where $H_{sys}$ denotes the systematic form of its parity check matrix defined as $[I_{12}|P]$.  
Fig. \ref{Golay_result} shows the decoding performance of the $(23,12,7)$ binary Golay codes in terms of iteration and the decoding complexity. Two decoders, AGD and TS-AGD, and two modifications of the parity check matrix are considered in Fig. \ref{Golay_result}. The number of iterations of the proposed TS-AGD algorithm can be reduced compared to the AGD as shown in Fig. \ref{Golay_result}(a). The decoding complexity of TS-AGD  with $H_m$ has the lowest value except for the midrange of the erasure probability as in Fig. \ref{EGolay_result}(b).

\begin{table}[]
\centering
\caption{The undecodable erasure patterns by the modified $H$ in the (23,12,7) binary Golay code}
\label{Golay_EP}
\begin{tabular}{|c|c|c|c|}
\hline
\textbf{The number}  & \textbf{Total number of}      & \textbf{TS-AGD and} & \textbf{TS-AGD and} \\
\textbf{of erasures}  & \textbf{erasure patterns}     & \textbf{AGD of $H_{sys}$} & \textbf{AGD of $H_{m}$}, \textbf{ML}  \\ \hline
\textbf{$\le 6$} &  		 & 0                          & 0                             \\ \hline
7      & 245157				   & 253                          & 253                    \\ \hline
8      & 490314 			 & 4554                         & 4554                   \\ \hline
9      & 817190    		 & 37973                       & 37950               \\ \hline
10     & 1144066 		 & 197754                       & 194810                \\ \hline
11      & 1352078    		& 700488                     & 656558             \\ \hline
\end{tabular}
\end{table}

\begin{figure*}[!t]
\centering
\subfloat[The average number of iterations]{\includegraphics[width=3.5in]{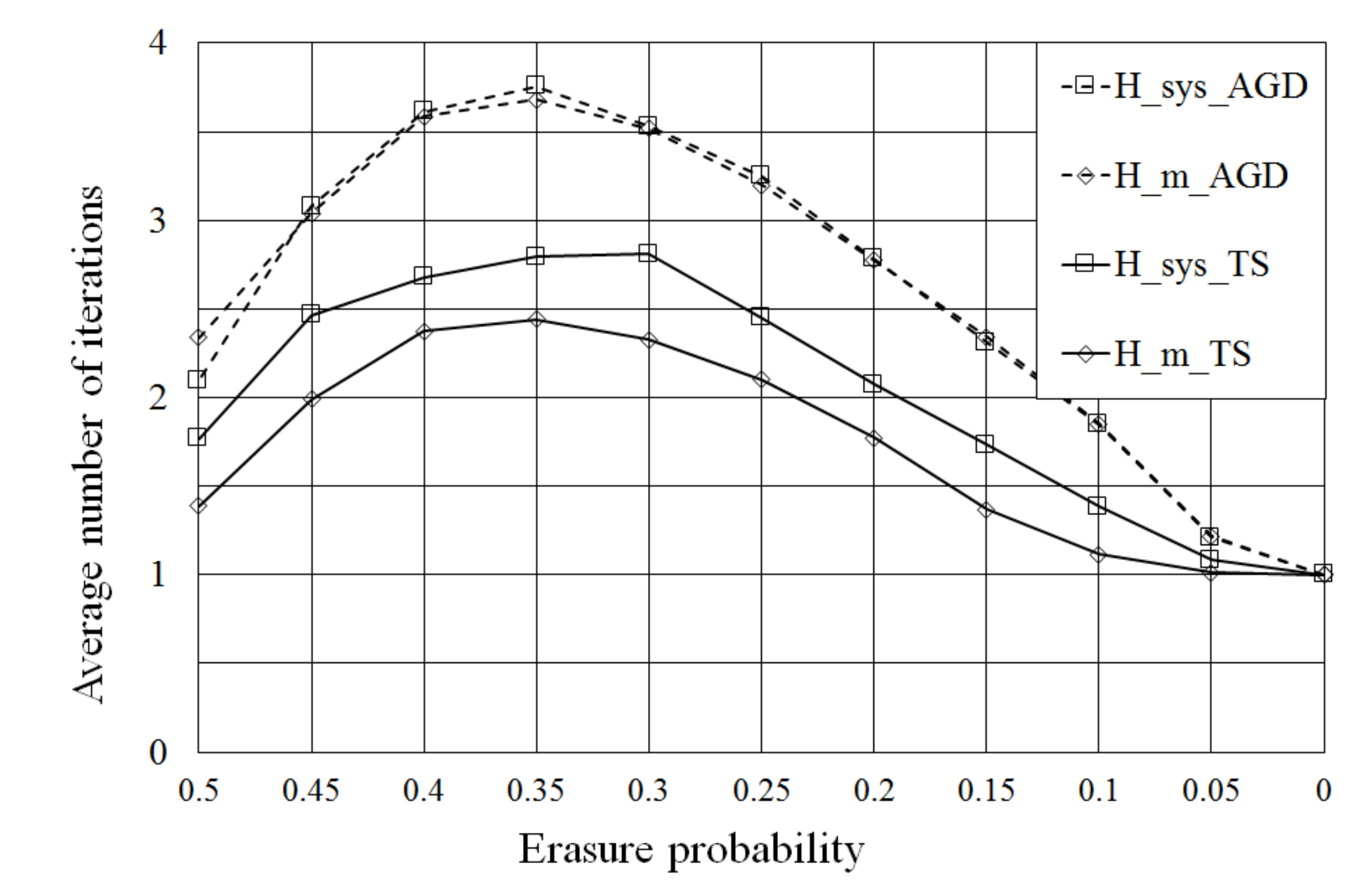}}%
\label{Golay_Iter_lb}
\hfil
\subfloat[Decoding complexity]{\includegraphics[width=3.5in]{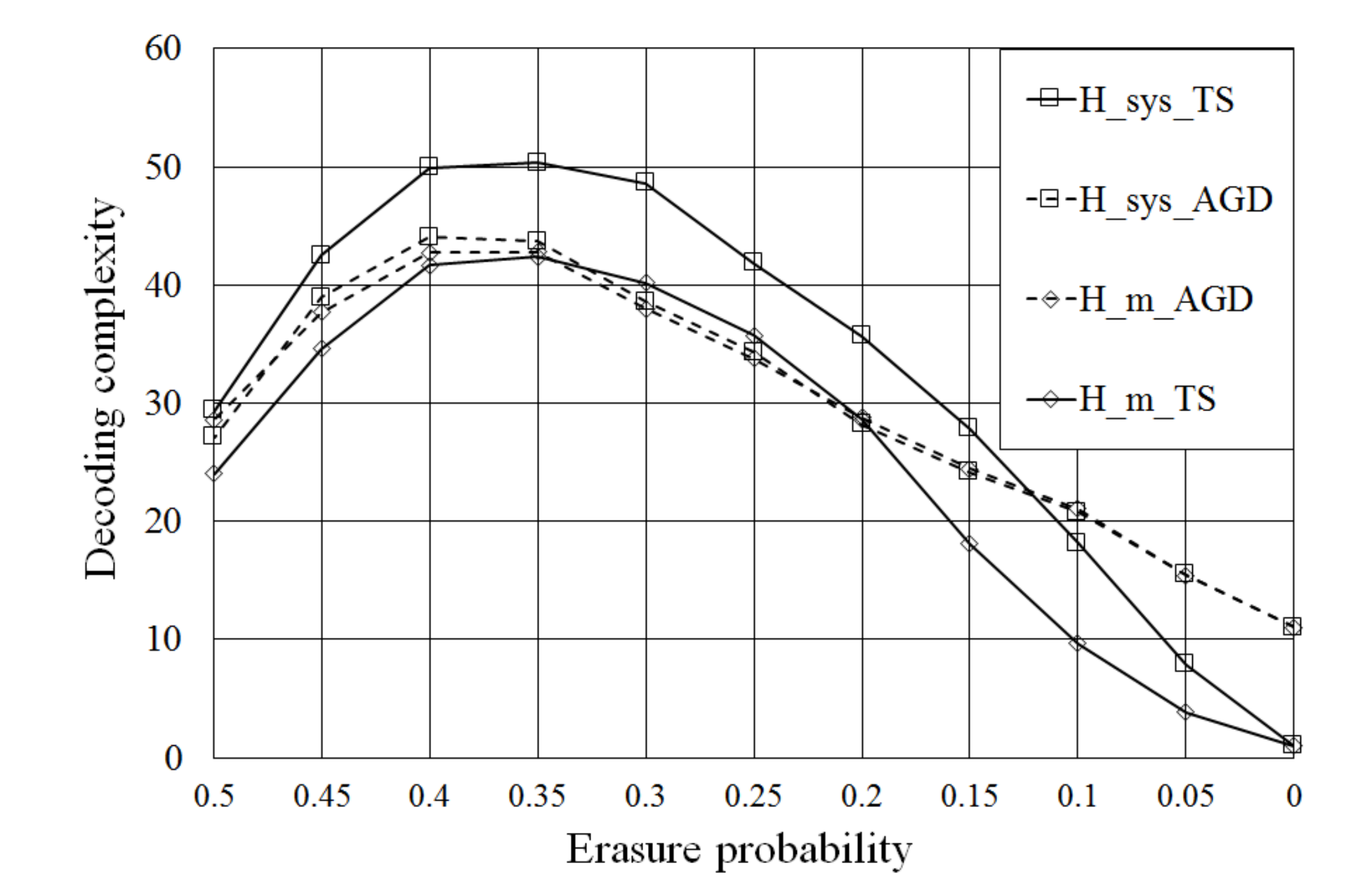}}%
\label{Golay_Comp_lb}
\caption{The average number of iterations and the decoding complexity of AGD and TS-AGD with modified $H$ for the $(23,12,7)$ Golay code.}
\label{Golay_result}
\end{figure*}

\item $(24, 12,8)$ binary extended Golay code:
In fact, while this is not a cyclic code, it is cyclic except for the last parity bit. Thus, we can apply the AGD algorithm and the proposed TS-AGD algorithm. Hehn modified the parity check matrix as in \cite{Hehn}

\begin{displaymath}
H_{Hehn} =
\left( \begin{array}{c}
1	1	1	0	0	0	0	0	1	0	0	1	1	0	0	0	0	0	1	0	0	0	0	1
\\1	1	0	0	0	0	1	0	0	0	0	1	0	0	1	1	1	0	0	0	0	0	0	1
\\1	1	0	1	0	0	1	0	1	0	1	0	0	1	0	0	0	0	0	0	0	0	0	1
\\1	1	1	0	0	0	1	1	0	0	0	0	0	0	0	0	0	0	0	1	0	1	0	1
\\1	1	0	0	0	1	0	0	0	1	0	1	0	1	0	0	0	0	0	0	0	1	0	1
\\1	1	0	1	0	0	0	1	0	1	0	0	0	0	0	0	1	0	1	0	0	0	0	1
\\0	1	1	0	0	0	1	0	0	0	0	1	0	1	0	0	0	1	0	0	1	0	0	1
\\1	1	0	1	1	0	0	0	0	0	0	1	0	0	0	0	0	0	0	1	1	0	0	1
\\1	1	1	1	0	1	0	0	0	0	0	0	0	0	1	0	0	1	0	0	0	0	0	1
\\1	1	0	0	1	0	0	0	0	0	1	0	0	0	1	0	0	0	1	0	0	1	0	1
\\0	0	1	1	0	0	1	0	1	0	0	1	0	0	1	0	0	0	0	1	0	0	0	1
\\0	0	1	1	0	1	0	1	0	0	1	1	0	0	1	0	0	0	0	0	0	0	1	0

\end{array} \right)
\end{displaymath}
and the systematic form of the parity check matrix is given as
\begin{displaymath}
\ \ H_{sys} \ \ =
\left( \begin{array}{c}
1	0	0	0	0	0	0	0	0	0	0	0	1	1	0	1	1	1	0	0	0	1	0	1
\\0	1	0	0	0	0	0	0	0	0	0	0	1	0	1	1	1	0	0	0	1	0	1	1
\\0	0	1	0	0	0	0	0	0	0	0	0	0	1	1	1	0	0	0	1	0	1	1	1
\\0	0	0	1	0	0	0	0	0	0	0	0	1	1	1	0	0	0	1	0	0	1	0	1
\\0	0	0	0	1	0	0	0	0	0	0	0	1	1	0	0	0	1	0	1	1	0	1	1
\\0	0	0	0	0	1	0	0	0	0	0	0	1	0	0	0	1	0	1	1	1	1	1	1
\\0	0	0	0	0	0	1	0	0	0	0	0	0	0	0	1	0	1	1	0	0	1	1	1
\\0	0	0	0	0	0	0	1	0	0	0	0	0	0	1	0	1	1	0	1	1	1	0	1
\\0	0	0	0	0	0	0	0	1	0	0	0	0	1	0	1	1	0	1	1	1	0	0	1
\\0	0	0	0	0	0	0	0	0	1	0	0	1	0	1	1	0	1	1	1	1	0	0	1
\\0	0	0	0	0	0	0	0	0	0	1	0	0	1	1	0	1	1	1	0	0	0	1	1
\\0	0	0	0	0	0	0	0	0	0	0	1	1	1	1	1	1	1	1	1	1	1	1	0
\end{array} \right).
\end{displaymath}

The modified parity check matrix based on the three proposed criteria can be given as
\begin{displaymath}
H_{m} =
\left( \begin{array}{c}
1	0	0	0	1	0	0	0	0	0	0	0	0	1	1	0	0	0	1	1	1	0	1	0\\
0	1	0	0	1	0	1	0	0	1	0	0	0	0	1	0	1	0	1	0	0	0	1	0\\
0	0	1	0	0	0	0	0	0	0	1	0	0	1	1	0	1	0	1	0	0	1	1	0\\
0	0	0	1	0	0	1	0	0	1	1	0	0	1	1	0	0	0	0	1	0	0	1	0\\
0	0	0	0	1	1	1	0	0	0	1	0	0	0	1	0	0	0	0	0	1	1	1	0\\
0	0	0	0	1	0	1	1	0	0	0	0	0	1	0	0	1	0	0	1	0	1	1	0\\
0	0	0	0	1	0	0	0	1	1	1	0	0	0	0	0	0	0	1	1	0	1	1	0\\
0	0	0	0	0	0	1	0	0	1	0	1	0	1	0	0	0	0	1	0	1	1	1	0\\
0	0	0	0	0	0	1	0	0	0	1	0	1	0	0	0	1	0	1	1	1	0	1	0\\
0	0	0	0	1	0	0	0	0	1	1	0	0	1	0	1	1	0	0	0	1	0	1	0\\
0	0	0	0	0	0	0	0	0	1	0	0	0	0	1	0	1	1	0	1	1	1	1	0\\
0	0	0	0	1	0	1	0	0	1	1	0	0	1	1	0	1	0	1	1	1	1	0	1
\end{array} \right)
\end{displaymath}
where the first 11 standard basis column indices are determined by the cyclic difference set with parameters $(23,12,6)$ as before and the last standard basis vector is located in the extended bit. The last row of $H_m$ has the Hamming weight of 12, which is larger than the minimum Hamming weight 8. Thus, we can further modify it by replacing the last row by sum of the first row and the last row as
\begin{displaymath}
H_{A} =
\left( \begin{array}{c}
1	0	0	0	1	0	0	0	0	0	0	0	0	1	1	0	0	0	1	1	1	0	1	0\\
0	1	0	0	1	0	1	0	0	1	0	0	0	0	1	0	1	0	1	0	0	0	1	0\\
0	0	1	0	0	0	0	0	0	0	1	0	0	1	1	0	1	0	1	0	0	1	1	0\\
0	0	0	1	0	0	1	0	0	1	1	0	0	1	1	0	0	0	0	1	0	0	1	0\\
0	0	0	0	1	1	1	0	0	0	1	0	0	0	1	0	0	0	0	0	1	1	1	0\\
0	0	0	0	1	0	1	1	0	0	0	0	0	1	0	0	1	0	0	1	0	1	1	0\\
0	0	0	0	1	0	0	0	1	1	1	0	0	0	0	0	0	0	1	1	0	1	1	0\\
0	0	0	0	0	0	1	0	0	1	0	1	0	1	0	0	0	0	1	0	1	1	1	0\\
0	0	0	0	0	0	1	0	0	0	1	0	1	0	0	0	1	0	1	1	1	0	1	0\\
0	0	0	0	1	0	0	0	0	1	1	0	0	1	0	1	1	0	0	0	1	0	1	0\\
0	0	0	0	0	0	0	0	0	1	0	0	0	0	1	0	1	1	0	1	1	1	1	0\\
1	0	0	0	0	0	1	0	0	1	1	0	0	0	0	0	1	0	0	0	0	1	1	1
\end{array} \right)
\end{displaymath}
where the last row has the minimum Hamming weight 8 but the first column is not a standard basis vector. The further modification is done by replacing the $i$-th row with the sum of the $i$-th row and the last row of $H_m$, $1 \le i \le 11$ and the last row with the first row of $H_m$ as

\begin{displaymath}
H_{B} =
\left( \begin{array}{c}
1	0	0	0	0	0	1	0	0	1	1	0	0	0	0	0	1	0	0	0	0	1	1	1\\
0	1	0	0	0	0	0	0	0	0	1	0	0	1	0	0	0	0	0	1	1	1	1	1\\
0	0	1	0	1	0	1	0	0	1	0	0	0	0	0	0	0	0	0	1	1	0	1	1\\
0	0	0	1	1	0	0	0	0	0	0	0	0	0	0	0	1	0	1	0	1	1	1	1\\
0	0	0	0	0	1	0	0	0	1	0	0	0	1	0	0	1	0	1	1	0	0	1	1\\
0	0	0	0	0	0	0	1	0	1	1	0	0	0	1	0	0	0	1	0	1	0	1	1\\
0	0	0	0	0	0	1	0	1	0	0	0	0	1	1	0	1	0	0	0	1	0	1	1\\
0	0	0	0	1	0	0	0	0	0	1	1	0	0	1	0	1	0	0	1	0	0	1	1\\
0	0	0	0	1	0	0	0	0	1	0	0	1	1	1	0	0	0	0	0	0	1	1	1\\
0	0	0	0	0	0	1	0	0	0	0	0	0	0	1	1	0	0	1	1	0	1	1	1\\
0	0	0	0	1	0	1	0	0	0	1	0	0	1	0	0	0	1	1	0	0	0	1	1\\
1	0	0	0	1	0	0	0	0	0	0	0	0	1	1	0	0	0	1	1	1	0	1	0
\end{array} \right).
\end{displaymath}
In fact, the first columns of $H_A$ and $H_B$ have Hamming weights 2. Then the parity check sequences of $H_p$, $H_A$, and $H_B$ are given as
\begin{displaymath}
s_{p,H_m}(t)=\left(\begin{array}{c}0	0	0	0	1	0	1	0	0	1	1	0	0	1	1	0	1	0	1	1	1	1	1	0\end{array} \right)
\end{displaymath}
\begin{displaymath}
s_{p,H_A}(t)=\left(\begin{array}{c}1	0	0	0	1	0	1	0	0	1	1	0	0	1	1	0	1	0	1	1	1	1	1	0\end{array} \right)
\end{displaymath}
\begin{displaymath}
\ s_{p,H_B}(t)= \left(\begin{array}{c}1	0	0	0	1	0	1	0	0	1	1	0	0	1	1	0	1	0	1	1	1	1	1	1\end{array} \right).
\end{displaymath}

In the $(24,12,8)$ extended Golay code, any of the modified parity check matrices cannot achieve the same performance as that of the ML decoder. However, the TS-AGD by adding redundant check equations to $H_B$ can give us the same decoding performance as the ML decoder, which is given as

\begin{displaymath}
H_{C} =
\left( \begin{array}{c}
H_B \\ H^{'}_{A}
\end{array} \right)
\end{displaymath}
where $H^{'}_{A}$ is a submatrix composed of nine rows out of the first 11 rows of $H_A$. Fig. \ref{Empi_Golay_Modif_lb} shows the relationship among the various modified parity check matrices. Table \ref{EGolay_EP} shows the decoding performance of the proposed TS-AGD and AGD with $H_{sys}$, $H_m$, $H_A$, $H_B$, $H_{Hehn}$, and $H_C$, where $H_C$ shows decoding performance identical to that of the ML decoder and better decoding performance than the decoding algorithm by Hehn.
Fig. \ref{EGolay_result}(a) shows the average number of iterations, where $H_c$ has the smallest number of iterations. However, the decoding complexity for $H_c$ is largest due to additional rows of the parity check matrix.

\begin{figure}[!t]
\centering
\includegraphics[width=5in]{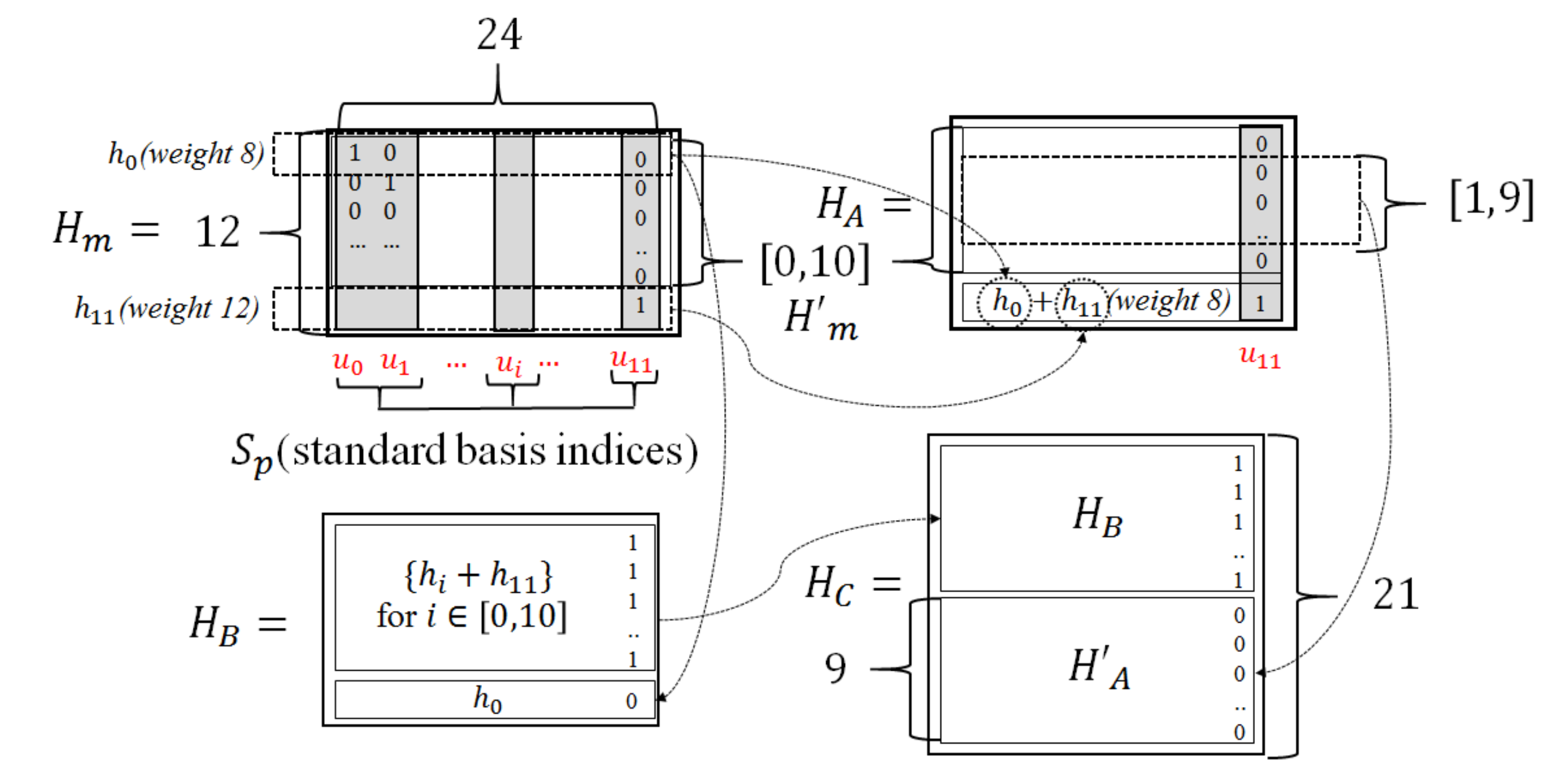}
\caption{Modifications of the parity check matrix in the $(24,12,8)$ extended binary Golay code.}
\label{Empi_Golay_Modif_lb}
\end{figure}

\begin{table*}[]
\centering
\caption{The undecodable erasure patterns by the modified $H$ for the $(24,12,8)$ binary extended Golay code}
\label{EGolay_EP}
\begin{tabular}{|c|c|c|c|c|c|c|}
\hline
\textbf{The}   & \textbf{Total number} & \textbf{TS-AGD} & \textbf{TS-AGD} & \textbf{TS-AGD} & \textbf{TS-AGD and} & {\textbf{TS-AGD and}} \\ 
\textbf{number of} & \textbf{of erasure} & \textbf{and AGD} & \textbf{and AGD} & \textbf{and AGD} & \textbf{AGD of $H_B$} & {\textbf{AGD of $H_C$ }}  \\
\textbf{erasures} & \textbf{patterns} & \textbf{of $H_{sys}$} & \textbf{of $H_{m}$} & \textbf{of $H_A$} & \textbf{and $H_{Hehn}$} & {\textbf{and ML}}  \\ \hline
\textbf{$\le 7$} &               & 0                & 0             & 0                      & 0                                           &        0                        \\ \hline
\textbf{8}       & 735471         & 759            & 759              & 759           & 759
& 759                                                                 \\ \hline
\textbf{9}       & 1307504         & 12144          & 12144            & 12144         & 12144
& 12144                                                           \\ \hline
\textbf{10}      & 1961256        & 92000          & 91080            & 91080         & 91080
& 91080                                                              \\ \hline
\textbf{11}      & 2496144         & 460253         & 426581          & 425178       & 425040                 
& 425040                                                            \\ \hline
\textbf{12}      & 2704156		  & 1515792        & 1344005          & 1325536       & 1322179    
& 1313116                                                              \\ \hline
\end{tabular}
\end{table*}

\begin{figure*}[!t]
\centering
\subfloat[The average number of iterations]{\includegraphics[width=3.5in]{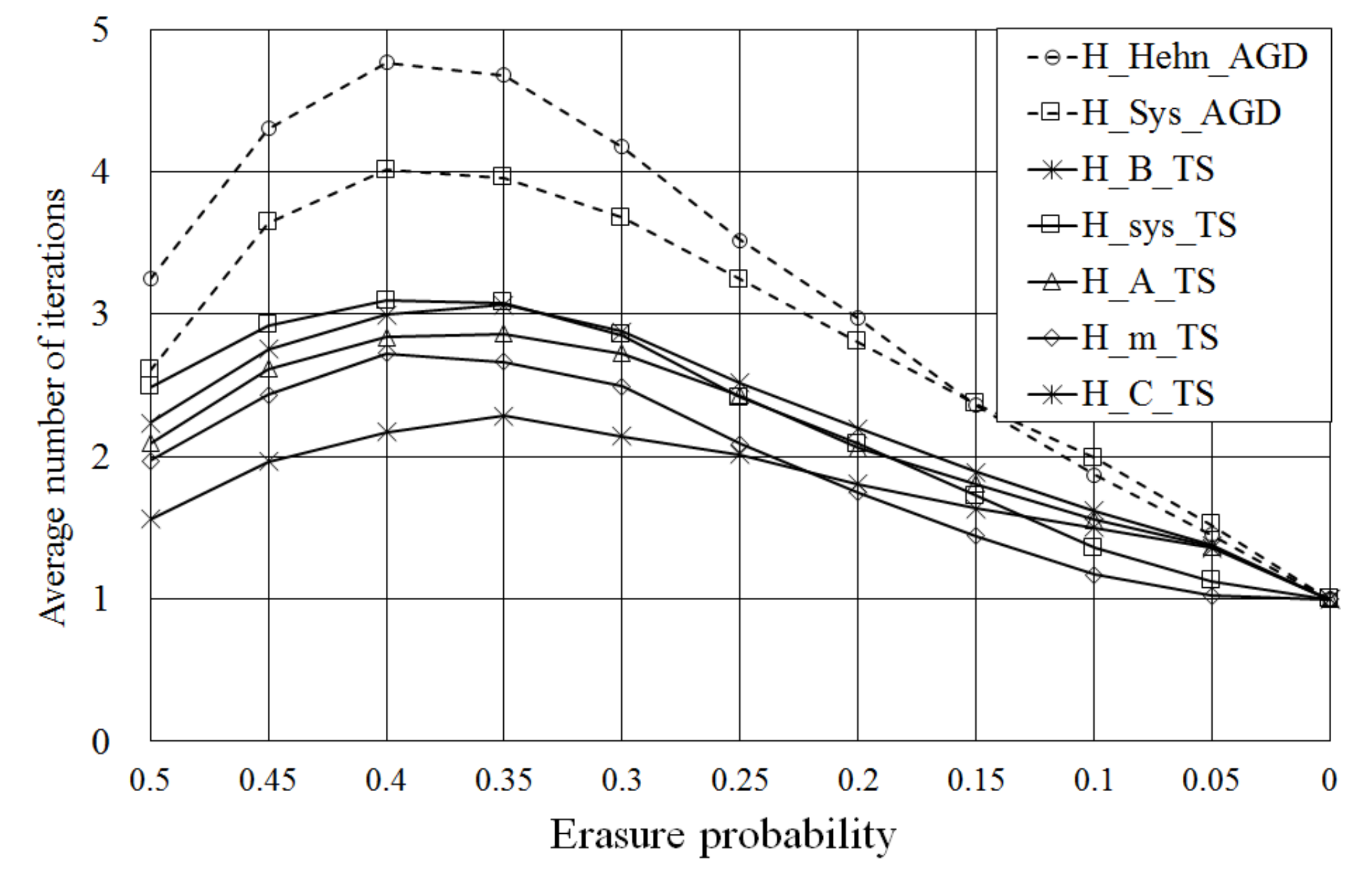}}%
\label{EGolay_Iter_lb}
\hfil
\subfloat[Decoding complexity]{\includegraphics[width=3.5in]{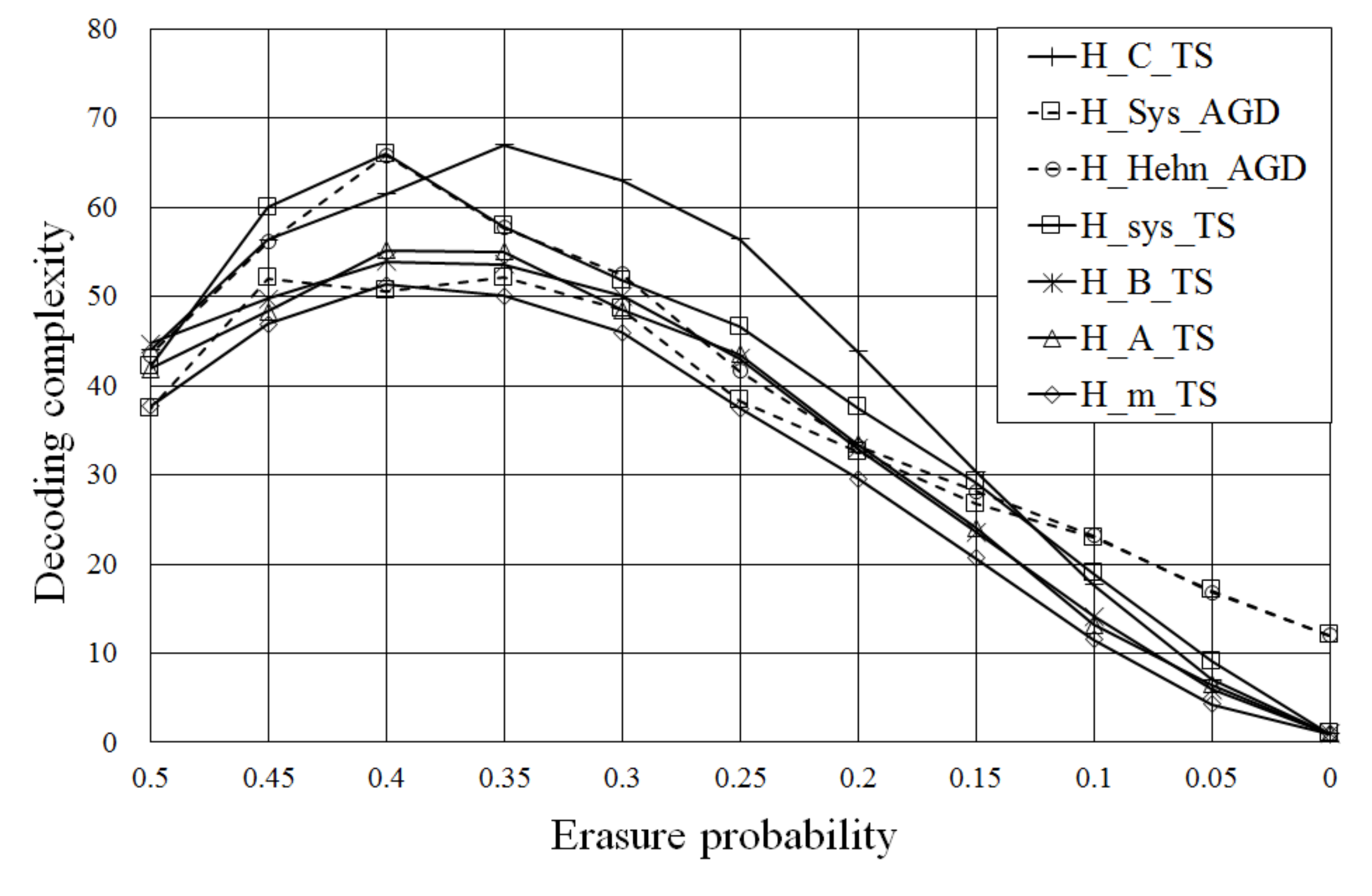}}%
\label{EGolay_Comp_lb}
\caption{The average number of iterations and the decoding complexity of AGD and TS-AGD with modified $H$ for the $(24,12,8)$ extended Golay code.}
\label{EGolay_result}
\end{figure*}

\item Ternary $(11,6,5)$ Golay code:
For the ternary $(11,6,5)$ Golay code, AGD and TS-AGD algorithms with the systematic and the modified form of their parity check matrices can achieve the ML decoding performance, whereas the number of iterations and the decoding complexity of the modified form are better than those of the systematic form. The systematic parity check matrix of the ternary $(11,6,5)$ Golay code is given as
\begin{displaymath}
H_{sys} =
\left( \begin{array}{c}
1	0	0	0	0	1	2	2	2	1	0\\
0	1	0	0	0	0	1	2	2	2	1\\
0	0	1	0	0	2	1	2	0	1	2\\
0	0	0	1	0	1	1	0	1	1	1\\
0	0	0	0	1	2	2	2	1	0	1
\end{array} \right).
\end{displaymath}

The modified form of the parity check matrix uses the parity check sequence constructed by the characteristic sequence of the cyclic difference set with parameters $(11,6,3)$ as
\begin{displaymath}
H_{m} =
\left( \begin{array}{c}
1	2	0	0	0	1	1	0	0	2	2\\
0	0	1	0	0	2	1	2	0	1	2\\
0	1	0	1	0	1	2	2	0	0	2\\
0	1	0	0	1	2	0	1	0	2	2\\
0	2	0	0	0	0	2	1	1	1	2
\end{array} \right)
\end{displaymath}

where the parity check sequence is given as
\begin{displaymath}
s_{p,m}(t)=\left(\begin{array}{c}0	1	0	0	0	1	1	1	0	1	1\end{array} \right).
\end{displaymath}

The average number of iterations and the decoding complexity of the ternary Golay codes are described in Fig. \ref{TGolay_result}, which shows performance similar to the previous cases for the Golay and extended Golay codes. By numerical analysis, there are no undecodable erasure patterns for the number of erasures $e<6$, there are 66 undecodable erasure patterns for $e=6$, and there are no decodable erasure patterns for $e>6$ for the modified form and the ML decoders.

\begin{figure*}[!t]
\centering
\subfloat[The average number of iterations]{\includegraphics[width=3.5in]{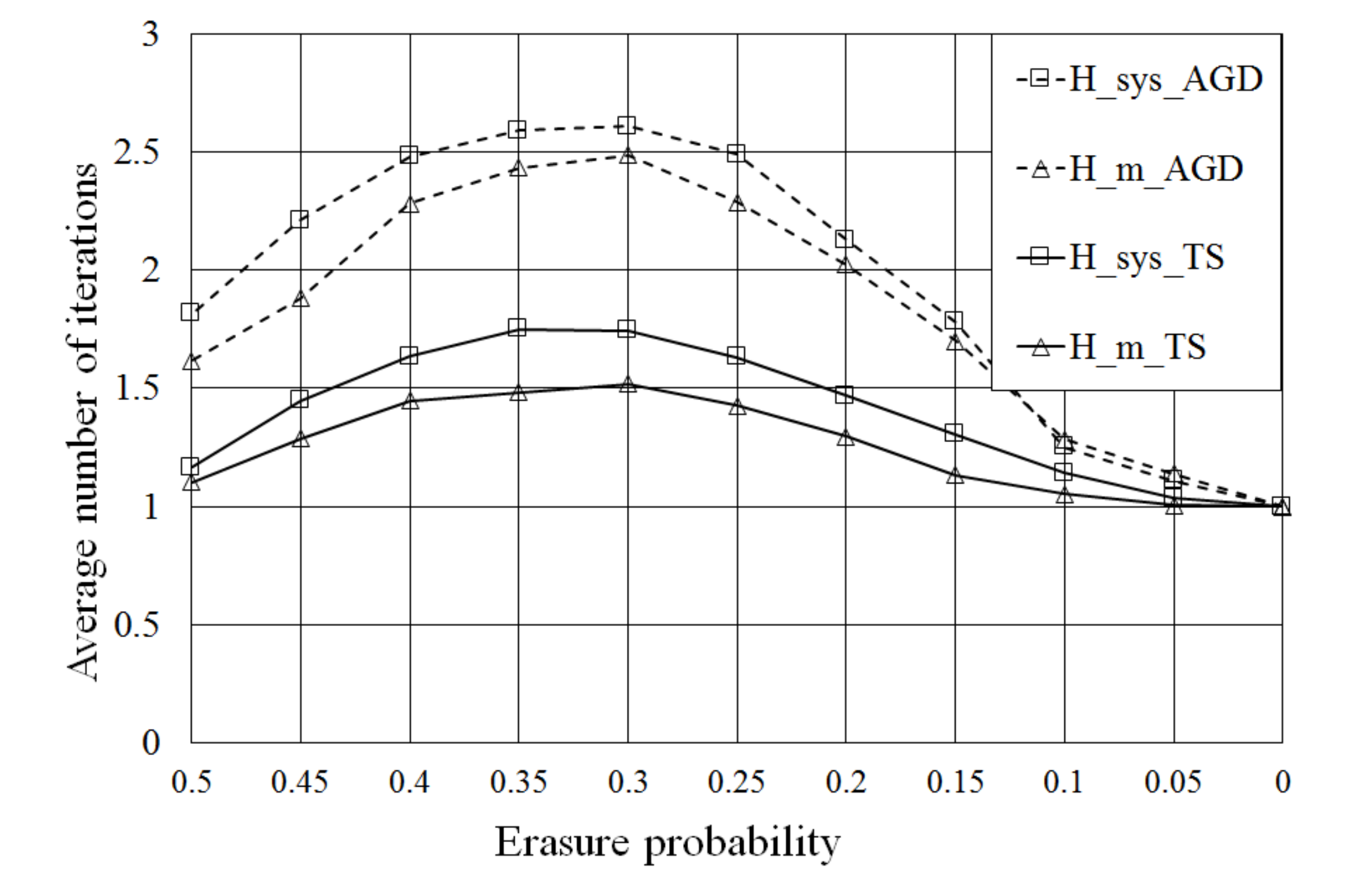}}%
\label{TGolay_Iter_lb}
\hfil
\subfloat[Decoding complexity]{\includegraphics[width=3.5in]{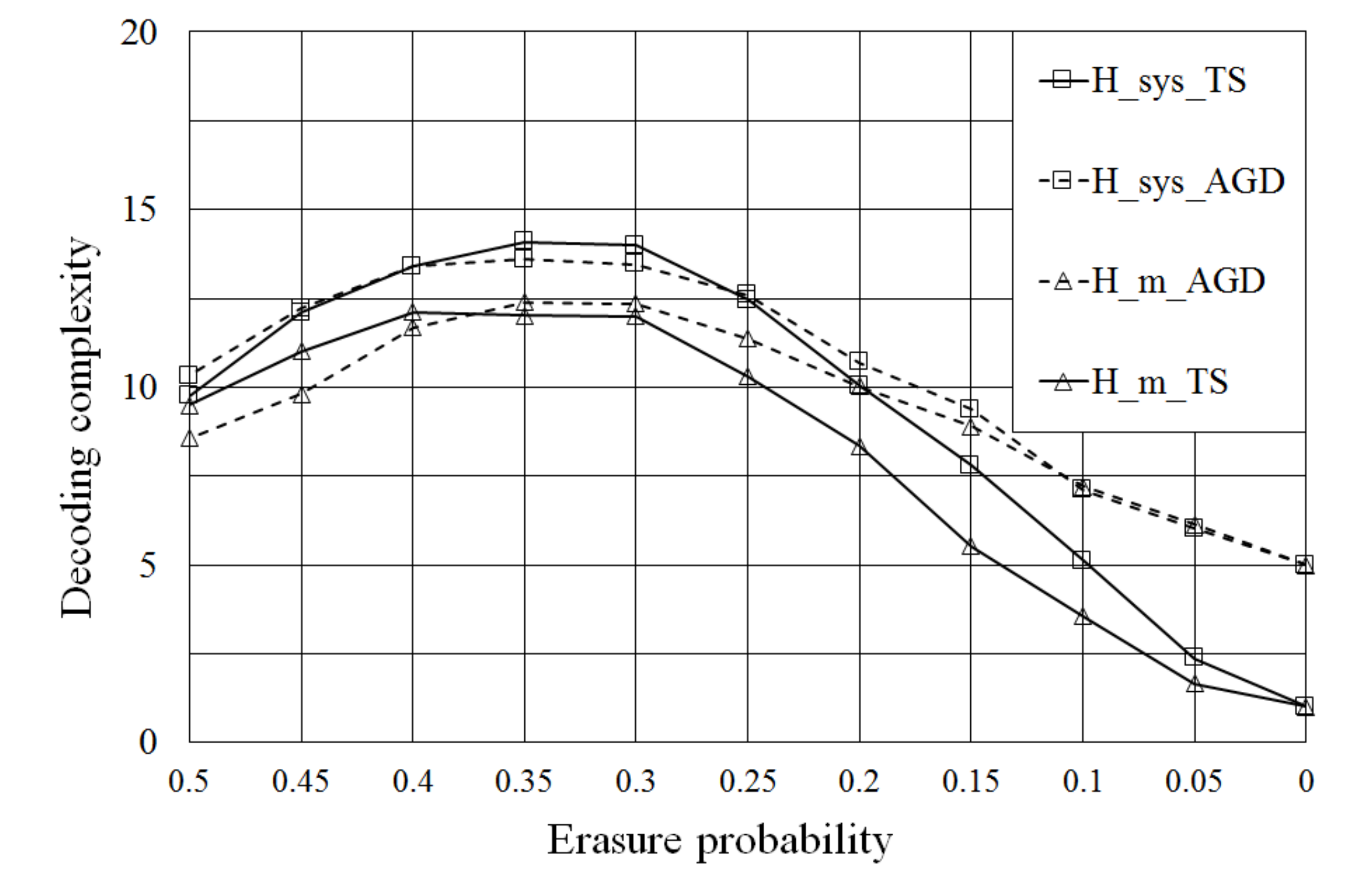}}%
\label{TGolay_Comp_lb}
\caption{The average number of iterations and the decoding complexity of AGD and TS-AGD with modified $H$ for the $(11,6,5)$ ternary Golay code.}
\label{TGolay_result}
\end{figure*}
\end{enumerate}

\subsubsection{Proposed TS-AGD Algorithms for Binary Primitive BCH Codes}
Binary primitive BCH codes are widely used due to their low-complexity encoding, large designed distance, and guaranteed decoding performance for certain number of erasures. However, BCH codes require inherently high decoding complexity and their decoding performance is degraded for large $n$ and $k$. The proposed TS-AGD can overcome the disadvantages of BCH codes by the low-complexity decoding with improved performance. Here, the proposed TS-AGD for the $(31,21,5),(31,16,7)$, and $(63,18,21)$ BCH codes are numerically analyzed in the erasure channel.
In general, $s_p(t)$ of the BCH code is generated by the cyclic difference set but there are some cases that the cyclic difference set does not exist for some parameters of the BCH code, that is, the $(31,21,5)$ and $(63,18,21)$ BCH codes. In these cases, $S_p$ can be constructed using the union of cyclotomic cosets of the finite field as an alternative construction method. In general, $s_p(t)$ does not have constant out-of-phase autocorrelation but has relatively low values of out-of-phase autocorrelation. Thus, this construction method of $s_p(t)$ also results in good decoding performance.
\begin{enumerate}[label=(\roman*)]
\item $(31,21,5)$ BCH code: 
AGD and TS-AGD with the parity check matrix of the systematic and modified forms are simulated. 
First, the systematic parity check matrix of the $(31,21,5)$ BCH code is given as 

\begin{displaymath}
H_{sys} =
\left( \begin{array}{c}
1	0	0	0	0	0	0	0	0	0	1	0	0	1	0	1	0	0	1	0	0	1	1	1	1	0	1	0	1	0	1\\
0	1	0	0	0	0	0	0	0	0	1	1	0	1	1	1	1	0	1	1	0	1	0	0	0	1	1	1	1	1	1\\
0	0	1	0	0	0	0	0	0	0	1	1	1	1	1	0	1	1	1	1	1	1	0	1	1	0	0	1	0	1	0\\
0	0	0	1	0	0	0	0	0	0	0	1	1	1	1	1	0	1	1	1	1	1	1	0	1	1	0	0	1	0	1\\
0	0	0	0	1	0	0	0	0	0	1	0	1	0	1	0	1	0	0	1	1	0	0	0	1	1	0	0	1	1	1\\
0	0	0	0	0	1	0	0	0	0	1	1	0	0	0	0	0	1	1	0	1	0	1	1	1	1	0	0	1	1	0\\
0	0	0	0	0	0	1	0	0	0	0	1	1	0	0	0	0	0	1	1	0	1	0	1	1	1	1	0	0	1	1\\
0	0	0	0	0	0	0	1	0	0	1	0	1	0	0	1	0	0	1	1	1	1	0	1	0	1	0	1	1	0	0\\
0	0	0	0	0	0	0	0	1	0	0	1	0	1	0	0	1	0	0	1	1	1	1	0	1	0	1	0	1	1	0\\
0	0	0	0	0	0	0	0	0	1	0	0	1	0	1	0	0	1	0	0	1	1	1	1	0	1	0	1	0	1	1
\end{array} \right).
\end{displaymath}

Note that there are no difference sets for the parity check sequence of the $(31,21)$ BCH code. Alternatively, we generate $s_p(t)$ using two cyclotomic cosets of the finite field $F_{2^5}$ including the elements $\alpha^7$ and $\alpha^{11}$ of the finite field $F_{2^5}$. The corresponding parity check matrix in the modified form is given as

\begin{displaymath}
H_{m} =
\left( \begin{array}{c}
1	0	0	0	1	1	0	0	1	0	1	0	1	0	1	1	0	1	0	0	1	0	1	0	0	0	0	0	0	1	0\\
0	0	1	0	1	1	0	0	0	0	1	0	0	1	0	0	0	0	0	0	1	1	1	0	1	0	0	1	0	1	1\\
0	0	0	1	1	0	0	0	0	0	1	1	0	1	0	1	1	1	1	0	0	1	1	0	0	0	0	0	0	1	0\\
0	0	0	0	0	1	1	0	1	0	1	1	1	1	0	0	1	1	0	0	0	0	0	0	1	0	0	0	0	1	1\\
0	1	0	0	1	1	0	1	1	0	0	1	0	0	0	0	1	1	1	0	0	1	0	0	1	0	0	0	1	0	0\\
0	0	0	0	0	1	0	0	0	1	1	1	1	0	1	0	0	0	1	0	0	1	0	0	1	0	0	1	1	0	1\\
0	1	0	0	1	0	0	0	1	0	0	0	1	0	0	1	1	0	1	1	0	0	1	0	0	0	0	1	1	1	0\\
0	1	0	0	0	1	0	0	1	0	0	1	0	0	1	1	0	1	0	0	0	0	0	1	0	0	0	1	1	1	1\\
0	1	0	0	0	0	0	0	1	0	1	0	0	0	1	1	0	0	1	0	1	0	1	0	1	1	0	1	0	0	1\\
0	1	0	0	1	0	0	0	0	0	0	1	1	1	0	1	0	0	1	0	1	1	0	0	1	0	1	1	0	0	0

\end{array} \right)
\end{displaymath}
and its parity check sequence is
\begin{displaymath}
s_{p}(t)=\left(\begin{array}{c}0	1	0	0	1	1	0	0	1	0	1	1	1	1	1	1	1	1	1	0	1	1	1	0	1	0	0	1	1	1	1\end{array} \right).
\end{displaymath}

\begin{table*}[]
\centering
\caption{The undecodable erasure patterns by the modified $H$ in (31, 21,5) BCH code}
\label{31_21_BCH_EP}
\begin{tabular}{|c|c|c|c|c|}
\hline
\textbf{The number}  & \textbf{Total number of}      & \textbf{TS-AGD and} & \textbf{TS-AGD and} & \textbf{ML}\\
\textbf{of erasures}  & \textbf{erasure patterns}     & \textbf{AGD of $H_{sys}$} & \textbf{AGD of $H_{m}$} & 	\\ \hline
\textbf{$\le 4$} & 0                          & 0      &   0    &                 \\ \hline
5       & 169911   		& 186             & 186       & 	186              \\ \hline
6        & 736281 	        & 5642            & 5642      &	5642             \\ \hline
7       & 2629575   	   & 83237           & 83235     &	 83235          \\ \hline
8       & 7888725  	   & 791027           & 790965    &  790965	         \\ \hline
9       & 20160075   	   & 5371029          & 5342850   &	 5340835         \\ \hline
10      & 44352165    	   & 26734183         & 26118709  &	  26030917          \\ \hline
\end{tabular}
\end{table*}

\begin{figure*}
\centering
\subfloat[The average number of iterations]{\includegraphics[width=3.5in]{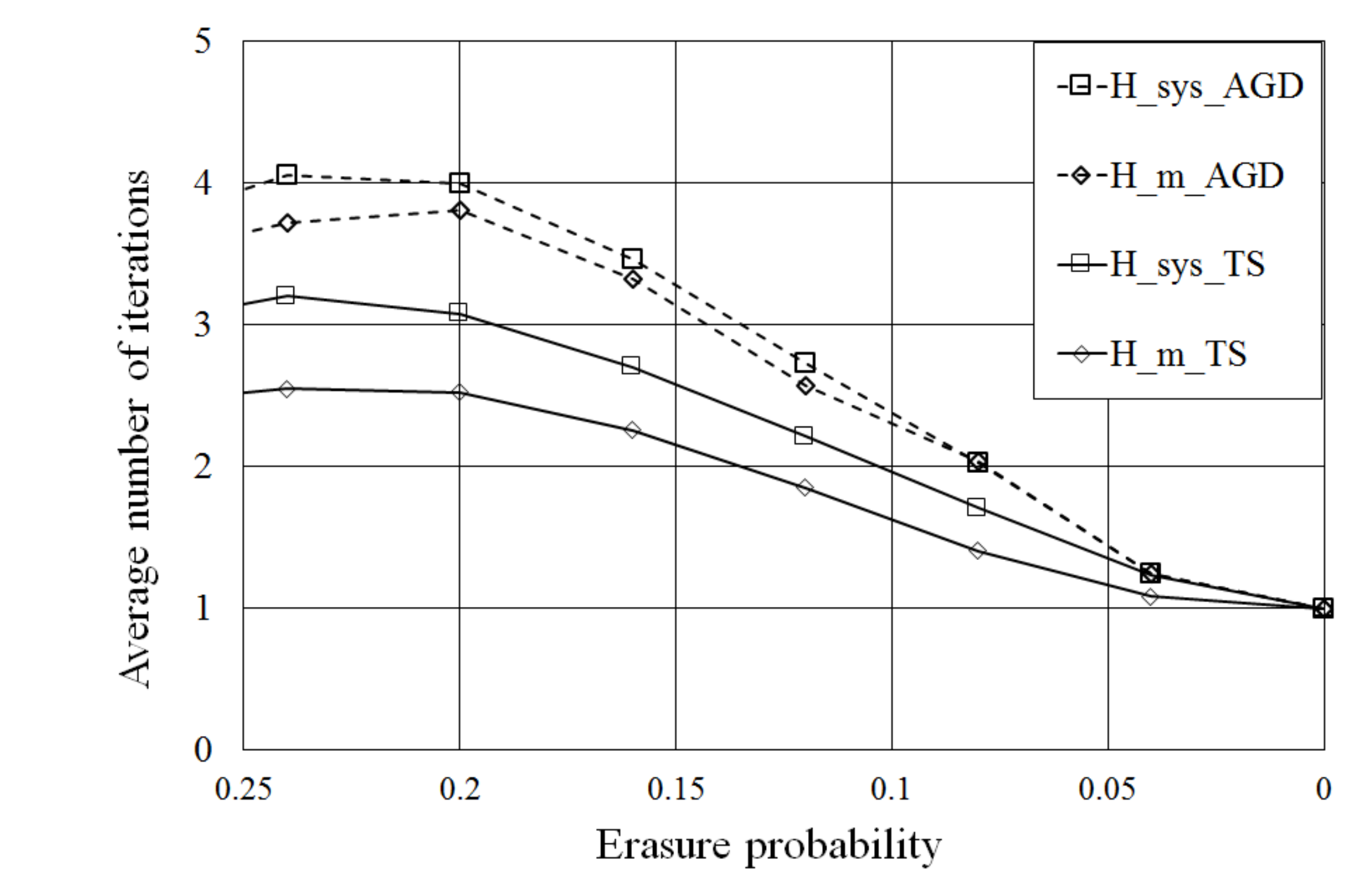}%
\label{31_21_BCH_Iter_lb}}
\hfil
\subfloat[Decoding complexity]{\includegraphics[width=3.5in]{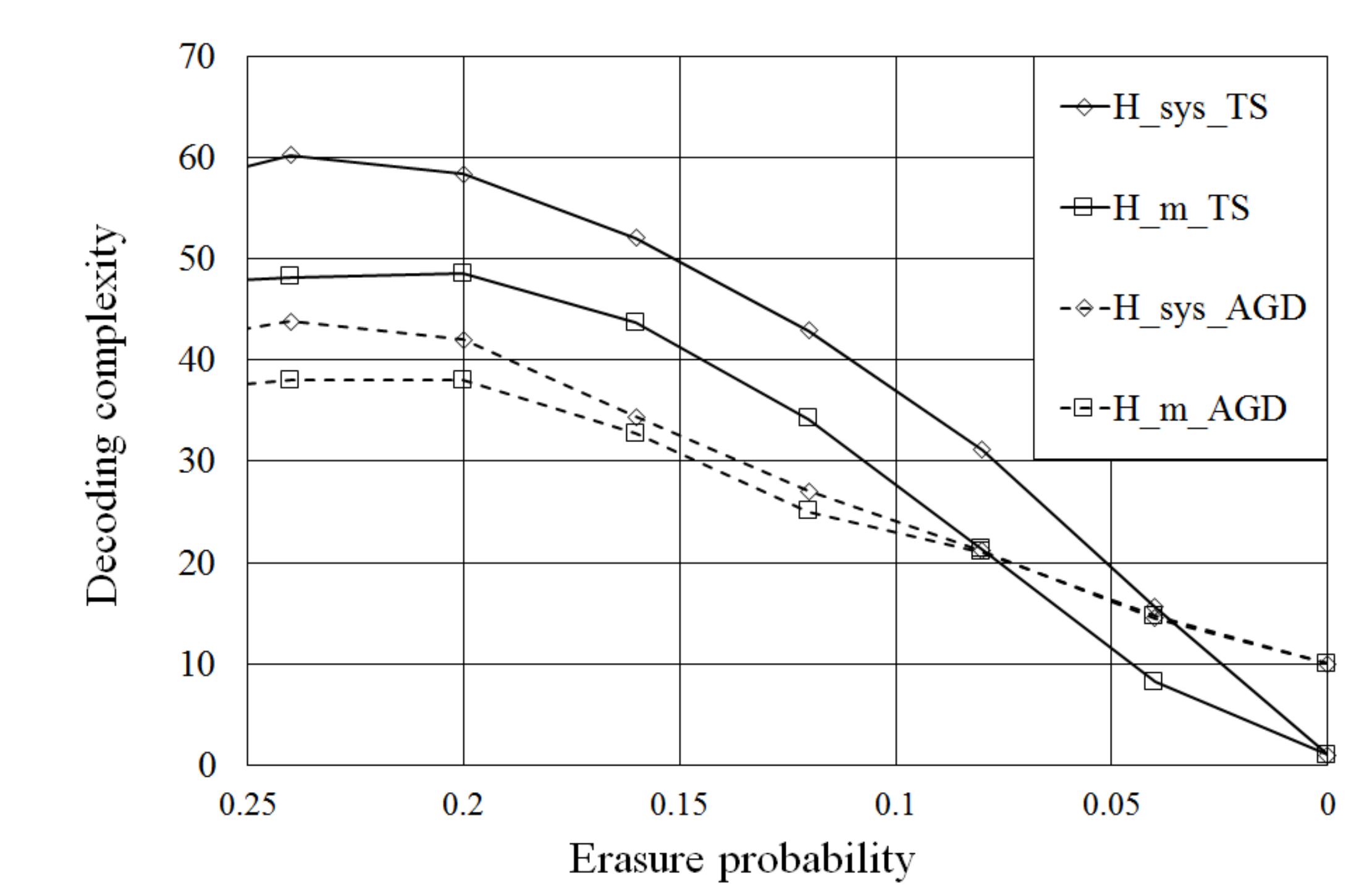}%
\label{31_21_BCH_Comp_lb}}
\caption{The average number of iterations and the decoding complexity of AGD and TS-AGD with modified $H$ for the $(31,21,5)$ binary BCH code.}
\label{31_21_BCH_label}
\end{figure*}
Table \ref{31_21_BCH_EP} shows the decoding performance of the proposed TS-AGD and AGD with $H_{sys}$ and $H_m$, which shows that the erasure decoding performance of $H_m$ is better than that of $H_{sys}$. For the average number of iterations, $H_m$ has a lower value than $H_{sys}$ and TS-AGD has a lower value than AGD. With regard to decoding complexity, TS-AGD is greater than AGD for high erasure probability because TS-AGD has a small number of $\tau$'s with $R_H(\tau)=0$ or $R_H(\tau)=1$. Therefore, TS-AGD is inefficient for a high erasure probability in that large computation is required for decoding.

\item $(31,16,7)$ BCH code:
For the $(31,16,7)$ BCH code, Hehn performed the simulation for the AGD but its parity check matrix is not given. Here, we construct its parity check matrix consisting of the same 15 cogs used in Hehn's construction as follows:
\begin{displaymath}
H_{Hehn} =
\left( \begin{array}{c}
1	0	0	0	0	0	0	0	0	0	0	0	0	0	0	1	1	0	0	1	0	0	0	0	1	1	1	1	0	0	0\\
0	1	1	1	0	1	0	0	0	0	0	1	0	0	1	0	0	0	0	0	0	0	0	1	0	0	0	0	1	0	0\\
0	1	1	1	0	0	1	0	0	0	0	0	1	0	0	0	0	1	0	0	0	0	0	1	0	0	0	0	0	0	1\\
0	0	1	1	1	0	0	0	0	0	1	0	0	0	0	0	0	0	0	0	1	0	1	0	1	0	1	0	0	0	0\\
0	1	1	1	0	0	0	0	0	0	1	0	0	1	0	0	0	0	1	0	1	0	0	1	0	0	0	0	0	0	0\\
0	1	1	0	1	1	0	0	0	1	0	1	0	0	0	0	0	0	0	1	0	0	0	0	0	0	0	1	0	0	0\\
0	1	1	0	1	0	0	1	0	0	0	0	0	0	0	0	0	0	1	0	1	0	0	0	0	0	1	1	0	0	0\\
0	1	1	0	1	0	0	0	0	0	0	0	1	0	0	0	1	1	0	0	0	0	0	0	0	0	0	1	0	1	0\\
0	1	1	0	1	0	0	0	0	0	0	0	0	0	1	0	0	0	0	0	0	1	0	0	0	1	0	1	1	0	0\\
0	1	1	0	0	1	1	0	0	0	0	1	0	0	0	0	0	0	1	0	1	0	0	0	0	0	0	0	0	0	1\\
0	1	1	0	0	1	0	0	1	0	0	1	0	0	0	0	0	0	0	0	0	1	0	0	1	1	0	0	0	0	0\\
0	1	1	0	0	0	1	0	0	0	1	0	0	1	1	0	0	0	0	0	0	0	0	0	0	0	0	0	1	0	1\\
1	1	0	0	0	0	1	0	0	0	0	1	0	0	0	0	1	0	0	0	1	0	0	0	1	1	0	0	0	0	0\\
0	0	1	0	1	0	1	0	0	0	1	0	0	0	0	1	0	0	0	0	0	0	1	0	0	1	0	0	1	0	0\\
0	1	0	1	0	0	1	0	0	0	0	1	0	1	0	0	0	1	0	0	1	0	0	0	0	0	0	0	1	0	0
\end{array} \right).
\end{displaymath}
The systematic parity check matrix of the above BCH code is given as
\begin{displaymath}
\ H_{sys} \ \ =
\left( \begin{array}{c}
1	0	0	0	0	0	0	0	0	0	0	0	0	0	0	1	1	0	0	1	0	0	0	0	1	1	1	1	0	0	0\\
0	1	0	0	0	0	0	0	0	0	0	0	0	0	0	0	1	1	0	0	1	0	0	0	0	1	1	1	1	0	0\\
0	0	1	0	0	0	0	0	0	0	0	0	0	0	0	0	0	1	1	0	0	1	0	0	0	0	1	1	1	1	0\\
0	0	0	1	0	0	0	0	0	0	0	0	0	0	0	0	0	0	1	1	0	0	1	0	0	0	0	1	1	1	1\\
0	0	0	0	1	0	0	0	0	0	0	0	0	0	0	1	1	0	0	0	1	0	0	1	1	1	1	1	1	1	1\\
0	0	0	0	0	1	0	0	0	0	0	0	0	0	0	1	0	1	0	1	0	1	0	0	0	0	0	0	1	1	1\\
0	0	0	0	0	0	1	0	0	0	0	0	0	0	0	1	0	0	1	1	1	0	1	0	1	1	1	1	0	1	1\\
0	0	0	0	0	0	0	1	0	0	0	0	0	0	0	1	0	0	0	0	1	1	0	1	1	0	0	0	1	0	1\\
0	0	0	0	0	0	0	0	1	0	0	0	0	0	0	1	0	0	0	1	0	1	1	0	0	0	1	1	0	1	0\\
0	0	0	0	0	0	0	0	0	1	0	0	0	0	0	0	1	0	0	0	1	0	1	1	0	0	0	1	1	0	1\\
0	0	0	0	0	0	0	0	0	0	1	0	0	0	0	1	1	1	0	1	0	1	0	1	0	1	1	1	1	1	0\\
0	0	0	0	0	0	0	0	0	0	0	1	0	0	0	0	1	1	1	0	1	0	1	0	1	0	1	1	1	1	1\\
0	0	0	0	0	0	0	0	0	0	0	0	1	0	0	1	1	1	1	0	0	1	0	1	1	0	1	0	1	1	1\\
0	0	0	0	0	0	0	0	0	0	0	0	0	1	0	1	0	1	1	0	0	0	1	0	0	0	1	0	0	1	1\\
0	0	0	0	0	0	0	0	0	0	0	0	0	0	1	1	0	0	1	0	0	0	0	1	1	1	1	0	0	0	1
\end{array} \right).
\end{displaymath}
The modified form of the parity check matrix for the proposed TS-AGD is given by the cyclic difference set with parameters $(31,16,8)$ as
\begin{displaymath}
H_{m} =
\left( \begin{array}{c}
1	1	0	1	0	0	1	0	0	1	1	0	0	1	0	0	0	1	1	0	1	1	0	0	0	0	1	0	0	0	0\\
1	0	1	1	0	1	1	0	0	1	0	1	1	0	0	0	0	0	1	0	1	1	0	0	0	0	1	0	0	0	0\\
1	0	0	0	1	1	1	0	0	1	1	1	1	0	0	0	0	0	1	0	0	1	1	0	1	0	0	0	0	0	0\\
1	0	0	0	0	0	0	1	0	0	1	0	0	1	0	0	0	0	1	0	1	0	1	0	0	0	1	0	0	0	0\\
1	0	0	0	0	1	0	0	1	0	1	1	1	1	0	0	0	1	1	0	1	0	1	0	1	0	0	0	0	0	0\\
1	0	0	0	0	1	0	0	0	1	0	0	0	1	1	0	0	0	0	0	1	1	0	0	0	0	1	0	0	0	0\\
0	0	0	1	0	1	1	0	0	1	1	0	0	0	0	1	0	0	0	0	0	0	1	0	1	0	0	0	0	0	0\\
1	0	0	1	0	1	0	0	0	1	1	0	0	1	0	0	1	1	0	0	1	0	1	0	1	0	1	0	0	0	0\\
1	0	0	0	0	1	0	0	0	1	1	1	0	1	0	0	0	0	0	1	0	0	1	0	0	0	0	0	0	0	0\\
0	0	0	1	0	1	0	0	0	0	0	1	1	0	0	0	0	1	1	0	1	0	0	1	0	0	0	0	0	0	0\\
1	0	0	0	0	1	0	0	0	0	0	1	0	0	0	0	0	0	1	0	1	1	1	0	0	1	0	0	0	0	0\\
0	0	0	0	0	0	1	0	0	1	1	0	0	0	0	0	0	1	1	0	0	1	0	0	1	0	0	1	0	0	0\\
1	0	0	0	0	0	0	0	0	1	1	1	0	0	0	0	0	0	1	0	0	1	0	0	0	0	1	0	1	0	0\\
0	0	0	1	0	1	0	0	0	1	0	0	1	0	0	0	0	0	0	0	1	0	0	0	1	0	1	0	0	1	0\\
0	0	0	0	0	0	1	0	0	0	1	0	1	1	0	0	0	1	1	0	1	0	0	0	0	0	0	0	0	0	1
\end{array} \right)
\end{displaymath}
and its parity check sequence is also given as
\begin{displaymath}
s_{p}(t)=\left(\begin{array}{c}1	0	0	1	0	1	1	0	0	1	1	1	1	1	0	0	0	1	1	0	1	1	1	0	1	0	1	0	0	0	0\end{array} \right).
\end{displaymath}

\begin{figure*}[!t]
\centering
\subfloat[BER and FER]{\includegraphics[width=3.5in]{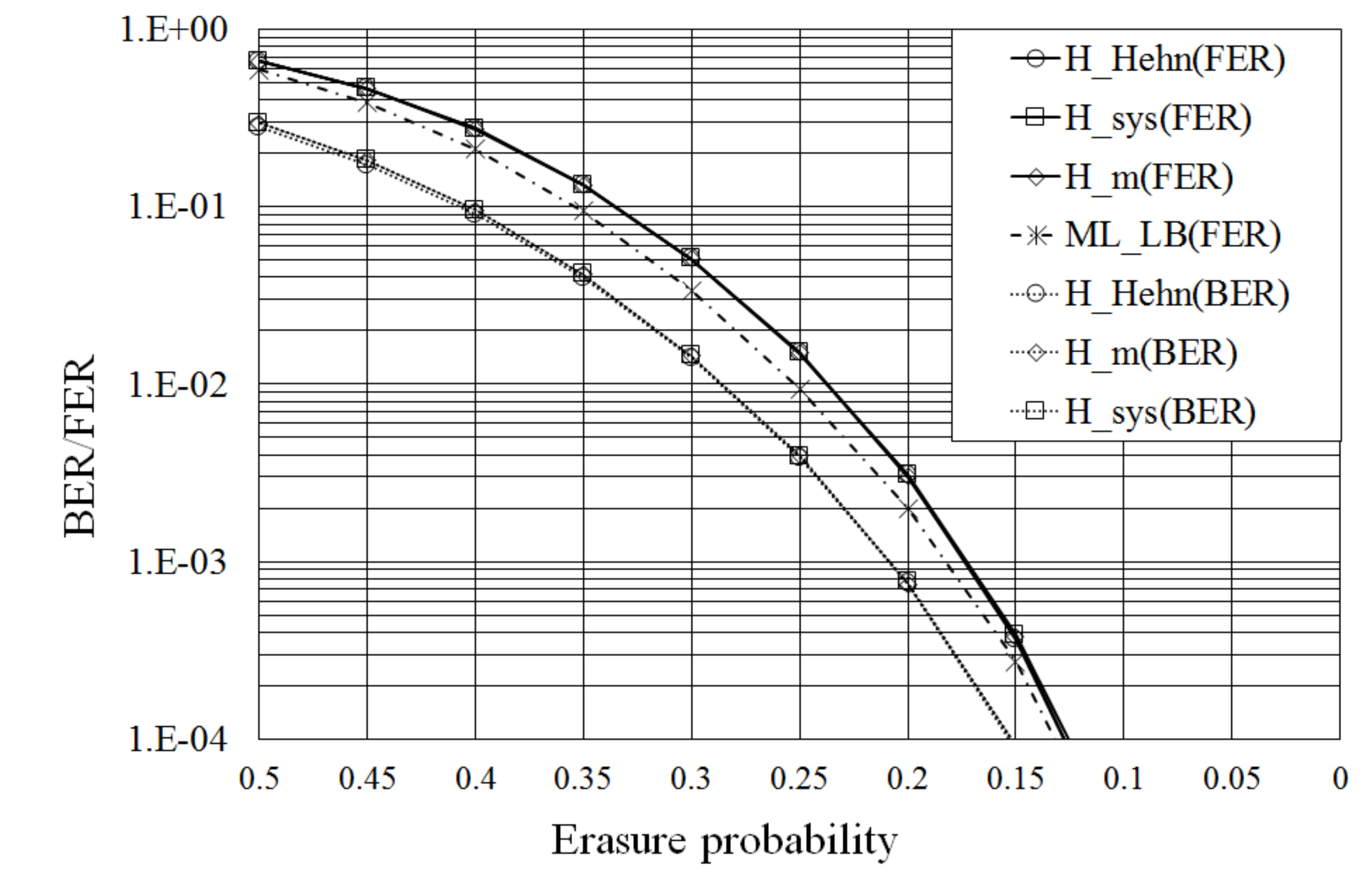}%
\label{31_16_BCH_FER_lb}}
\hfil
\subfloat[The average number of iterations]{\includegraphics[width=3.5in]{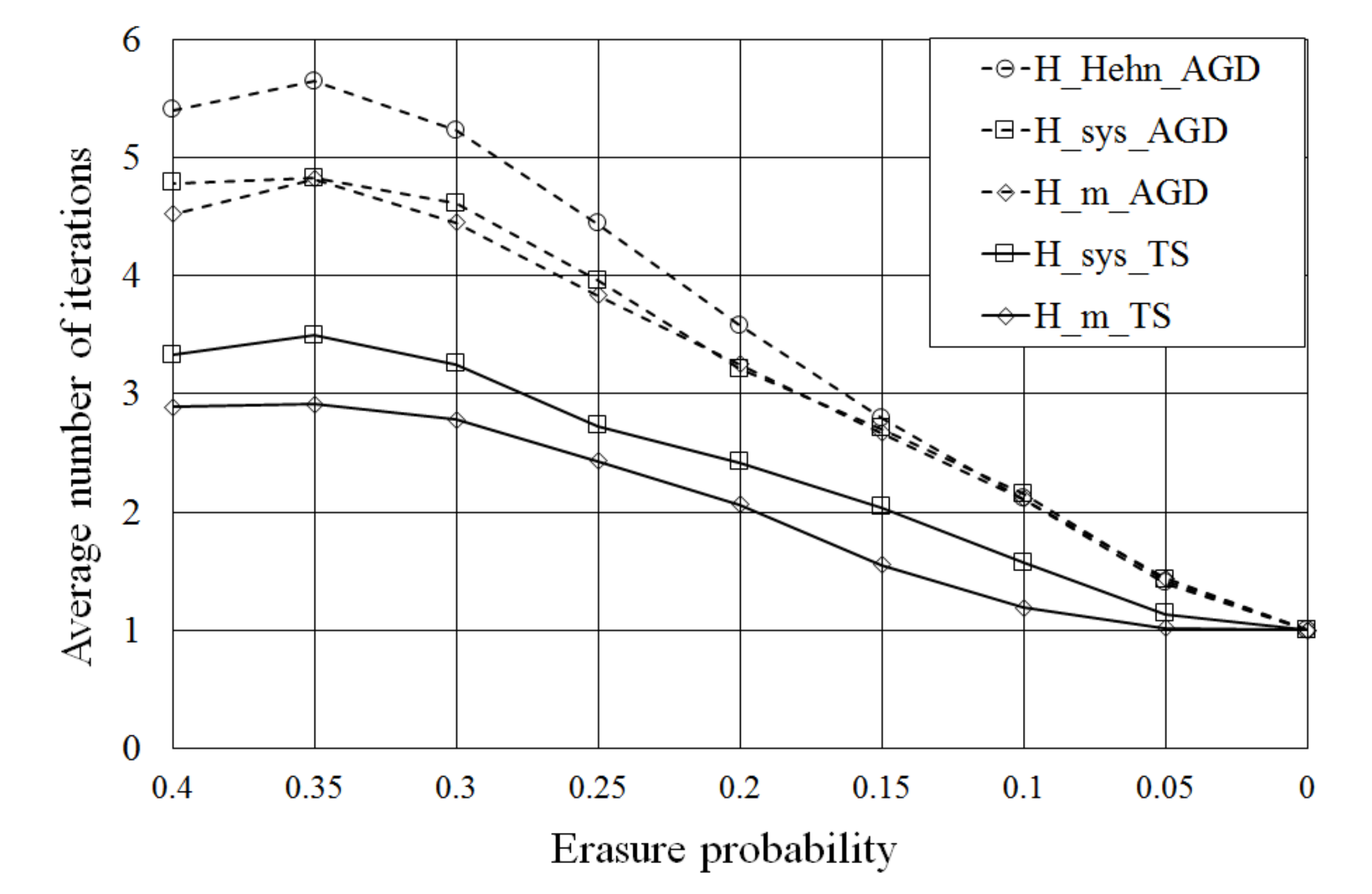}%
\label{31_16_BCH_Iter_lb}}
\hfil
\subfloat[Decoding complexity]{\includegraphics[width=3.5in]{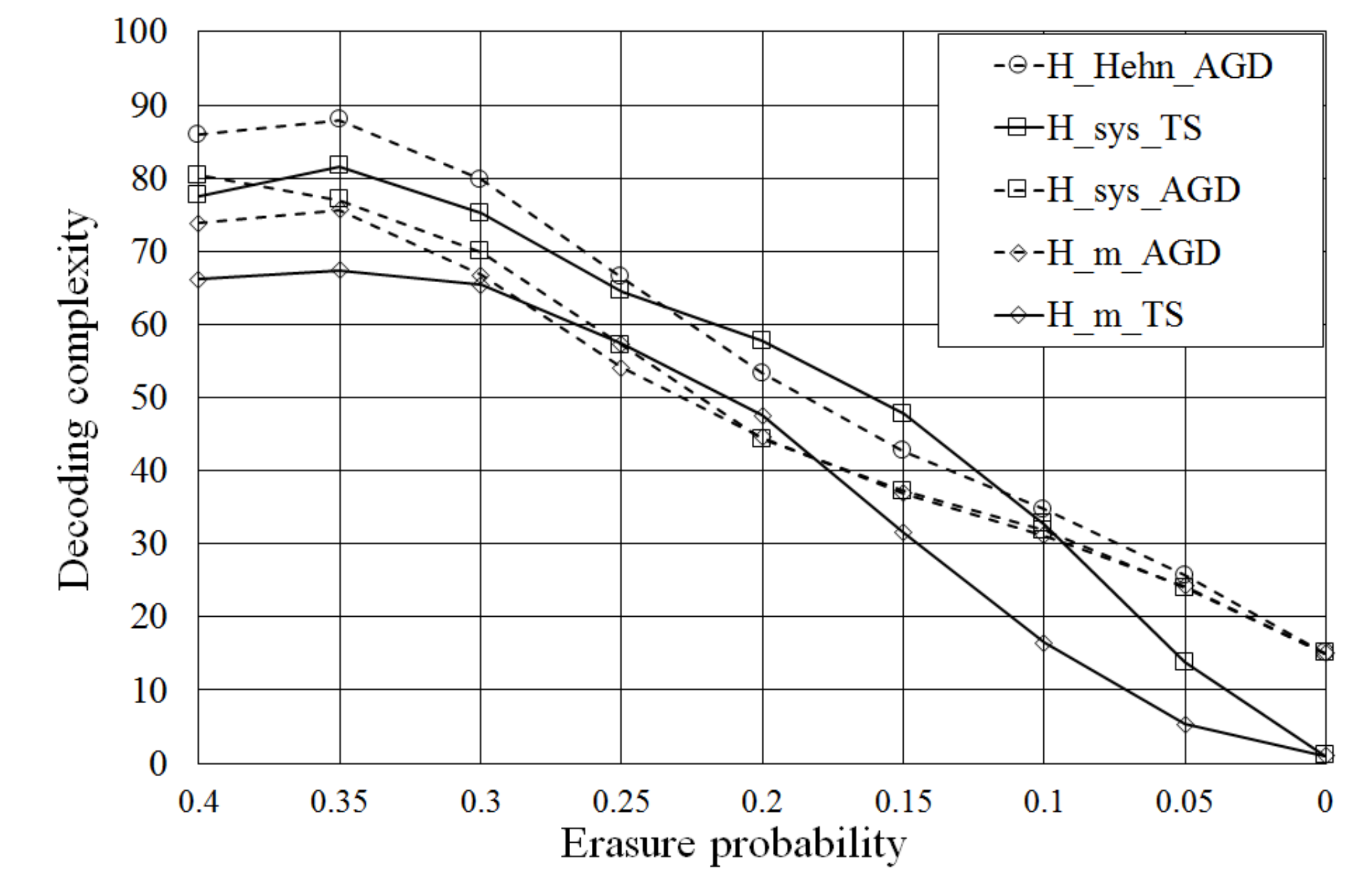}%
\label{31_16_BCH_Comp_lb}}
\caption{Decoding performance, the average number of iterations, and the decoding complexity of AGD and TS-AGD with modified $H$ for the $(31,16,7)$ binary BCH code.}
\label{31_16_BCH_label}
\end{figure*}

Henceforth, the decoding performance of the $(31,16,7)$ BCH code is numerically analyzed in three aspects; the bit error rate (BER) and frame error rate (FER), the average number of iterations, and the decoding complexity. For large values of $n$ and $k$, it is difficult to estimate the ML decoding performance by counting all of the erasure patterns. Instead, the lower bound of the ML decoding performance by $ML\_LB(FER)$ is described in Fig. \ref{31_16_BCH_label}.
For BER and FER, the decoding performances of $H_{sys}$, $H_{Hehn}$, and $H_{m}$ are nearly identical.
 
For the average number of iterations, $H_m$ also has a lower value than $H_{sys}$ and TS-AGD has significantly lower value than AGD. For AGD, $H_{sys}$ has fewer number of iterations than $H_{Hehn}$, which means that modification of the parity check matrix can also decrease the number of iterations. It is also shown that modifying the parity check matrix can decrease the decoding complexity.

\item $(63,18,21)$ BCH code:
For the parameters of large $n$ and lower coding rates, it is difficult to obtain $n-k$ cogs to generate $H$ by Hehn's method. Here, the systematic form and the proposed modification can be used to obtain the parity check matrices not only for the decoding complexity and delay reductions but also for the erasure decoding performance improvement as well.
There is no difference set for the parameters of the $(63,18,21)$ BCH code and thus the proposed modification of $H$ is done using the cyclotomic cosets of the coset leaders of the finite field $F_{2^6}$ in $\{\alpha^3, \alpha^5, \alpha^7, \alpha^{11}, \alpha^{13}, \alpha^{15}, \alpha^{23}, \alpha^{27}\}$. The corresponding parity check sequence is given as
\begin{displaymath}
s_{p}(t)=\begin{array}{c}(1	1	1	0	1	0	0	0	1	1	0	0	0	0	0	0	1	0	1	0	0	1	0	0	0	0	0	0	0	0	0	1\\ \, 	1	0	0	0	1	0	0	0	0	0	1	0	0	0	0	1	0	0	0	0	0	0	0	1	0	0	0	1	0	1	1).
\end{array}
\end{displaymath}

\begin{figure*}[!t]
\centering
\subfloat[BER and FER]{\includegraphics[width=3.5in]{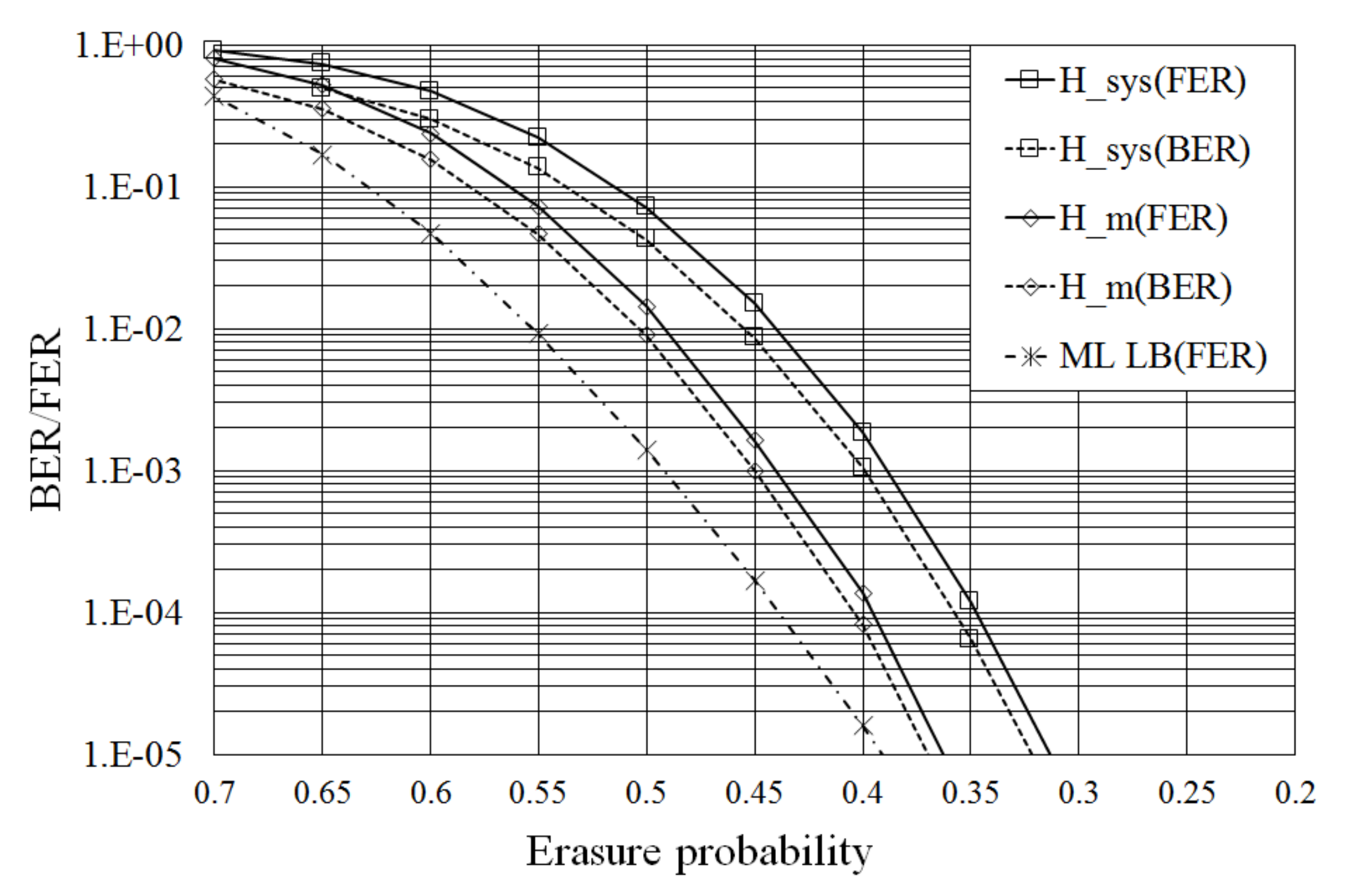}%
\label{63_18_BCH_FER_lb}}
\hfil
\subfloat[The average number of iterations]{\includegraphics[width=3.5in]{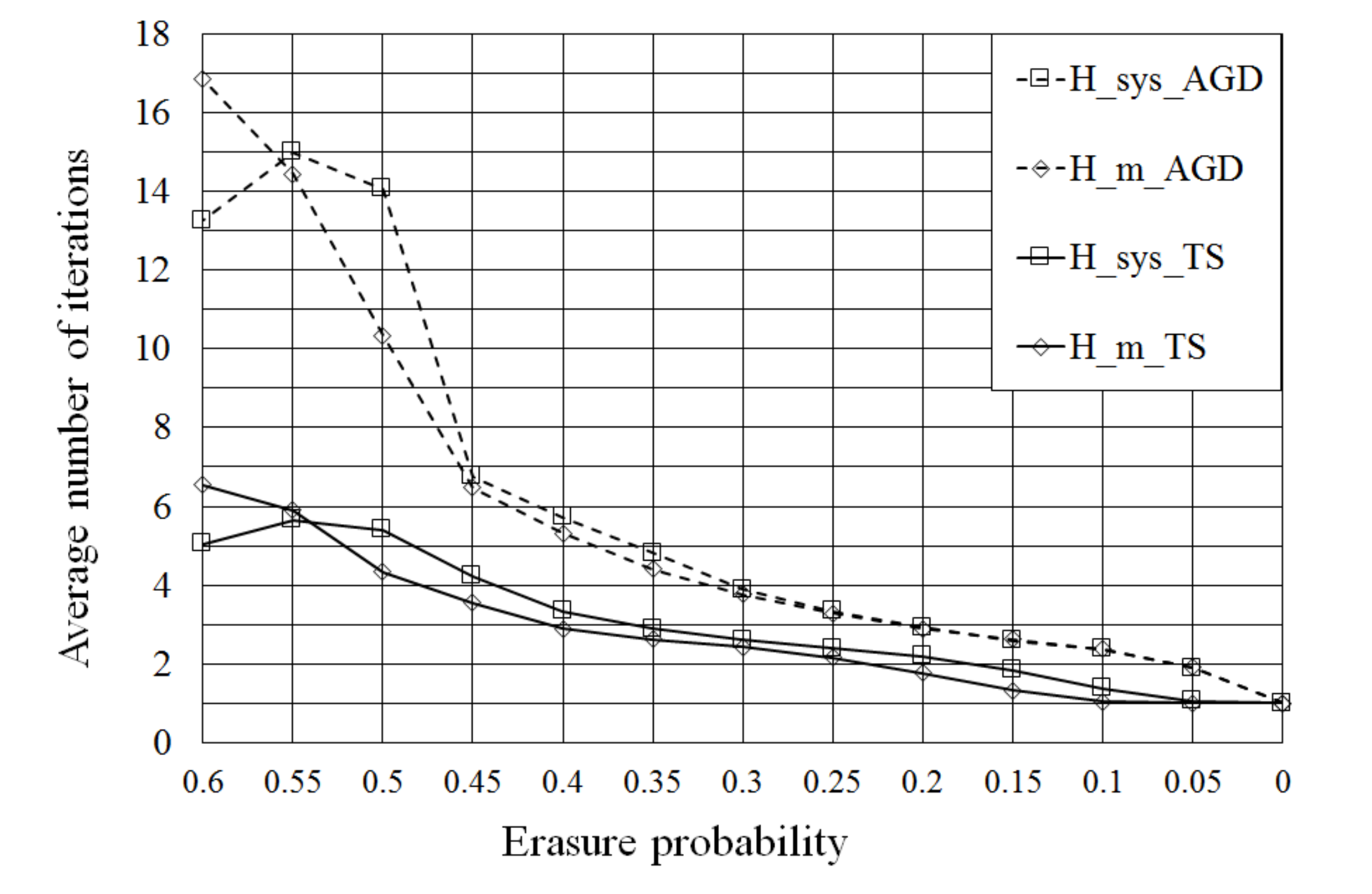}%
\label{63_18_BCH_Iter_lb}}
\hfil
\subfloat[Decoding complexity]{\includegraphics[width=3.5in]{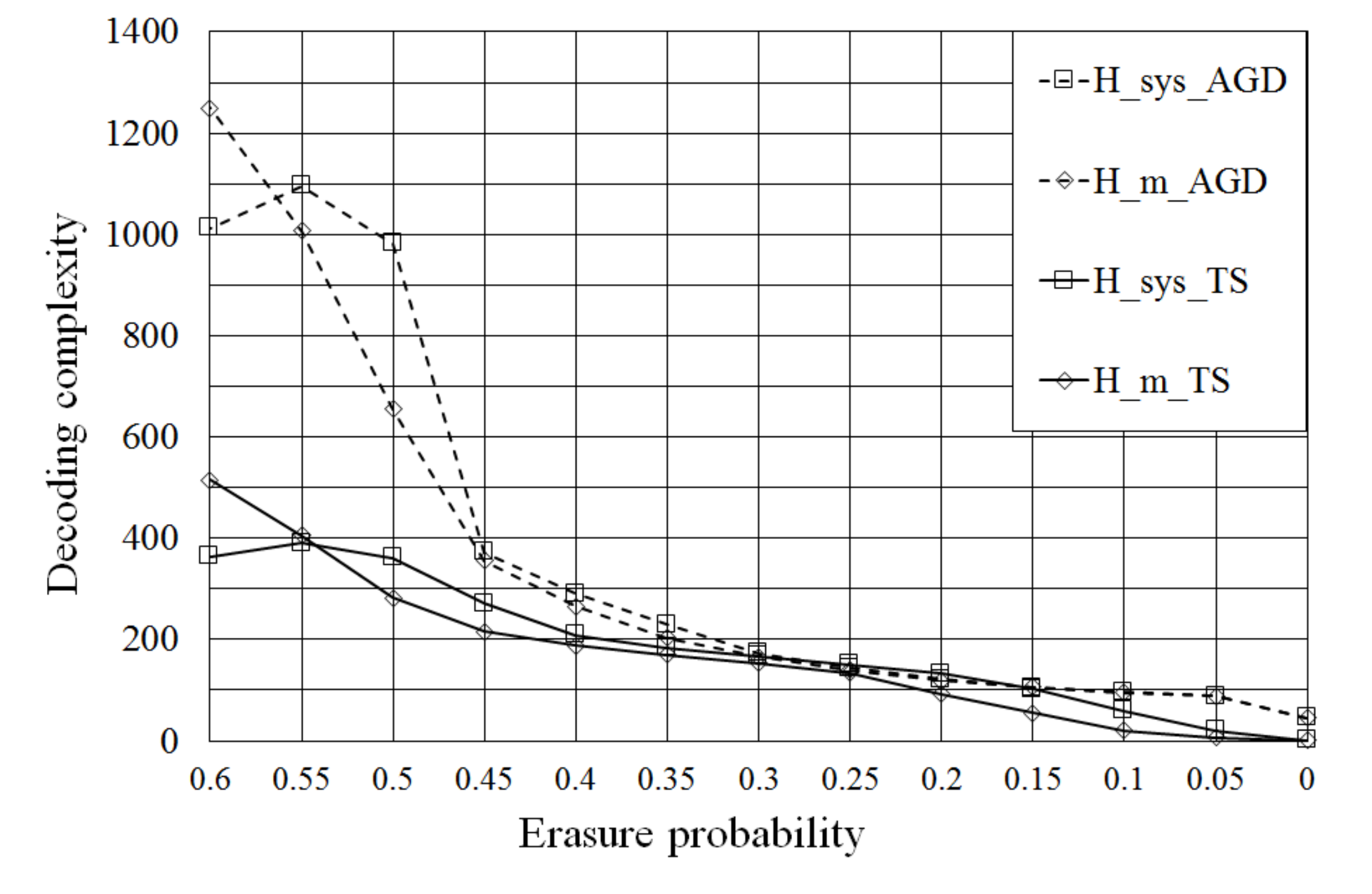}%
\label{63_18_BCH_Comp_lb}}
\caption{Decoding performance, the average number of iterations, and the decoding complexity of AGD and TS-AGD with modified $H$ for the $(63,18,21)$ binary BCH code.}
\label{63_18_BCH_label}
\end{figure*}

The decoding performance of the above BCH code is shown in Fig. \ref{63_18_BCH_label}. For BER and FER, $H_{m}$ is better than $H_{sys}$. Regarding the average number of iterations and the decoding complexity, TS-AGD is better than AGD for both $H_m$ and $H_{sys}$.

\end{enumerate}

\section{Two-Stage AGD for Cyclic MDS Codes}
\label{Prop_MDS}
In this section, the proposed TS-AGD is applied to cyclic MDS codes. In order to achieve the perfect decoding, stopping redundancy and submatrix inversion are also used for the TS-AGD for cyclic MDS codes.

\subsection{Modification of the Parity Check Matrix for Cyclic MDS Codes}
The criteria for the modification of the parity check matrices in Section \ref{Prop} can be simplified for the TS-AGD of cyclic MDS codes from the properties of the MDS codes.
\begin{proposition}[The first and third criteria for cyclic MDS codes]
For the parity check matrix of the $(n,k)$ MDS codes, $n-k$ standard basis vectors can be made in any columns of the parity check matrix and the Hamming weight of all rows is $k+1$, which is the minimum Hamming weight of its dual codes.
\end{proposition} 
\begin{IEEEproof}
This can be easily proved from the theorems in Section 2 of Chapter 11 in \cite{MacWil}.
\end{IEEEproof}

Thus, the first and third criteria can always be satisfied in the parity check matrix of the MDS codes but for the second criterion, we have to make the magnitude of the out-of-phase Hamming autocorrelation of the parity check sequence as low as possible. 

In order to improve the decoding performance of AGD and IED, the expanded parity check matrix is proposed by expanding the rows of the parity check matrix. That is, the $(n-k)\times n$ parity check matrix can be expanded to a $b(n-k) \times n$ matrix, which is composed of $b$ distinct parity check matrices. Note that each $(n-k)\times n$ parity check matrix has its own parity sequence. Then, the TS-AGD using the expanded parity check matrix decodes the received codeword by the first $(n-k)\times n$ parity check matrix. If it fails, successful decoding is possible using the subsequent parity check matrices. Note that if the perfect decoding is possible by the expanded parity check matrix, the number of rows in the expanded parity check matrix is called the \textit{stopping redundancy}.

\subsection{Proposed TS-AGD for Cyclic MDS Codes}
\subsubsection{TS-AGD Algorithm for Cyclic MDS Codes}
The procedure of the proposed TS-AGD for MDS codes is nearly identical to that of the binary codes introduced in the previous section but the detailed decoding procedure is slightly different. For the binary codes in Fig. \ref{TS_AGD_Tau_1_case}, there is a case that the erasure symbols in the non-standard basis part cannot be successfully decoded at the first iteration for $R_H(\tau) = 1$. Unlike the binary codes, TS-AGD for the MDS codes can always successfully decode the cyclically shifted received codewords with $\tau$ such that $R_H(\tau)\le 1$ because the non-standard basis columns of the parity check matrix always consist of nonzero elements. Therefore, the maximum number of iterations is reduced to 2 if there exists $\tau$ which meets the condition of $R_H(\tau)\le 1$. However, the proposed TS-AGD cannot decode the received codewords of the cyclic MDS codes for the cases of $R_H(\tau) \ge 2$.
\subsubsection{Performance Analysis of Cyclic MDS Codes and LRCs}
For $(n,k)$ cyclic MDS codes, their minimum distance is the largest value $n-k+1$, which means that the best ML decoding performance of the MDS codes can be obtained in the erasure channel. However, since the minimum Hamming weight of rows in the parity check matrix of the MDS codes is the largest value $k+1$, this degrades the decoding performance for AGD or IED compared to the binary codes due to the third modification criterion of the parity check matrix.

In order to mitigate the degradation of the decoding performance due to the third criterion without expansion of the parity check matrix, we can also consider cyclic locally repairable codes (LRCs) \cite{Tamo}, which can be constructed by slightly modifying the MDS codes as follows. LRC is originally used to reduce the decoding complexity of the repair process in distributed storage systems. LRCs have slightly shorter minimum Hamming distances than MDS codes, which reduces the decoding performance gap between AGD and the ML decoder. In this subsection, the proposed TS-AGD decoding algorithm can be applied to LRCs as well as cyclic MDS codes in order to achieve the ML decoding performance. For $(d_L^\perp-1) | k$ and $ d_L^\perp | n $ , the generator polynomial of the optimal cyclic LRC is given as
\begin{displaymath}
\label{Defining_Set_LRC_lb}
g(x)=\prod_{i\in\{L\cup M\}}{(x-\alpha^i)}
\end{displaymath}
where $L=\{l | l \text{ mod } {d^{\perp}_{L}} = 0\}$ and $M=\{0,1,2,...,n-\frac{k}{d^{\perp}_{L}-1}{d^{\perp}_{L}} \}$.
For the code parameters $(n,k)=(15,8)$, there exist a $(15,8,8)$ MDS code, a $(15,8,7)$ cyclic LRC with $d^{\perp}_{min}=5$, and a $(15,8,5)$ cyclic LRC with $d^{\perp}_{min}=3$. From (\ref{Defining_Set_LRC_lb}), the generator polynomial of the $(15,8,7)$ cyclic LRC has the zeros $\{ 1,\alpha^1, \alpha^2, \alpha^3, \alpha^4, \alpha^5, \alpha^{10} \}$. Similarly, the generator polynomial of the $(15,8,5)$ cyclic LRC has the zeros $\{ 1,\alpha^1, \alpha^2, \alpha^3, \alpha^6, \alpha^9, \alpha^{12} \}$. The characteristic sequence of the cyclic difference set with parameters $(15,8,4)$, that is, an $m$-sequence of period 15 can be used for the parity check sequence as
\begin{displaymath}
s_{p}(t)=(0	0	0	1	0	0	1	1	0	1	0	1	1	1	1).
\end{displaymath} 
Then, the corresponding masks $A$ of the parity check matrices of the $(15,8,8)$ MDS code, and the $(15,8,7)$ and $(15,8,5)$ cyclic LRCs are given as
\begin{displaymath}
\ \ \ A_{MDS} \ \ \ = 
\left( \begin{array}{c}
1	0	0	1	0	0	1	1	0	1	0	1	1	1	1\\
0	1	0	1	0	0	1	1	0	1	0	1	1	1	1\\
0	0	1	1	0	0	1	1	0	1	0	1	1	1	1\\
0	0	0	1	1	0	1	1	0	1	0	1	1	1	1\\
0	0	0	1	0	1	1	1	0	1	0	1	1	1	1\\
0	0	0	1	0	0	1	1	1	1	0	1	1	1	1\\
0	0	0	1	0	0	1	1	0	1	1	1	1	1	1
\end{array} \right)
\end{displaymath}
\begin{displaymath}
A_{LRC_{(15,8,7)}} \ =
\left( \begin{array}{c}
1	0	0	1	0	0	1	0	0	1	0	0	1	0	0\\
0	1	0	1	0	0	1	1	0	1	0	1	1	1	1\\
0	0	1	1	0	0	0	1	0	1	0	1	1	1	1\\
0	0	0	1	1	0	1	1	0	1	0	1	1	1	1\\
0	0	0	1	0	1	1	1	0	1	0	1	1	1	1\\
0	0	0	1	0	0	1	1	1	1	0	1	1	1	1\\
0	0	0	1	0	0	1	1	0	1	1	1	1	1	1
\end{array} \right)
\end{displaymath}
\begin{displaymath}
A_{LRC_{(15,8,5)}} \ \ =
\left( \begin{array}{c}
1	0	0	1	0	0	1	1	0	1	0	1	1	1	1\\
0	1	0	0	0	0	1	0	0	0	0	1	0	0	0\\
0	0	1	0	0	0	0	1	0	0	0	0	1	0	0\\
0	0	0	0	1	0	0	0	0	1	0	0	0	0	1\\
0	0	0	1	0	1	1	1	0	1	0	1	1	1	1\\
0	0	0	1	0	0	0	0	1	0	0	0	0	1	0\\
0	0	0	1	0	0	1	1	0	1	1	1	1	1	1
\end{array} \right).
\end{displaymath}
\begin{table*}[]
\centering
\caption{The undecodable erasure patterns for the $(15, 8)$ cyclic MDS code and LRCs}
\label{15_8_MDS_EP}

\begin{tabular}{|c|c|c|c|c|c|c|c|}
\hline
\textbf{The} & \textbf{Total number } & \textbf{TS-AGD} & \textbf{TS-AGD} & \textbf{TS-AGD, AGD} & \textbf{TS-AGD and} & \textbf{ML} & \textbf{ML} \\
\textbf{number of} & \textbf{of erasure} & \textbf{and AGD} & \textbf{and AGD} & \textbf{and ML with} & \textbf{AGD with} & \textbf{with} & \textbf{with}\\
\textbf{erasures} & \textbf{patterns} & \textbf{with $H_{sys}$} & \textbf{with $H_{MDS}$} & \textbf{$H_{{LRC}_{(15,8,5)}}$} & \textbf{ $H_{{LRC}_{(15,8,7)}}$} & \textbf{$H_{{LRC}_{(15,8,7)}}$} & \textbf{$H_{MDS}$}\\ \hline
$\le$ 3 &  & 0 & 0 & 0 & 0 & 0 & 0 \\ \hline
4 & 1365 & 90 & 0 & 0 & 0 & 0 & 0 \\ \hline
5 & 3003 & 1128 & 168 & 60 & 3 & 0 & 0 \\ \hline
6 & 5005 & 3520 & 2380 & 820 & 400 & 0 & 0 \\ \hline
7 & 6435 & 5820 & 5680 & 3600 & 3570 & 405 & 0 \\ \hline
\end{tabular}
\end{table*}
\begin{figure*}[!t]
\centering
\subfloat[The average number of iterations]{\includegraphics[width=3.5in]{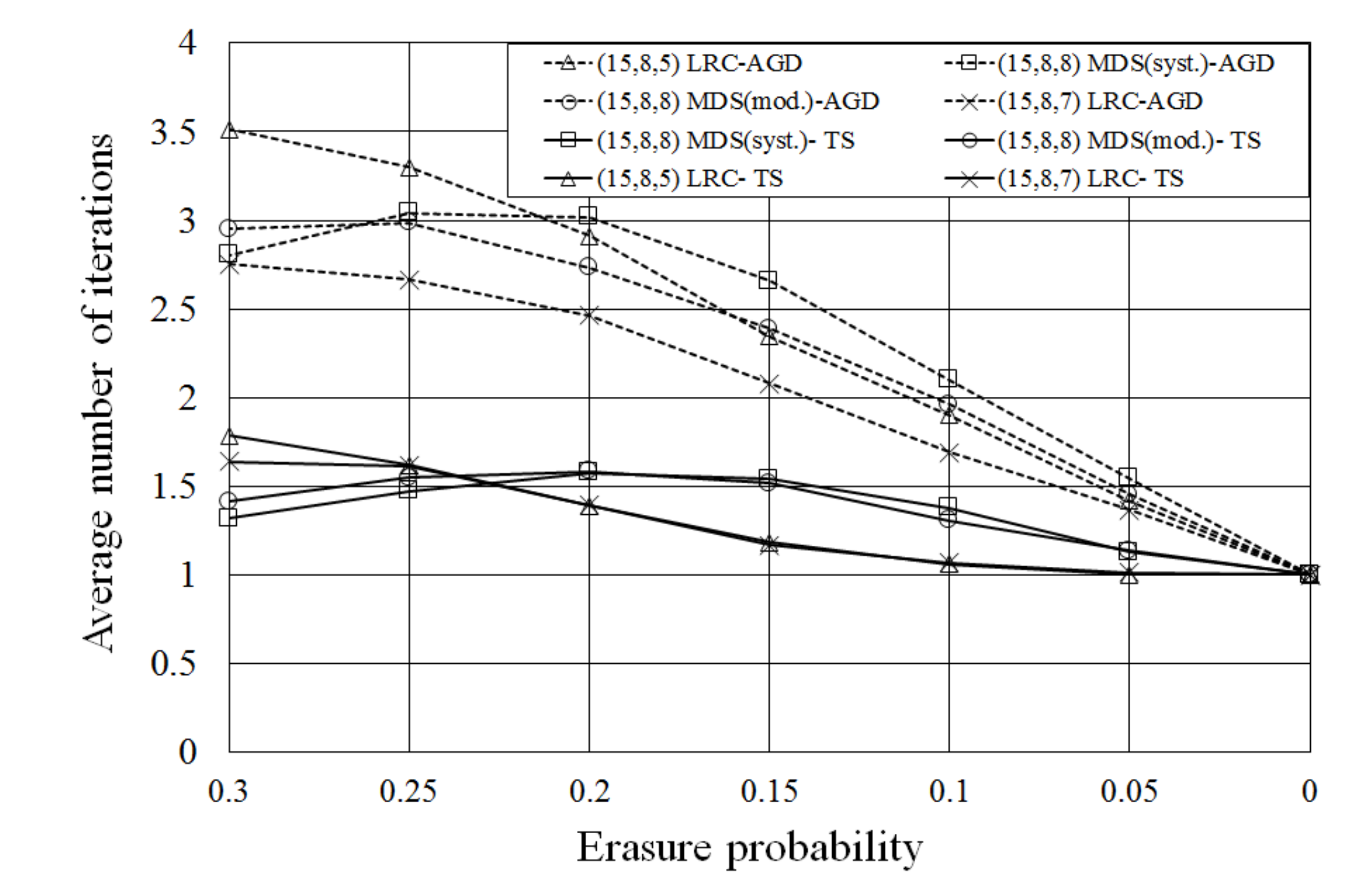}%
\label{15_8_MDS_iter_lb}}
\hfil
\subfloat[Decoding complexity]{\includegraphics[width=3.5in]{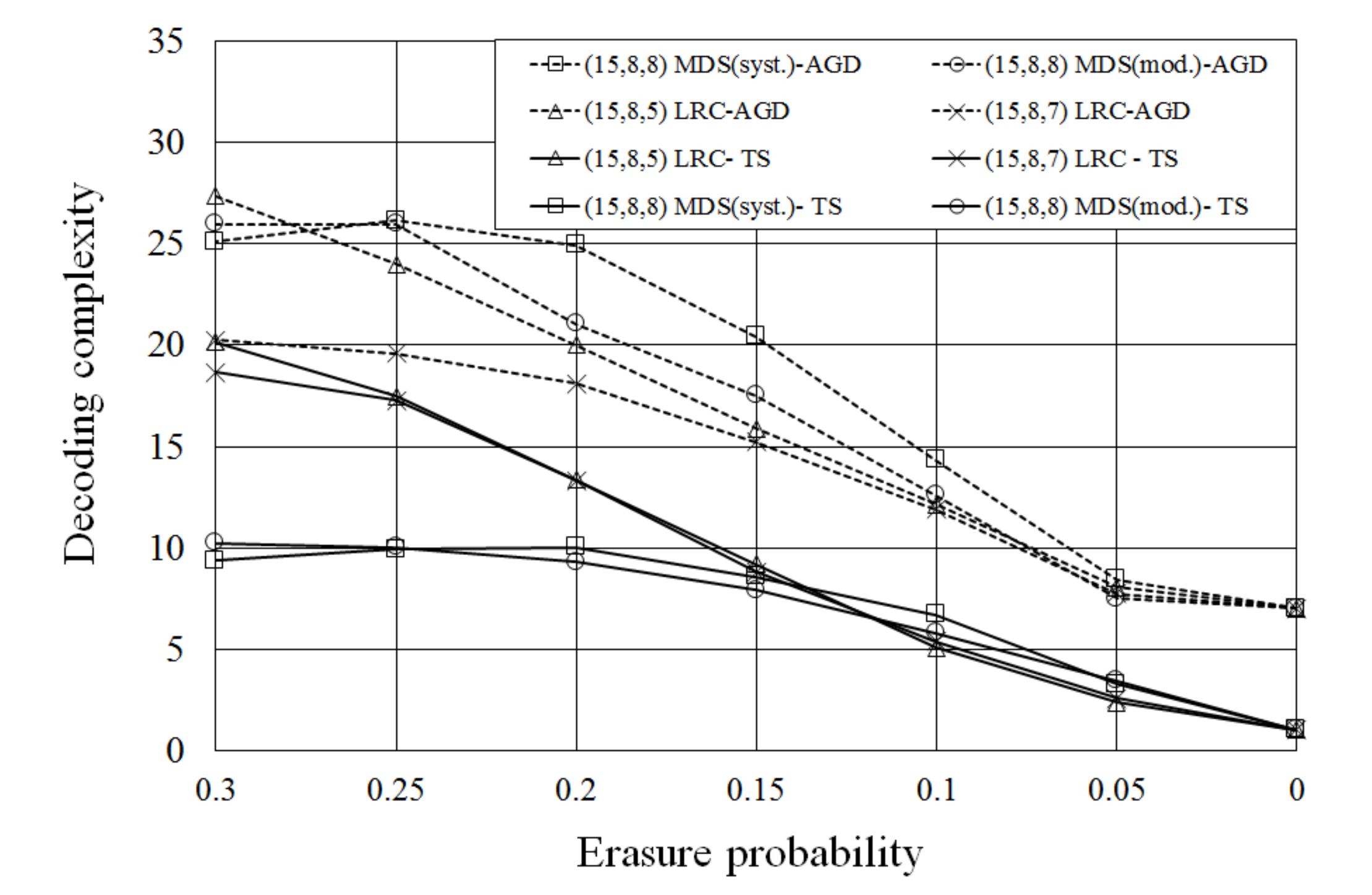}%
\label{15_8_MDS_Comp_lb}}
\caption{The average number of iterations and the decoding complexity of the $(15,8)$ cyclic MDS code and cyclic LRCs.}
\label{15_8_MDS_label}
\end{figure*}
The erasure decoding performance of the above three $(15,8)$ codes is shown in Table \ref{15_8_MDS_EP}. Clearly, the ML decoding performance of LRC with lower $d^{\perp}_{L}$ is degraded compared to that of MDS codes but the performance gap between TS-AGD and the ML decoder becomes smaller. For the $(15,8,5)$ cyclic LRC, TS-AGD performs the perfect decoding, that is, there is no difference in the decoding performance between TS-AGD and the ML decoder. TS-AGD has the best performance for the $(15,8,7)$ cyclic LRC which can replace the $(15,8,8)$ MDS code. Fig.\ref{15_8_MDS_label} shows that TS-AGD has lower decoding complexity and the fewer iterations than those of AGD for the MDS code and LRCs. For TS-AGD, MDS codes has lower decoding complexity than LRCs because MDS code uses the TS-AGD algorithm. That is, TS-AGD decoding for LRC is not always successful when $R_H(\tau)=1$, because LRC is not an MDS code and some elements of the non-standard basis columns of the parity check matrix of LRC are zero.

Instead of mitigating the strict condition of the cyclic MDS codes by the cyclic LRCs, the expanded parity check matrix can be used to enhance the erasure decoding performance of the cyclic MDS codes. Numerical analysis of the expanded parity check matrix using $m$-sequences is introduced in the following subsection. 

\subsubsection{Performance Analysis of TS-AGD With Expanded Parity Check Matrix for Cyclic MDS Codes}
To analyze the erasure decoding performance of TS-AGD with expanded parity check matrix for cyclic MDS codes, it is necessary to know the Hamming auto- and cross-correlations of the parity check sequences of the expanded parity check matrix. By counting the number of decodable $(n-k)$-erasure patterns by Lemma \ref{Bonferroni_lb}, the decoding performance of the TS-AGD with expanded parity check matrix can be estimated. Each term of the expanded parity check matrix in (\ref{Bonferroni_ineq}) can be modified as in the following proposition and theorems.
\begin{proposition}[The first term in the Bonferroni inequality in the expanded parity check matrix]
\label{FT_EPCM_lb}
The first term in (\ref{Bonferroni_ineq}) is modified in the expanded parity check matrix as follows:
\begin{displaymath}
{\sum_{I\subset A, |I|=1}|E_i|}=bn(k(n-k)+1).
\end{displaymath}
\begin{IEEEproof}
Suppose that the expanded parity check matrix has $b$ parity check sequences. The $\tau$-shifted parity check sequence $s_p(t+\tau)$ can correct $n-k$ erasure symbols in the following two cases:
\begin{enumerate}
\item $R_H(\tau)=0$: $n-k$ erasure symbols are located in the $n-k$ standard basis indices and the decoder can correct the ${n-k \choose n-k}=1$ erasure pattern.
\item $R_H(\tau)=1$: $n-k-1$ erasure symbols are located in the standard basis  indices and the decoder can correct the ${n-k \choose n-k-1}{k \choose 1}=(n-k)k$ erasure patterns.
\end{enumerate}
Each parity check sequence of the expanded parity check matrix has up to $n$ cyclically equivalent parity check sequences and therefore, it can correct up to $n(k(n-k)+1)$ erasure patterns.
\end{IEEEproof}
\end{proposition}
\begin{theorem}[The second term in the Bonferroni inequality in the expanded parity check matrix]
\label{ST_EPCM_lb}
The second term in (\ref{Bonferroni_ineq}) can also be modified in the expanded parity check matrix as 
\begin{multline}
\label{ST_EPCM_Eqn_lb}
{\sum_{I\subset V, |I|=2}\left|\bigcap_{i \in I}{E_i}\right|} = \sum_{\tau_1,\tau_2=0}^{n-1}{\sum_{1 \le i < j \le b}{}}
 {\left(4F_{s_{p,i}(t+\tau_1),s_{p,j}(t+\tau_2)}(k-2)+nF_{s_{p,i}(t+\tau_1),s_{p,j}(t+\tau_2)}(k-1)\right)} 
\end{multline}
where $F_{s_{p,i}(t+\tau_1),s_{p,j}(t+\tau_2)}(\gamma)$ returns 1 if $\sum_{t=0}^{n-1}(s_{p,i}(t+\tau_1)s_{p,j}(t+\tau_2))=\gamma$ and 0, otherwise.
\begin{IEEEproof}
The proof is similar to that of Theorem \ref{selectionpcps}. For the $i$-th and $j$-th cyclically shifted parity check sequences, the number of doubly counted decodable erasure patterns is expressed as
\begin{equation}
\label{BFIQ_LB}
{\sum_{I\subset V, |I|=2}\left|\bigcap_{i \in I}{E_i}\right|} =\sum_{\tau_1, \tau_2=0}^{n-1}{\sum_{1 \le i < j \le b}{\sum_{\gamma=0}^{k}{c_{\gamma}F_{s_{p,i}(t+\tau_1),s_{p,j}(t+\tau_2)}(\gamma)}}}
\end{equation}
where $c_\gamma$ is the number of doubly counted decodable erasure patterns from $s_{p,i}(t+\tau_1)$ and $s_{p,j}(t+\tau_2)$. Equation (\ref{BFIQ_LB}) partitions the number of doubly counted decodable erasure patterns by $\tau_1$, $\tau_2$, $s_{p_i}(t)$, $s_{p_j}(t)$, and $\gamma$. The remaining problem is to determine $c_{\gamma}$. For the given $\tau$, $s_{p_i}(t)$, and $s_{p_j}(t)$, the doubly counted decodable erasure patterns can be computed as follows: 

\begin{enumerate}
\item If $\gamma \le k-3$: A doubly counted decodable erasure pattern does not occur because there should be $n-k-3$ erasure symbols in $A_{00}$ and the remaining three erasure symbols cannot be decoded regardless of their locations of $A_{01}$, $A_{10}$, and $A_{11}$, where the Hamming cross-correlation values of one parity check sequence and the erasure sequence are larger than or equal to 2.
\item If $\gamma = k-2$: We have $|A_{11}|=k-2$, $|A_{10}|=|A_{01}|=2$, and $|A_{00}|=n-k-2$. Then, doubly counted decodable erasure patterns occur when one erasure symbol is located in $A_{01}$, one erasure is in $A_{10}$, and $n-k-2$ erasure symbols are in $A_{00}$. Therefore, $c_{\gamma}$ is 4.
\item If $\gamma = k-1$: We have $|A_{11}|=k-1$, $|A_{10}|=|A_{01}|=1$, and $|A_{00}|=n-k-1$. Then, the doubly counted decodable erasure patterns can occur when one erasure symbol is located in $A_{01}$, one erasure symbol is in $A_{10}$, and $n-k-2$ erasure symbols are in $A_{00}$, where $c_\gamma=n-k-1$. In addition, doubly counted decodable erasure patterns occur when one erasure symbol is located in $A_{11}$ and the other $n-k-1$ erasure symbols are in $A_{00}$, $A_{01}$, or $A_{10}$, where $c_\gamma=k+1$. The sum of the two cases gives us $c_\gamma=n$.
\end{enumerate}
Thus, the theorem is proved.
\end{IEEEproof}
\end{theorem}
The distribution of the Hamming auto- and cross-correlation values of the parity check sequences can be used to count the first and the second terms in the Bonferroni inequality by Proposition \ref{FT_EPCM_lb} and Theorem \ref{ST_EPCM_lb}. The Hamming auto- and cross-correlations of pseudorandom sequences, especially the $m$-sequences of period $n=2^m-1$, have been researched. There can be used to analyze the erasure decoding performance of TS-AGD. In this subsection, TS-AGD with the expanded parity check matrix for $(n, \frac{n+1}{2})$ MDS codes is analyzed, where the parity check sequences for $(n-k)\times n$ parity check matrices are constructed using the $m$-sequence and its decimated sequences.
\begin{enumerate}[label=(\roman*)]
\item $m \le 3$: 
For $m=3$, only one $(n-k)\times n$ parity check matrix with a parity check sequence constructed by the $m$-sequence of period 7 can achieve the perfect decoding. It can be easily shown by the numerical analysis.
\item $m=4$: 
There are two $m$-sequences, $s_{p_1}(t)$ and $s_{p_2}(t)$ of period $n=15$. The distribution of their Hamming cross-correlation values for $\tau\in[0,n-1]$ can be given as \cite{Niho}
\begin{displaymath}
\sum_{t=0}^{n-1}{s_{p_1}(t)s_{p_2}(t+\tau)}= \left\{ \begin{array}{ll}
3, & \textrm{4 times}\\
4, & \textrm{5 times}\\
5, & \textrm{4 times}\\
6  & \textrm{2 times}.\\
\end{array} \right.
\end{displaymath}

In this case, we can derive the number of the decodable 8-erasure patterns for the expanded parity check matrix with $b=2$ by the inclusion-exclusion principle. For the first term in (\ref{ST_EPCM_Eqn_lb}), the number of doubly counted erasure patterns is computed as $2\times5(7\times8+1)=1710$. For the second term, it is given as
\begin{displaymath}
\begin{split}
\sum_{I\subset V, |I|=2}\left|\bigcap_{i \in I}{E_i}\right| = \sum_{\tau=0}^{14}{\sum_{1 \le i < j \le 2}{}}\Big(4F_{s_{p,i}(t),s_{p,j}(t+\tau)}(5)
+15F_{s_{p,i}(t),s_{p,j}(t+\tau)}(6)\Big)
\\=\sum_{\tau=0}^{14}{(4F_{s_{p,i}(t),s_{p,j}(t+\tau)}(5))}
=15\times4\times2=120,
\end{split}
\end{displaymath} 
which makes at least 1590 decodable erasure patterns and it is the exact value because it has no triply or more counted erasure pattern. Note that the total number of 8-erasure patterns is ${15 \choose 8} = 6435$.
\item $m \ge 5$:
\begin{table}[]
\centering
\caption{Hamming cross-correlation distribution of $m$-sequence of period $31$ and its decimated sequences}
\label{31_15_m_sequence_relation_lb}
\begin{tabular}{|c|c|c|}
\hline
Decimation  & The number of Hamming & Types\\ 
& cross-correlations (values) &  \\ \hline
3 & 3(6,8,10) & Gold \cite{Gold}, Kasami \cite{Kasami}\\ \hline
5 & 3(6,8,10) & Gold \cite{Gold} \\ \hline
7 & 3(6,8,10) & Welch \cite{Welch}\\ \hline
11 & 3(generally, 5)(6,8,10) & Boston and McGuire \cite{Boston}\\ \hline
15 & 6(6,7,8,9,10,11) & \\ \hline
\end{tabular}
\end{table}
For $m=5$, there are six $m$-sequences of period 31, whose Hamming cross-correlation distributions are listed in Table \ref{31_15_m_sequence_relation_lb}. For these cases, the peak correlation values are either 10 or 11, which means that there are no doubly counted decodable erasure patterns because there are no Hamming correlation values larger than $k-2=14$. Therefore, any expanded parity check matrix with $b\le 6$ has erasure decoding performance achieving the upper bound. The maximum value of the Hamming cross-correlation of the $m$-sequence and its decimated sequences can be derived as $\left\lfloor\frac{{2^m}+2^{\frac{(m+2)}{2}}+3}{4}\right\rfloor$ \cite{Dowling}. Thus, for $m \ge 5$, we have
\begin{displaymath}
\left\lfloor\frac{{2^m}+2^{\frac{(m+2)}{2}}+3}{4}\right\rfloor < 2^{m-1}-2=k-2. 
\end{displaymath} 
Thus, it is easily checked that there are no doubly counted erasure patterns for construction of the expanded parity check matrices for any combinations of an $m$-sequence and its decimated sequences for $m\ge 5$. Therefore, the total number of decodable erasure patterns of TS-AGD with expanded parity check matrix constructed by $m$-sequences can be maximized for the cyclic MDS codes. However, the performance by TS-AGD is worse than that of the perfect decoding for cyclic MDS codes.

\end{enumerate}

\subsection{Perfect Decoding by TS-AGD With Expanded Parity Check Matrix for Cyclic MDS Codes}
In order to achieve the perfect decoding by TS-AGD with the expanded parity check matrix for cyclic MDS codes, the required stopping redundancy $\rho=b(n-k)$ is grown exponentially as $n$ and $k$ increase. It is known to be NP-hard to calculate or approximate the exact value $\rho$ for the perfect decoding \cite{McGregor}. For small values of $n$ and $k$ of the cyclic MDS codes, it will be shown that we can find the optimal $\rho$ which meets the lower bound. In this paper, we only consider the case of $\rho \le 3(n-k)$ and we propose a construction method of the expanded parity check matrix for the perfect decoding in this subsection. First, three lower bounds on the stopping redundancy are proposed.

\subsubsection{Lower Bounds on $\rho$ for the Perfect Decoding by TS-AGD}
The first lower bound is similar to the Gilbert (sphere packing) bound as in the following theorem.

\begin{theorem}[Gilbert-like lower bound]
\label{Gil_Lik_LB}
\begin{displaymath}
\rho \ge \left \lceil \frac{\binom{n}{n-k}}{n((n-k)k+1)} \right \rceil (n-k)
\end{displaymath}
\end{theorem}
\begin{IEEEproof}
Suppose that an expanded parity check matrix has $b$ parity check sequences. If there are no doubly counted decodable erasure patterns, the number of decodable erasure patterns is $bn((n-k)k+1)$ from Proposition \ref{FT_EPCM_lb}, which is larger than or equal to ${n \choose {n-k}}$. Thus, the theorem is proved.
\end{IEEEproof}

This bound can be improved by lotto designs \cite{Colbourn} and the Bonferroni inequality \cite{Bonferroni}.

\begin{definition}[Lotto design \cite{Colbourn}]
An $(n,k,p,t)$-lotto design is an $n$-set $V$ of elements and a set $\cal{B}$ of $k$-element subsets (blocks) of $V$, such that for any $p$-subset $P$ of $V$, there is a block $B\in\cal{B}$, for which $|P\cap B|\ge t$. $L(n,k,p,t)$ denotes the smallest number of blocks in any $(n,k,p,t)$-lotto design.
\end{definition}

By using the above lotto design, we can obtain more improved lower bounds on $\rho$ as follows.

\begin{theorem}[Lower bound by the lotto design]
\label{lotto_LB}
\begin{displaymath}
\rho \ge \left \lceil \frac{L(n,n-k,n-k,n-k-1)}{n} \right \rceil (n-k)
\end{displaymath}
\end{theorem}
\begin{IEEEproof}
In order to decode the cyclic MDS codes, it is necessary for the Hamming correlation values to be less than or equal to 1, i.e., $R_H(\tau)\le 1$. It also means that the intersection between the standard basis indices and the support set of erasure sequence is larger than or equal to $n-k-1$. Then, the minimum number of the parity check sequences in expanded parity check matrix is lower bounded by  $\frac{L(n,n-k,n-k,n-k-1)}{n}$.
\end{IEEEproof}

The lotto design improves the lower bound in Theorem \ref{lotto_LB}. Moreover, the lower bound for $\rho$ can also be improved by the Bonferroni inequality as follows.

\begin{theorem}[Lower bounds by the Bonferroni inequality]
\label{Impv_bound_Bonferroni}
\begin{equation}
\label{Impv_bound_Bonferroni_eqn_lb}
\rho \ge \left \lceil \frac{{n \choose n-k}-4A(n,6,n-k)}{n(k(n-k)-3)} \right\rceil(n-k)
\end{equation}
where $A(n,d,w)$ denotes the maximum number of codewords for the $(n,d,w)$ constant weight codes.
\end{theorem}
\begin{IEEEproof}
For an expanded parity check matrix with $b$ parity check sequences, the number of decodable erasure patterns follows (\ref{Bonferroni_ineq}), whose right hand side can be used as an upper bound. In this approach, the second term is calculated as in Theorem \ref{ST_EPCM_lb} if the Hamming auto- and cross-correlations of the parity check sequences are known. If cyclically shifted parity check sequences are considered, we have $bn$ distinct parity check sequences, which can be considered as constant weight codewords. Now, we have to count the number of two codewords with Hamming distance less than or equal to 4. By definition, $A(n,6,k)$ is the maximum number of $n$-tuple binary codewords which have a weight of $k$ and the minimum distance 6. Then, for each codeword, there exist at least $bn-A(n,6,n-k)$ codewords which have Hamming distance less than or equal to 4. Thus, the total number of pair of codewords with Hamming distance less than or equal to 4 is at least $\frac{bn}{2}\left(bn-A(n,6,n-k)\right)$ because all pairs are counted twice. The minimum value of the second term in the RHS of (\ref{Bonferroni_ineq}) can be computed for $n=k-2$, that is, a Hamming distance 4. Thus, we have
\begin{multline}
\sum_{I\subset V, |I|=2}\left|\bigcap_{i \in I}{E_i}\right|\ge \sum_{\tau_1,\tau_2=0}^{n-1}{\sum_{1 \le i < j \le b}{4R_{s_{p,i}(t+\tau_1),s_{p,j}(t+\tau_2)}(k-2)}}  \\ \ge 4 \times \frac{b}{2}(bn-A(n,6,n-k)).
\label{DS_49}
\end{multline}
From Proposition \ref{FT_EPCM_lb}, (\ref{DS_49}), $\left|\bigcup_{i\in V} E_i \right|={n \choose {n-k}}$, and $|V|=bn$, the right inequality in (\ref{Bonferroni_ineq}) can be modified as (\ref{Impv_bound_Bonferroni_eqn_lb}).
\end{IEEEproof}
The value of $A(n,d,w)$ is not exactly known in general and its upper bounds are used in this paper.

\subsubsection{Examples of the Perfect Decoding for $\rho \le 3(n-k)$}
Table \ref{nk_points_lb} lists the required values $b$ for the perfect decoding by TS-AGD with expanded parity check matrix for $(n,k)$ cyclic MDS codes. The underlined values denote the maximum values among the previously derived three lower bounds and the values in parenthesis refer to the lower bounds on $b$, which are different from the numerically obtained values of $b$.

Algorithm \ref{GreSystAlg} shows one of the simple construction method of the expanded parity check matrix for $(n,k)$ cyclic MDS codes using the set of $(n-k)$-erasure patterns. Using Algorithm 2, the values of $b$ for the perfect decoding are numerically derived for $(n,k)$ cyclic MDS codes in Table \ref{nk_points_lb}.

To obtain specific values of the lower bounds, the upper bounds of $A(n,d,w)$ in \cite{CW_Upper} and the lower bounds of $L(n,k,p,t)$ in \cite{LottoTable} are used.
\begin{algorithm}
\caption{Greedy algorithm for the construction of the expanded parity check matrix $H$}
\label{GreSystAlg}
\begin{algorithmic} 
\REQUIRE $b(n-k) \times n$ expanded parity check matrix $H$, the set of all $(n-k)$ erasure sequences $E$, $S = \phi$, $b = 1$, and $\tau=0$
\WHILE{$E\setminus S \neq \phi$}
\STATE{$v \in E\setminus S$}
\STATE{$s_{p,b}(t) \gets \bar{v}$}
\FOR {$\tau=0$ \TO $n-1$}
\STATE{$C \gets \{ s_e(t) |\sum_{t=0}^{n-1}{s_e(t)s_{p,b}(t+\tau)} \le 1, s_e(t) \in E \} $}
\STATE{$S \gets S \cup C$}
\ENDFOR
\STATE{$b \gets b+1$}
\ENDWHILE
\end{algorithmic}
\end{algorithm}

\begin{table*}[]
\centering
\caption{Required $b \le 3$ for perfect decoding with expanded parity check matrix for $(n,k)$ cyclic MDS codes with $3\le k \le 8$ and $8 \le n \le 20$}
\label{nk_points_lb}
\begin{tabular}{|c|c|c|c|c|c|c|c|c|c|c|c|c|c|}
\hline
$k$/$n$ & 8 & 9 & 10 & 11 & 12 & 13 & 14 & 15 & 16 & 17 & 18 & 19 & 20 \\ \hline
3 & 1 & 1 & 1 & 1 & 1 & \textbf{2(1)} & 1 & {\ul 2} & {\ul 2} & {\ul 2} & {\ul 2} & 2 & \textbf{3(2)} \\ \hline
4 & 1 & 1 & {\ul 2} & \textbf{3(2)} & \textbf{3(2)} & {\ul 3} & {\ul \textbf{4(3)}} & \textbf{4(3)} &  &  &  &  &  \\ \hline
5 & 1 & 1 & {\ul 2} & 2 & {\ul 3} & \textbf{5(3)} &  &  &  &  &  &  &  \\ \hline
6 & 1 & 1 & {\ul 2} & 2 & \textbf{4(3)} &  &  &  &  &  &  &  &  \\ \hline
7 & 1 & 1 & 1 & 2 & {\ul 3} &  &  &  &  &  &  &  &  \\ \hline
8 & 1 & 1 & 1 & 1 & \textbf{3(2)} & \textbf{5(3)} &  &  &  &  &  &  &  \\ \hline
\end{tabular}
\end{table*}

Some $(n,k)$ MDS codes in Table \ref{nk_points_lb} can be analyzed as follows.

\begin{enumerate}[label=(\roman*)]
\item $(n,k)=(10,5)$: The lower bound by Theorem \ref{Impv_bound_Bonferroni} shows a stricter bound compared to the other bounds. The values $b$ by Theorems \ref{Gil_Lik_LB} and \ref{lotto_LB} are computed as
\begin{displaymath}
b_{Thm.3}=\left\lceil\frac{{10 \choose 5}}{10(5\times5+1)}\right\rceil=\left\lceil0.969\right\rceil=1
\end{displaymath}
\begin{displaymath}
b_{Thm.4}=\left\lceil\frac{L(10,5,5,4)}{10}\right\rceil=\left\lceil\frac{10}{10}\right\rceil=1
\end{displaymath}
whereas Theorem \ref{Impv_bound_Bonferroni} gives us a tighter lower bound as
\begin{displaymath}
b_{Thm.5}=\left\lceil\frac{{10\choose 5}-4\times7}{10(5\times5-3)}\right\rceil=\lceil1.0181\rceil=2.
\end{displaymath}
Using Algorithm \ref{GreSystAlg}, the expanded parity check matrix can be constructed with two parity check sequences as
\begin{displaymath}
\begin{array}{c}
s_{p,1}(t)=(1	0	1	0	0	1	1	0	1	0)$ $\\
s_{p,2}(t)=(0	1	1	1	1	0	0	1	0	0).
\end{array}
\end{displaymath}

\item $(n,k)=(11,5)$: The values $b$ of the three lower bounds are equal to 2. Construction of the expanded parity check matrix can be realized by the characteristic sequences of the cyclic difference sets with parameters $(11,5,2)$ as
\begin{displaymath}
\begin{array}{c}
s_{p,1}(t)=(0	1	0	1	1	1	0	0	0	1	0)$ $\\
s_{p,2}(t)=(0	0	1	0	0	0	1	1	1	0	1).
\end{array}
\end{displaymath}

\item $(n,k)=(13,4)$: The values of $b$ by Theorems \ref{Gil_Lik_LB}, \ref{lotto_LB}, and \ref{Impv_bound_Bonferroni} are given as
\begin{displaymath}
b_{Thm.3}=\left\lceil\frac{{13 \choose 4}}{13(4\times9+1)}\right\rceil=\left\lceil1.486\right\rceil=2
\end{displaymath}
\begin{displaymath}
b_{Thm.4}=\left\lceil\frac{L(13,4,4,3)}{13}\right\rceil=\left\lceil2.153\right\rceil=3
\end{displaymath}
\begin{displaymath}
b_{Thm.5}=\left\lceil\frac{{13\choose 4}-4\times13}{13(4\times9-3)}\right\rceil=\lceil1.545\rceil=2.
\end{displaymath}
Using Algorithm \ref{GreSystAlg}, the optimal expanded parity check matrix of the $(13,4)$ cyclic MDS code can be constructed by the following three parity check sequences as
\begin{displaymath}
\begin{array}{c}
s_{p,1}(t)=(0	0	1	1	1	0	0	0	0	0	0	0	1)$ $\\
s_{p,2}(t)=(0	0	0	0	1	0	0	0	1	0	1	1	0)$ $\\
s_{p,3}(t)=(0	1	0	1	0	0	0	0	1	1	0	0	0).
\end{array}
\end{displaymath}
\end{enumerate}

\subsection{TS-AGD With Submatrix Inversion for Cyclic MDS Codes}

Matrix inversion is not widely used in the erasure decoding but for some codes in the erasure channel, it is permissible for small submatrix inversion. In particular, raptor codes \cite{Mackay} or regenerating codes for distributed storage systems \cite{Rashimi} often use an inversion operation of a small submatrix for decoding. The conventional assumption of stopping redundancy for IED is not an inversion-based decoding, but it requires lots of additional check nodes for a large value of $n$. However, TS-AGD allowing submatrix inversion up to a $u\times u$ matrix dramatically reduces the stopping redundancy for the perfect decoding. The operation of submatrix inversion in the proposed TS-AGD for cyclic MDS codes is always guaranteed by the following proposition.

\begin{proposition}[The nonsingularity of parity check matrix of cyclic MDS codes]
\label{Nonsingular_block}
For any square submatrix of the modified parity check matrix for MDS codes is nonsingular.
\end{proposition}
\begin{IEEEproof}
It can be proved by Theorem 8 in Chapter 11.4 in \cite{MacWil}.
\end{IEEEproof}

Thus, Algorithm \ref{TwostAGD} becomes Algorithm \ref{TSAGD_Mtx_inv_lb} for the perfect decoding by TS-AGD with expanded parity check matrix and submatrix inversion for cyclic MDS codes. In Algorithm \ref{TSAGD_Mtx_inv_lb}, the $u$ elements of the syndrome vector with indices $i_{j_1}, i_{j_2},..., i_{j_u},$ where for $k\in[1,u]$, $j_k$ is in the $S_e\cap \bar{S}_p$ can be computed as
\begin{displaymath}
s_{i_{j_k}}=e_{j_1}h_{i_{j_k}, j_1}+e_{j_2}h_{i_{j_k}, j_2}+...+e_{j_u}h_{i_{j_k}, j_u}+a_{i_{j_k}}=0, \text{for } k\in[1,u]
\end{displaymath}
where $a_{i_{j_k}}$ denotes the symbols recovered by the received codeword and the parity check matrix in columns whose indices are not in $S_e \cap \bar{S}_p$. By solving the above system of linear equations by submatrix inversion, the erasure symbols $e_{j_1}, e_{j_2},...,e_{j_u}$ can be recovered. Then, the remaining erasure symbols are decoded by the inversionless VNU.

Three lower bounds on $b$ for the perfect decoding by TS-AGD with expanded parity check matrix and submatrix inversion for the cyclic MDS codes are derived.

\begin{algorithm}
\caption{TS-AGD for the expanded parity check matrix with submatrix inversion}
\begin{algorithmic} 
\label{TSAGD_Mtx_inv_lb}
\REQUIRE $b(n-k)\times n$ parity check matrix $H$, parity check sequences $s_{p,i}(t)$, erasure sequence $s_e(t)$

\FOR {$i=0$ \TO $u$}
\FOR {$j=0$ \TO $b$}
\FOR {$\tau=0$ \TO $n-1$}
\IF{$\sum_{t=0}^{n-1}{s_e(t+\tau) s_{p,j}(t)}=i$}
\IF{$i \le 1$}
\STATE Follow Algorithm \ref{TwostAGD} for cyclic MDS codes
\STATE STOP
\ELSE
\STATE Select columns of $H$ with indices in $S_e \cap \bar{S_p}$
\STATE Select $|S_e \cap \bar{S_p}|$ rows whose indices are indices of $``1"$ in the $j$-th column of the standard basis vector, $j\in \bar{S_e}\cap S_p$
\STATE Invert $|S_e \cap \bar{S_p}| \times |S_e \cap \bar{S_p}|$ submatrix
\STATE Find erasure symbols with indices in $S_e \cap \bar{S_p}$
\STATE Decode the other $|S_e \cap S_p|$ erasure symbols by additional iterations without inversion
\STATE STOP
\ENDIF
\ENDIF
\ENDFOR
\ENDFOR
\ENDFOR
\end{algorithmic}
\end{algorithm}

\subsubsection{Bonferroni Inequality for TS-AGD With Expanded Parity Check Matrix and Submatrix Inversion for Cyclic MDS Codes}
The Bonferroni inequality in (\ref{Bonferroni_ineq}) can be modified as in the following theorems.

\begin{theorem}[The first term of the Bonferroni inequality with submatrix inversion]
The first term in (\ref{Bonferroni_ineq}) is modified in the expanded parity check matrix with submatrix inversion as
\begin{displaymath}
{\sum_{I\subset V, |I|=1}|E_i|}=bn\sum_{i=0}^{u}{\binom{n-k}{i}\binom{k}{i}}.
\end{displaymath}
\begin{IEEEproof}
Suppose that the expanded parity check matrix has $b$ parity check sequences. The $\tau$-shifted parity check sequence $s_p(t+\tau)$ can correct $n-k$ erasure symbols if $R_H(\tau)\le u$. If $R_H(\tau)=i$, $n-k$ erasure symbols are in the $n-k-i$ standard basis indices and the decoder can correct ${n-k \choose n-k-i}{k \choose i}$ erasure patterns. The number of decodable erasure patterns is the sum of all $i\in[0,u]$, which proves the theorem.
\end{IEEEproof}
\end{theorem}

\begin{theorem}[The second term in the Bonferroni inequality with submatrix inversion]
\label{ST_EPCM_MI_lb}
The second term can also be modified in (\ref{Bonferroni_ineq}) in the expanded parity check matrix with submatrix inversion as
\begin{displaymath}
\begin{split}
\label{(31,15)_Optimal_lb}
{\sum_{I\subset V, |I|=2}\left|\bigcap_{i \in I}{E_i}\right|} = \sum_{1 \le i < j \le b}{\sum_{\mu=0}^{u-1}{\sum_{\tau_1,\tau_2=0}^{n-1}{\sum_{0\le \zeta+\eta_1\le u, 0\le \zeta+\eta_2\le u}{}}}}
{{k-2u+\mu \choose \zeta}{2u-\mu \choose \eta_1}} {2u-\mu \choose \eta_2}
\\{n-k-2u+\mu \choose n-k-\eta_1-\eta_2-\zeta} F_{s_{p,i}(t+\tau_1),s_{p,j}(t+\tau_2)}(k-2u+\mu)
\end{split}
\end{displaymath} 
where $F_{s_{p,i}(t+\tau_1),s_{p,j}(t+\tau_2)}(\gamma)$ returns 1 if $\sum_{t=0}^{n-1}(s_{p,i}(t+\tau_1)s_{p,j}(t+\tau_2))=\gamma$ and otherwise, 0.
\begin{IEEEproof}
The proof is the generalization of that of Theorem \ref{ST_EPCM_lb}. For the $i$-th and the $j$-th parity check sequences cyclically shifted by $\tau_1$ and $\tau_2$, the function $F_{s_{p,i}(t),s_{p,j}(t+\tau_2)}(\gamma)$ is computed as follows. If $\gamma=k-2u+\mu$ for $\mu\in[0,u]$, we have $|A_{11}|=k-2u+\mu$, $|A_{10}|=|A_{01}|=2u-\mu$, and $|A_{00}|=n-k-2u+\mu$. Let $\zeta$, $\eta_1$, and $\eta_2$ be the numbers of erasure symbols in $A_{00}$, $A_{10}$, and $A_{01}$. To decode the received codeword in two parity check sequences, the Hamming correlation of each parity check sequences is less than or equal to $u$, where $\zeta+\eta_1 \le u$ and $\zeta+\eta_2 \le u$. This provides the proof.
\end{IEEEproof}
\end{theorem}

\subsubsection{Lower Bounds of the Stopping Redundancy for TS-AGD in an Expanded Parity Check Matrix With Submatrix Inversion}

The three lower bounds on $b$ for TS-AGD with expanded parity check matrix and $u\times u$ submatrix inversion for the cyclic MDS codes can be modified as in the following theorems.
\begin{theorem}[Gilbert-like lower bound of TS-AGD with expanded parity check matrix and submatrix inversion]
\label{Gilbert_Inv_lb}
\begin{displaymath}
\rho \ge \left \lceil \frac{\binom{n}{n-k}}{n\sum_{i=0}^{u}{\binom{n-k}{i}\binom{k}{i}}} \right \rceil (n-k)
\end{displaymath}
\end{theorem}
\begin{IEEEproof}
It manifests from Theorem \ref{Gil_Lik_LB}.
\end{IEEEproof}

\begin{theorem}[Lower bound by the lotto design for the TS-AGD with expanded parity check matrix and submatrix inversion]
\label{Lottery_Inv_lb}
\begin{displaymath}
\rho \ge \left \lceil \frac{L(n,n-k,n-k,n-k-u)}{n} \right \rceil (n-k).
\end{displaymath}
\end{theorem}
\begin{IEEEproof}
It manifests from Theorem \ref{lotto_LB}.
\end{IEEEproof}

\begin{theorem}[Lower bound by the Bonferroni inequality for the TS-AGD with expanded parity check matrix and submatrix inversion]
\label{Impv_bound_Bonferroni_Inv}

\begin{displaymath}
\rho \ge \left\lceil\frac{{n \choose k}-{2u \choose u}^2A(n,4u+2,n-k)}{n(\sum_{i=0}^{u}{{n-k \choose i}{k \choose i}-{2u \choose u}^2})}\right\rceil(n-k)
\end{displaymath}
where $A(n,d,w)$ is the maximum number of codewords for $(n,d,w)$ constant weight codes.
\end{theorem}

\begin{IEEEproof}
The proof is the generalization of that of Theorem \ref{Impv_bound_Bonferroni}. Two parity check sequences that have Hamming correlation less than $k-2u$ have no doubly counted decodable erasure patterns, because two parity check sequences cannot be simultaneously decoded regardless of their locations of erasure symbols for $|A_{00}|\le n-k-2u-1$. For $|A_{00}|= n-k-2u$, the doubly counted decodable erasure patterns exist only when $u$ erasure symbols are located in $A_{10}$ and $A_{01}$, respectively, where $|A_{10}|=|A_{01}|=2u$. Then, the number of cases is ${2n \choose n}^2$. The remaining part is similar to the proof of Theorem \ref{Impv_bound_Bonferroni}.
\end{IEEEproof}

\section{Conclusion}
\label{Conc}
In this paper, TS-AGD algorithms for cyclic binary and cyclic MDS codes are proposed by modifying and expanding the parity check matrix. Modification criteria of the parity check matrix are proposed and the proposed TS-AGD algorithms are shown to be able to reduce the average number of iterations and the decoding complexity. The perfect codes, BCH codes, and MDS codes are considered for the proposed TS-AGD algorithms, where some of them achieve the perfect decoding. For the MDS codes, the modified decoding algorithm with expanded parity check matrix and submatrix inversion for perfect decoding is discussed. It is shown that some cyclic codes achieve the perfect decoding by the proposed TS-AGD with the expanded parity check matrix and submatrix inversion.

\appendices

\section*{Appendix A : Proof of Maximization of the Upper Bounds in (\ref{Tau_0_pf_eqn}) and (\ref{Tau_1_pf_eqn})}
The objective functions to be minimized are as follows:

\begin{enumerate}
\item For $R_H(\tau)=0$, the objective function is $\sum_{\tau_1,\tau_2}{2|S_p|+a(\tau_1,\tau_2)-n \choose |S_e|}$.
\item For $R_H(\tau)=1$, the objective function is
\begin{multline}
\label{initial_tau_1_eqn}
\sum_{\tau_1,\tau_2}{} {n-|S_p|-a(\tau_1,\tau_2) \choose 1}^2{2|S_p|+a(\tau_1,\tau_2)-n \choose |S_e|-2}\\+{a(\tau_1,\tau_2) \choose 1} {2|S_p|+a(\tau_1,\tau_2)-n \choose |S_e|-1}.
\end{multline}
\end{enumerate}
It is easy to check that the following constraints are used for optimization:
\begin{enumerate}[label=(\roman*)]
\item $\text{For all } \tau_1\text{ and }\tau_2,  0 \le a(\tau_1,\tau_2) \le n-|S_p|$.
\item $\text{For any $\tau_2$}$, $\sum_{\tau_1=0}^{n-1}{ a(\tau_1, \tau_2) }=(n-|S_p|)^2$.
\item $\text{For any $\tau$}, a(\tau, \tau) = n-|S_p|$.
\item $|S_e| \le |S_p|$.
\end{enumerate}
Let $g(x,y)$ be a function defined by
\begin{displaymath}
g(x,y) = \left\{ \begin{array}{ll}
\prod_{i=0}^{y-1}{\frac{x-i}{i+1}}, & \textrm{if $x\ge y+1$}\\
0, & \textrm{otherwise}
\end{array} \right.
\end{displaymath}
where $x$ and $y$ are real numbers. In fact, we have that $g(x,y)={x \choose y}$ for $x,y\in Z^+$. It is easy to check that $g(x,y)$ is a convex function. First, the objective function for $R_H(\tau)=0$ is convex because $g(2|S_p|-n+a(\tau_1,\tau_2), |S_e|)={2|S_p|-n+a(\tau_1,\tau_2) \choose |S_e|}$.

At this point, we will prove that the objective function for $R_H(\tau)=1$ is convex for $\frac{1}{9}<\frac{|S_e|}{|S_p|}\le 1$ and $|S_e|\ge 3$ because if $\frac{|S_e|}{|S_p|} \le \frac{1}{9}$ or $|S_e|\le 2$, most of the received codewords can successfully be decoded by the proposed TS-AGD. Clearly, the convexity of (\ref{initial_tau_1_eqn}) can be proved by the convexity of summands. Then, the summand of (\ref{initial_tau_1_eqn}) can be rewritten as 
\begin{multline}
\label{tedious_eqn_bf}
a(\tau_1,\tau_2)g(2|S_p|-n+a(\tau_1,\tau_2),|S_e|-1) +
 (n-|S_p|-a(\tau_1,\tau_2))^2 g(2|S_p|-n+a(\tau_1,\tau_2),|S_e|-2).
\end{multline}

\begin{figure}[!t]
\centering
\includegraphics[width=4in]{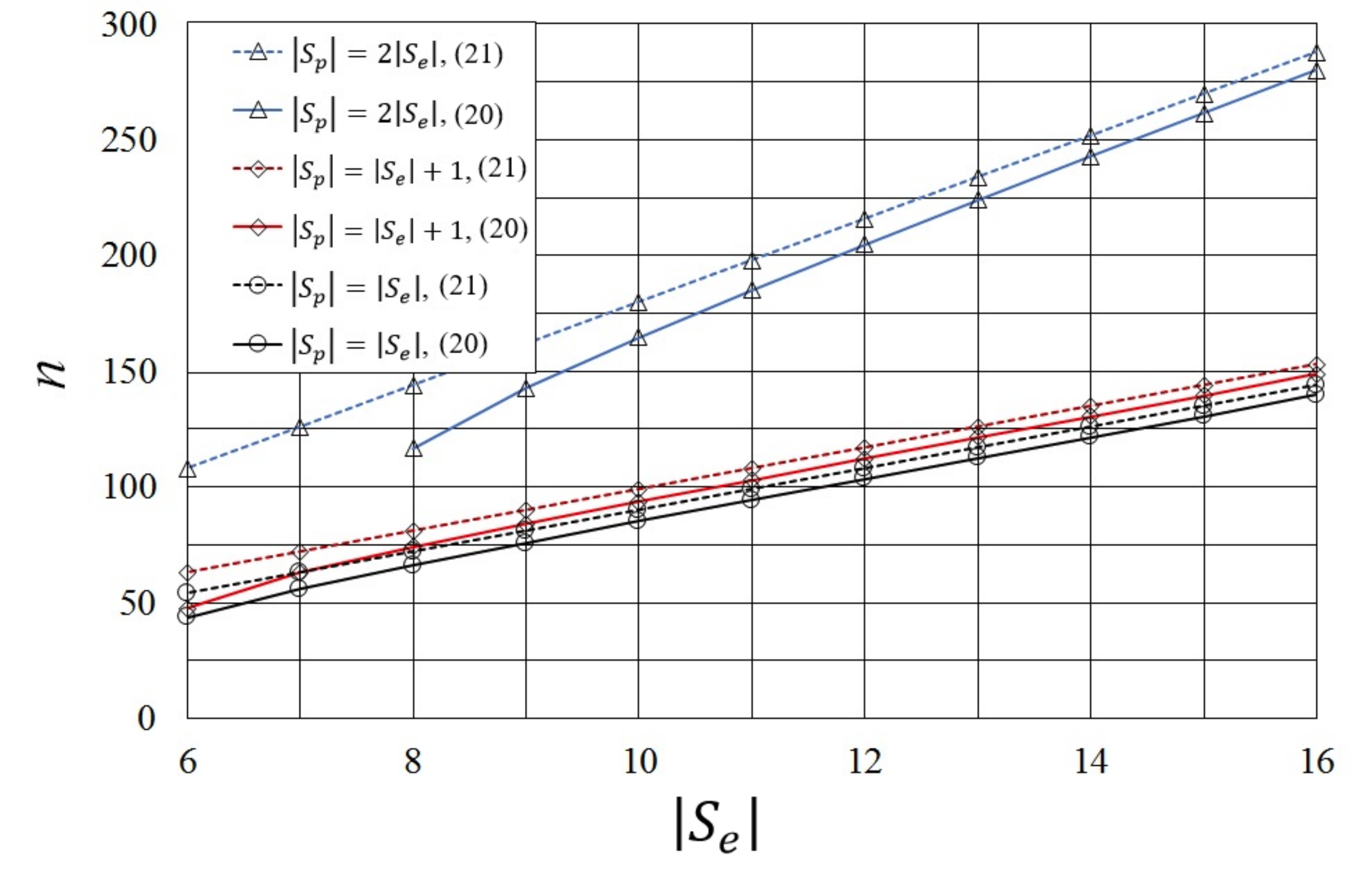}
\caption{The upper bound of  the convexity region by the inequalities (\ref{conv_reg_real}) and (\ref{conv_reg_asymp}) when $6\le |S_e| \le16$.}
\label{TS_AGD_Proof_Tau_1_Guaranteed_Lb}
\end{figure}

Using $g(x,y) = \frac{x-y+1}{y}g(x,y-1)$ for $x\ge y-1$, (\ref{tedious_eqn_bf}) can be modified as 
\begin{multline}
\label{tedious_eqn}
(a(\tau_1,\tau_2)(2|S_p|+a(\tau_1,\tau_2)-n-|S_e|+2) + 
\\(|S_e|-1)(n-|S_p|-a(\tau_1,\tau_2))^2)g(2|S_p|-n+a(\tau_1,\tau_2),|S_e|-2).
\end{multline}
The convexity of (\ref{tedious_eqn}) can be proved by its second derivative. Let 
\begin{displaymath}
\begin{split}
f(a)=a(\tau_1,\tau_2)(2|S_p|+a(\tau_1,\tau_2)-n-|S_e|+2) 
+ (|S_e|-1)(n-|S_p|-a(\tau_1,\tau_2))^2.
\end{split}
\end{displaymath}
Then, (\ref{tedious_eqn}) can be expressed as the product of $f$ and $g$. Then the convexity of (\ref{tedious_eqn}) can be proved by deriving the following inequality
\begin{displaymath}
\label{TS_AGD_Proof_convexity}
(fg)''=f''g+2f'g'+fg''\ge 0.
\end{displaymath}
It is not difficult to derive the $s$-derivative of $g(x,y)$ in terms of $x$ as
\begin{displaymath}
g^{(s)}(x,y)=\sum_{S, |S|=s}{\prod_{i\in{[0,y-1]\setminus[S]}}{\frac{x-i}{i+1}}}.
\end{displaymath} 
Using the geometric-harmonic mean inequality
\begin{displaymath}
\label{TS_AGD_Proof_Simp}
{(x_1x_2...x_n)}^{\frac{1}{n}}\ge \frac{n}{\frac{1}{x_1}+\frac{1}{x_2}+...+\frac{1}{x_n}}
\end{displaymath}
with $x_i=(x-i+1)$ and $n=y$, we have

\begin{equation}
\label{deriv_eqn_1}
(g(x,y))^{\frac{1}{y}}\ge \frac{bg(x,y)}{g'(x,y)} 
\end{equation}
\begin{equation}
\label{deriv_eqn_2}
\frac{y}{(g(x,y))^{\frac{1}{y}}}g(x,y)\le  g'(x,y).
\end{equation}
In general, $g^{(s)}(x,y)$ is the summation of polynomials factored into $y-s+1$ polynomials of degree one. Using (\ref{deriv_eqn_1}) and (\ref{deriv_eqn_2}), (\ref{tedious_eqn}) can be modified as

\begin{equation}
\label{TS_AGD_Proof_Simp_2}
\frac{b-s+1}{(g(a,b-s+1))^{\frac{1}{b-s+1}}} g^{(s-1)}(a,b) \le  g^{(s)}(a,b).
\end{equation}
Using (\ref{TS_AGD_Proof_Simp_2}), we have
\begin{displaymath}
\begin{split}
(fg)''=f''g+2f'g'+fg'' \ge 
\\  \frac{(|S_e|-3)^2}{g(2|S_p|-n+a(\tau_1,\tau_2),|S_e|-3)^{\frac{2}{|S_e|-3}}} f 
+ \frac{2(|S_e|-3)}{g(2|S_p|-n+a(\tau_1,\tau_2),|S_e|-3)^{\frac{1}{|S_e|-3}}}f'+f''g 
\\ \ge \bigg( \frac{(|S_e|-3)^2}{g(|S_p|,|S_e|-3)^{\frac{2}{|S_e|-3}}} f+\frac{2(|S_e|-3)}{g(|S_p|,|S_e|-3)^{\frac{1}{|S_e|-3}}}f'+f''\bigg) g.
\end{split}
\end{displaymath}
Let $w=\frac{2(|S_e|-3)}{g(|S_p|,|S_e|-3)^{\frac{1}{|S_e|-3}}}$. Then, it is enough to show that
\begin{equation}
\label{Simp_Eqn_for_Thm1}
{w}^2 f+2wf'+f''\ge0.
\end{equation}
It is easy to check that $w$ is an increasing function for $|S_e|$ and $\frac{|S_e|}{|S_p|}$ and  a decreasing function for $|S_p|$. Then, left hand side of (\ref{Simp_Eqn_for_Thm1}) can be rewritten as
\begin{displaymath}
\begin{split}
L(a)={w}^2\Big( (|S_e|-1)\left(n-|S_p|-a(\tau_1,\tau_2)\right)^2+
a(\tau_1,\tau_2)(-n+2|S_p|-|S_e|+a(\tau_1,\tau_2)+2)\Big)+
\\2w(-2n|S_e|+n+|S_e|(2|S_p|+2a(\tau_1,\tau_2)-1)+2)+2|S_e|.
\end{split}
\end{displaymath}
At this stage, it is necessary to prove that $L(0)>0$ and that its discriminant is negative in terms of $a$. It is easy to check that $L(a)$ is linear in terms of $|S_e|$ with a negative slope. Thus, $L(a)$ has its minimum value at the maximum value of $|S_e|$. If $|S_e|=|S_p|$, we have
\begin{multline}
L(0)={w}^2(|S_p|-1)(n-|S_p|)^2+{w}(n(2-4|S_p|)+
4|S_p|^2-2|S_p|+4)+2|S_p| \ge 0.
\label{SP_Important_1}
\end{multline}
Let $z=\frac{|S_p|}{n}$. Then for sufficiently large values of $n$ and $p$, (\ref{SP_Important_1}) can be written as
\begin{multline}
\label{Thm1_L_0_eqn}
\frac{L(0)}{n^2w}=w(|S_p|-1)(1-z)^2-4|S_p|+4{|S_p|}^2 
 \ge \left((w(|S_p|-1)+4)z-w(|S_p|-1)\right)(z-1) 
\\ =(w(|S_p|-1)+4)\left(z-\frac{w(|S_p|-1)}{w(|S_p|-1)+4}\right)(z-1).
\end{multline}
Clearly, (\ref{Thm1_L_0_eqn}) is positive for a sufficiently large $p$. Thus, we have $L(0)\ge0$. Next, the discriminant is written as
\begin{multline}
\label{conv_reg_real}
D={w}^4 n^2-10 {w}^4 n |S_p|+4 {w}^4 n+9 {w}^4 |S_p|^2
-4 {w}^4 |S_p|+4 {w}^4+8 {w}^2 |S_p|^2<0.
\end{multline}
It can also be reduced with sufficiently large values of $n$ and $p$, whose simplified inequality is given as
\begin{equation}
\label{conv_reg_asymp}
(9{w}^4+8{w}^2)z^2-10{w}^4z+{w}^4<0.
\end{equation}
For $\frac{1}{9}<z<1$, it is easy to derive $D<0$ for a large value of $w$. Fig. \ref{TS_AGD_Proof_Tau_1_Guaranteed_Lb} shows the upper bound of convexity region by (\ref{conv_reg_real}) and (\ref{conv_reg_asymp}), which shows that the two bounds become identical as $|S_e|$ becomes larger. Thus we prove the convexity of (\ref{tedious_eqn}) for the proposed convexity region.

Using the solution of the optimization program cvx for (\ref{tedious_eqn}), its minimum value occurs at
\begin{displaymath}
 a(\tau_1,\tau_2) = \frac{(n-|S_e|+1)^2-n-|S_e|+1}{n-1} \text{ }\text{for all } \tau_1 \text{ and } \tau_2,
\end{displaymath}
which means that the out-of-phase autocorrelation values of $s_p(t)$ are constant.

\section*{Acknowledgment}

The authors would like to thank...

\ifCLASSOPTIONcaptionsoff
  \newpage
\fi




\begin{thebibliography}{99}
\bibitem{RC}
A. G. Dimakis, P. B. Godfrey, Y. Wu, M. Wainwright, and K. Ramchandran, ``Network coding for distributed storage systems,'' \emph{IEEE Trans. Inf. Theory}, vol. 56, no. 9, pp. 4539-4551, Sep. 2010.
\bibitem{LRC}
D. S. Papailiopoulos and A. G. Dimakis, ``Locally repairable codes,'' \emph{Proc. IEEE Int. Symp. Inf. Theory (ISIT)}, pp. 2771-2775, Jul. 2012.
\bibitem{OSMD}
I. S. Reed, ``A class of multiple-error-correcting codes and the decoding scheme,'' \emph{IRE Trans. Inform. Theory}, vol. IT-4, pp. 38-49, Sep. 1954.
\bibitem{PD}
F. J. MacWilliams, ``Permutation decoding of systematic codes,'' \emph{Bell Syst. Tech. J.,} vol. 43, pp. 485-505, 1964.
\bibitem{RS1}
J. Bellorado and A. Kavcic, ``Low-complexity soft-decoding algorithms for Reed-Solomon codes Part I: An algebraic soft-in hard-out Chase decoder,'' \emph{IEEE Trans. Inf. Theory}, vol. 56, no. 3, pp. 945-959, Mar. 2010.
\bibitem{RS2}
J. Bellorado, A. Kavcic, M. Marrow, and L. Ping, ``Low-complexity soft-decoding algorithms for Reed-Solomon codes Part II: Soft-input soft-output iterative decoding,'' \emph{IEEE Trans. Inf. Theory,} vol. 56, no. 3, pp. 960-967, Mar. 2010.
\bibitem{Hollmann}
H. D. L. Hollmann and L. M. G. M. Tolhuizen, ``On parity-check collections for iterative erasure decoding that correct all correctable erasure patterns of a given size,'' \emph{IEEE Trans. Inf. Theory,} vol. 53, no. 2, pp. 823-828, Jan. 2007.
\bibitem{Hehn} 
T. Hehn, O. Milenkovic, S. Laendner, and J. Huber, ``Permutation decoding and the stopping redundancy hierarchy of cyclic and extended cyclic codes,'' \emph{IEEE Trans. Inf. Theory,} vol. 54, no. 12, pp. 5308-5331, Dec. 2008.
\bibitem{Hehn2}
T. Hehn, J. B. Huber, O. Milenkovic, and S. Laendner, ``Multiple-bases belief-propagation decoding of high-density cyclic codes,'' \emph{IEEE Trans. Commun.,} vol. 58, no. 1, pp. 1-8, Jan. 2010.
\bibitem{Chen}
C. Chen, B. Bai, X. Yang, L Li, and Y Yang, ``Enhancing iterative decoding of cyclic LDPC codes using their automorphism groups,'' \emph{IEEE Trans. Commun.,}  vol. 61, no. 6, pp. 2128-2137, Apr. 2013.
\bibitem{Liu}
K. Liu, S. Lin, and K. Abdel-Ghaffar, ``A revolving iterative algorithm for decoding algebraic cyclic and quasi-cyclic LDPC codes,'' \emph{IEEE Trans. Commun.,}, vol. 61, no. 12, pp. 4816-4827, Dec. 2013.
\bibitem{SR_first}
M. Schwartz and A. Vardy, ``On the stopping distance and the stopping redundancy of codes,'' \emph{IEEE Trans. Inf. Theory}, vol. 52, no. 3, pp.922-932, Mar. 2006.
\bibitem{SR1}
T. Etzion, ``On the stopping redundancy of Reed-Muller codes,'' \emph{IEEE Trans. Inf. Theory,} vol. 52, no. 11, pp. 4867-4879, Sep. 2006.
\bibitem{SR2}
J. Han, P. H. Siegel, and R. M. Roth, ``Single-exclusion number and the stopping redundancy of MDS codes,'' \emph{IEEE Trans. Inf. Theory}, vol. 55, no. 9, pp. 4155-4166, Sep. 2009.
\bibitem{SR3}
J. Zhang, F. W. Fu, and D. Wan, ``Stopping sets of algebraic geometry codes,'' \emph{IEEE Trans. Inf. Theory,} vol. 60, no. 3, pp. 1488-1495, Mar. 2014.
\bibitem{Bonferroni}
K. Dohmen, \emph{Improved Bonferroni Inequalities via Abstract Tubes.}, Berlin, Germany: Springer-Verlag, 2003.
\bibitem{Tamo}
I. Tamo, A. Barg, S. Goparaju, and R. Calderbank, ``Cyclic LRC codes and their subfield subcode," \emph{IEEE Int. Symp. Inf. Theory (ISIT)}, pp. 1262-1266, Jun. 2015.
\bibitem{Colbourn}
C. J. Colbourn and J. H. Dinitz, \emph{Handbook of Combinatorial Designs.}, New York, NY, USA: CRC Press, 2006.
\bibitem{Niho}
Y. Niho, ``Multivalued cross-correlation functions between two maximal linear recursive sequence,'' Ph.D. dissertation, Univ. Southern Calif., Los Angeles, 1970.
\bibitem{Gold}
R. Gold, ``Maximal recursive sequences with 3-valued recursive cross-correlation functions (corresp.),'' \emph{IEEE Trans. Inform. Theory}, vol. 14, no. 1, pp. 154-156, 1968.
\bibitem{Kasami}
T. Kasami, ``The weight enumerators for several classes of subcodes of the 2nd order binary Reed-Muller codes,'' \emph{Inf. Control}, vol. 18, no. 4, pp. 369-394, 1971.
\bibitem{Welch}
A. Canteanut, P. Charpin, and H. Dobbertin, ``Binary $m$-sequences with three-valued crosscorrelation: a proof of Welch's conjecture,'' \emph{IEEE Trans. Inform. Theory}, vol. 46, no. 1, pp. 4-8, 2000.
\bibitem{Boston}
N. Boston and G. McGuire, ``The weight distributions of cyclic codes with two zeros and zeta functions,'' \emph{J. Symbolic Comput.}, vol. 45, no. 7, pp. 723-733, 2010.
\bibitem{Dowling}
T. A. Dowling and R. J. McEliece, ``Cross-correlation of reverse maximal-length shift register sequences,'' JPL Space Programs Summary 37-53, vol. 3, pp.192-193, 1968.
\bibitem{CW_Upper}
E. Agrell, A. Vardy, and K. Zeger, ``Upper bounds for constant-weight codes,'' \emph{IEEE Trans. Inf. Theory,} vol. 46, no. 7, pp. 2373-2395, Nov. 2000.
\bibitem{LottoTable}
P. C. Li and G. H. J. Van Rees, "Lotto design tables," \emph{Journal of Combinatorial Designs}, vol. 10, no. 5, pp.335-359, Aug. 2002.
\bibitem{Mackay}
D. Mackay, ``Fountain codes,'' \emph{IEE Proceedings-Communications}, vol. 152, no. 6, pp. 1062-1068, Dec. 2005.
\bibitem{Rashimi}
K. V. Rashmi, N. B. Shah, and P. V. Kumar, ``Optimal exact-regenerating codes for distributed storage at the MSR and MBR points via a product-matrix construction,''  \emph{IEEE Trans. Inform. Theory}, vol. 57, no. 8, pp. 5227-5239, Jul. 2011.
\bibitem{MacWil}
F. J. MacWilliams and N. J. A. Sloane, \emph{The Theory of Error Correcting Codes.}, New York, NY, USA: North-Holland, 1977.
\bibitem{McGregor}
A. McGregor and O. Milenkovic, ``On the hardness of approximating stopping and trapping sets,''  \emph{IEEE Trans. Inform. Theory}, vol. 56, no. 4, pp. 1640-1650, Mar. 2010.
\end{thebibliography}
%

\begin{IEEEbiography}{Chanki Kim}
Biography text here.
\end{IEEEbiography}

\begin{IEEEbiography}{Jong-Seon No}
Biography text here.
\end{IEEEbiography}





\end{document}